# Biofabrication for neural tissue engineering applications

L. Papadimitriou[a,c], P. Manganas[a,c], A. Ranella[a*] and E. Stratakis[a,b*]

[a] Institute of Electronic Structure and Laser (IESL), Foundation for Research and Technology-Hellas (FORTH), Heraklion, 71003, Greece

[b] Physics Department, University of Crete, Heraklion, 71003 Crete, Greece

[c] These authors contributed equally to this work.

[*] Corresponding authors: ranthi@iesl.forth.gr, stratak@iesl.forth.gr

## Abstract

Unlike other tissue types, the nervous tissue extends to a wide and complex environment that provides a plurality of different biochemical and topological stimuli which in turn define the advanced functions of that tissue. As a consequence of such complexity, the traditional transplantation therapeutic methods are quite ineffective; therefore, the restoration of peripheral and central nervous system injuries has been a continuous scientific challenge. Tissue engineering and regenerative medicine in the nervous system have provided new alternative medical approaches. These methods use external biomaterial supports, known as scaffolds, in order to create platforms for the cells to migrate to the injury site and repair the tissue. The challenge in neural tissue engineering (NTE) remains the fabrication of scaffolds with precisely controlled, tunable topography, biochemical cues and surface energy, capable of directing and controlling the function of neuronal cells towards the recovery from neurological disorders and injuries. At the same time, it has been shown that neural tissue engineering provides the potential to model neurological diseases *in vitro*, mainly via lab-on-a-chip systems, especially in cases for which it is difficult to obtain suitable animal models. As a consequence of the intense research activity in the field, a variety of synthetic approaches and 3D fabrication methods have been developed for the fabrication of NTE scaffolds, including soft lithography and self-assembly, as well as subtractive (top-down) and additive (bottom-up) manufacturing. This article aims at reviewing the existing research effort in the rapidly growing field related to the development of biomaterial scaffolds and lab-on-a-chip systems for NTE applications. Besides presenting recent advances achieved by NTE strategies, this work also delineates existing limitations and highlights emerging possibilities and future prospects in this field.

## Table of Contents

## 1. Introduction

The nervous tissue consists of the central and the peripheral nervous systems (CNS and PNS) and is the most complex system in the body. Injuries to the human nervous system affect more than 1 billion people around the world, with 6.8 million dying as a result of them each year [1], and have been associated with a wide variety of disorders including neurodegenerative diseases, as well as brain and spinal cord traumatic injuries and stroke [2]. The central nervous tissue does not regenerate under normal conditions and, to date, there is no treatment modality with clinically documented efficacy to actively improve CNS repair. Current medical approaches focus primarily on stabilization and prevention, e.g. orthopedic fixation of an unstable spine, and consequently on rehabilitation and the preparation of prosthetics. On the contrary, the management of a PNS injury is much simpler. The currently applied treatments involve nerve autografts and allografts, however there are many difficulties, including shortage of donor nerves, donor site morbidity, aberrant regeneration, infectious diseases, as well as immunological issues [3]. It is therefore understood that there is a vital need for engineered alternatives to autograft application [4].

In view of the ineffectiveness of current therapeutic methods, the restoration of the damaged PNS and CNS has been a continuous challenge for neurologists and neurobiologists. As a result, novel treatment strategies for the injured nervous system have been pursued. Tissue engineering and regenerative medicine in the nervous system have provided new medical approaches as alternatives to traditional transplantation methods. These methods use external biomaterial supports, known as scaffolds, in order to create a platform for the cells to migrate to the injury site and repair the tissue. 3D scaffold models have been found to be critical for mimicking the exact microcellular environment and cell–cell interactions. Pioneering works in this domain have identified and characterized 3D matrices as vital for cell anchorage and precise replication of the cellular microenvironment and have enabled the creation of living tissues from a source of cells. Such scaffolds are often loaded with cells and/or growth factors to hasten the differentiation of cells to preferred types of lineage in order to promote new tissue formation.

In recent years, neural tissue engineering (NTE) has significantly contributed to the research efforts devoted towards the identification of suitable strategies for the recovery from neurological disorders and injuries. Based on these efforts, it has been realized that the nervous tissue is undoubtedly the most complex system of the human anatomy, comprising a complex multilayer environment whose topographical features display a large spectrum of morphologies and size scales. As a result, it demands far more intricate tissue-engineered scaffolds and architectures [5, 6]. At the same time, the physicochemical characteristics of NTE constructs are critical for neuronal cell function and viability. Besides this, it has been shown that NTE provides the potential to model neurological diseases *in vitro*, mainly via lab-on-chip systems, especially in cases for which it is difficult to obtain suitable animal models. As a consequence of the intense research activity in the field, a variety of synthetic approaches and 3D fabrication methods have been developed for the fabrication of NTE scaffolds, including soft lithography and self-assembly, as well as top-down (subtractive) and bottom-up (additive) manufacturing.

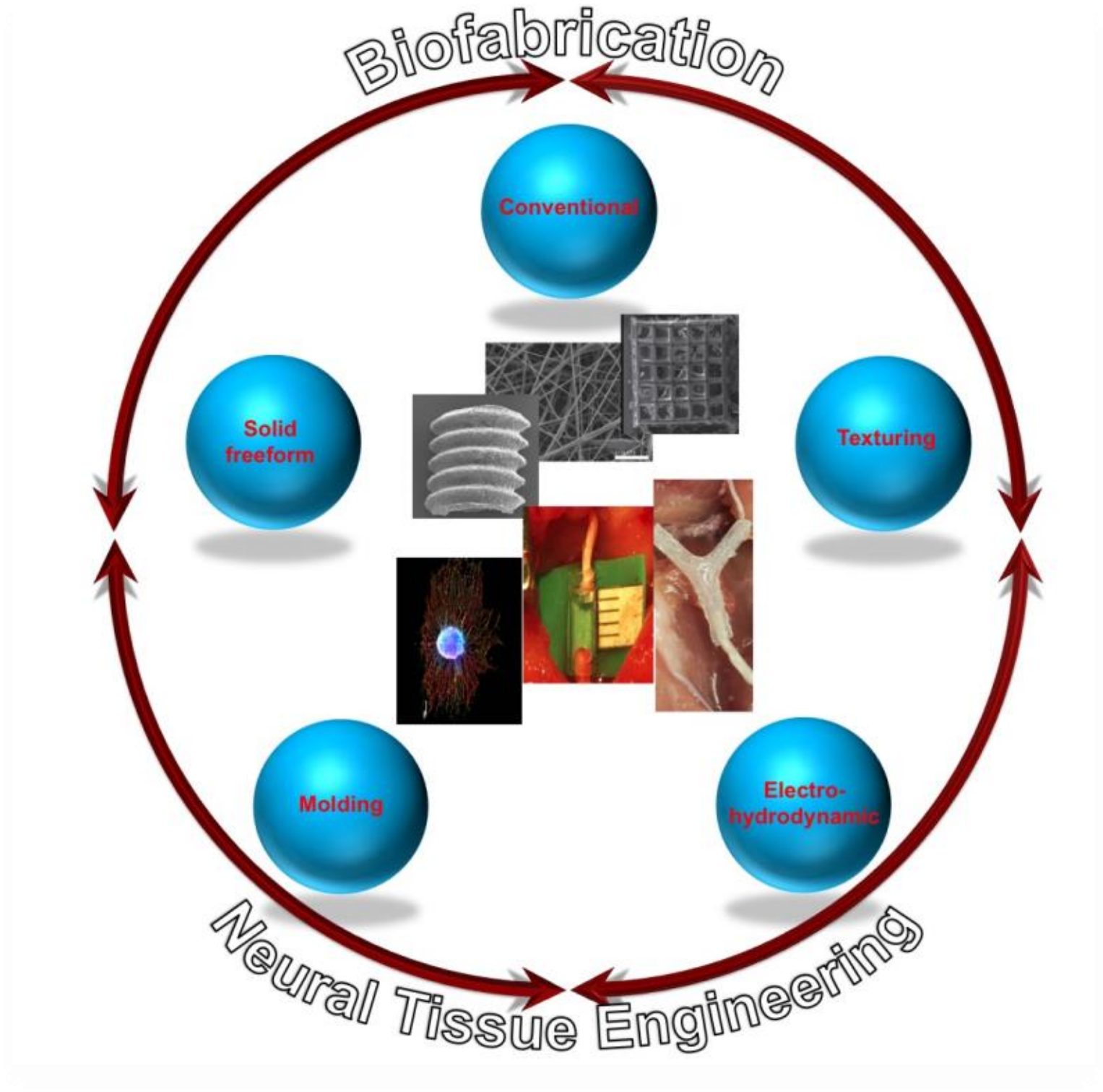

**Figure 1:** The main biofabrication methods for the development of biomaterial scaffolds for neural tissue engineering applications.

The present review article aims at reviewing the existing body of literature in the rapidly growing field related to the development of biomaterial scaffolds and lab-on-chip systems for NTE applications (Fig. 1). In particular, the first part reviews the various methods used for the fabrication of NTE constructs of various sizes, shapes and physicochemical characteristics. The second part is devoted to the most important reports employed to address CNS and PNS injury issues, via the application of NTE scaffolds and lab-on-chip systems. Besides presenting recent advances achieved by NTE strategies, this work also delineates existing limitations and highlights emerging possibilities and future prospects in this field.

## 2. Biofabrication of neural tissue engineering components

The creation of tissue engineering components, including scaffolds and integrated systems, has gone through a lengthy process, with several techniques being developed in order to create complex structures from a variety of natural and synthetic components. From the beginning, the aim has been the fabrication of ideal scaffolds to match the physical, chemical and mechanical properties of the tissue. These scaffolds should bear several desirable characteristics such as pores, fibers and channels, which are crucial for applications in NTE.

In this section, we will present the most important biofabrication methods that have been used for the creation of NTE constructs and explain the principles on which they are based [7, 8]. Each of these methods is able to produce a different range of shapes and structures to fulfil the needs and the requirements of the corresponding application and has its own advantages and disadvantages. With the term "Biofabrication", we refer to all methods used to produce constructs with biological function for use in tissue engineering and regenerative medicine applications [8]. As such, we have started from the older and more "traditional" conventional methods and continued with the advanced technologies that have been developed over the last few decades. Table 1 summarizes the characteristics of the fabrication methods that have been applied in NTE, the structural characteristics of the resulting scaffolds, the advantages and the disadvantages of each method, as well as their potential impact in the future of NTE.

**Table 1:** Methods of fabrication of tissue engineering systems, particularly applied for NTE

| Techniques | Structural characteristics | Advantages | Disadvantages | Potential Impact in NTE |
|---|---|---|---|---|
| **Conventional methods** | | | | |
| Solvent casting / Particulate leaching [9] | Through the control of the amount of porogen added, as well as its size and shape, these scaffolds usually have an average pore size of ~500μm with ~95% porosity | - Simple, easy, inexpensive<br>- Does not require large equipment<br>- Controllable pore size<br>- Less amount of polymer required for scaffold creation | - Scaffolds may retain some toxicity<br>- Time consuming<br>- Pore shape and inter-pore openings cannot be controlled<br>- Can only be used to produce thin membranes (≤3mm thickness) | Fabrication of biocompatible scaffolds for peripheral nerve injury repair, in combination with molding techniques |
| Phase separation (Nonsolvent induced phase separation NIPS [10] and Thermally induced phase separation -TIPS [11]) | Porous scaffolds with possible integration of bioactive molecules | - Can produce porous scaffolds with integrated bioactive molecules, due to the low temperatures used during fabrication<br>- No decrease in the activity of the molecule<br>- Can be easily combined with other fabrication methods (ie particulate leaching, rapid prototyping) | - Difficult to control the precise scaffold morphology<br>- Limited material selection | Fabrication of scaffolds, in conjunction with molding techniques |
| Self-assembly [12] | Nanofibers with amino acid residues that can be modified by the addition of bioactive molecules | - Production of very thin nanofibers<br>- The nanofibers have amino acid residues that can be chemically modified by the addition of bioactive moieties<br>- Does not require organic solvents<br>- Reduction of cytotoxicity due to the use of aqueous salt solutions or physiological media | - Complicated and elaborate process<br>- Poor mechanical stability makes it difficult to create stable 3D structures<br>- The engineered nanofibers can be fragmented and are susceptible to endocytosis<br>- High cost of synthesis | Formation of injectable materials for nerve regeneration |
| Freeze-drying [13] | Porous scaffolds without the presence of potentially harmful solvents | - Controllable pore size through controlling the freezing rate and the pH<br>- Does not require high temperatures<br>- Does not require a separate leaching step | - Small pore sizes<br>- Long processing times<br>- Difficult to produce scaffolds with hierarchical structures (e.g. vascularized systems) | Design of scaffolds for nerve repair using biocompatible and biodegradable materials |
| Gas foaming [14] | Creation of porous scaffolds with pore sizes ranging from 100-500μm | - Does not require the use of organic solvents and high temperatures | - Limited mechanical property control | Fabrication of scaffolds, in conjunction with other techniques |
| | | - Controllable porosity dependent on the amount of gas dissolved in the polymer | - Possibility of creating a non-porous external surface | (i.e. molding, phase separation) |
| Hydrogel formation [15] | Hydrogels can have a range of different properties that depend on the type of polymeric material used and the method of crosslinking used | - Tunable mechanical strength<br>- Mostly biocompatible and biodegradable<br>- Can offer controllable drug release rates<br>- Incorporation of biological material (DNA, proteins, cells) | - Potential lack of biodegradability for some materials<br>- Difficulty with drug loading in cases of non-hydrophilic drugs<br>- Potential toxicity from unreacted crosslinking molecules | Fabrication of scaffolds using bioprinting methods and design of bioinks for brain delivery and regeneration |
| **Molding and texturing methods** | | | | |
| Compression molding / Injection molding [9] | Scaffolds with controllable porosity through the use of porogens with different sizes and chemical properties | - Control of pore size, interconnectivity and geometry<br>- Does not require organic solvents | - High temperatures required for non-amorphous polymers<br>- Possibility of residual porogen | Construction of in vitro brain models for drug screening and efficacy testing (such as lab-on-a-chip devices). Should be used in combination with other methods |
| Photolithography [16] | Scaffolds with details in the nano- and micrometer scale printed on photoresists | - Large scale patterning<br>- High resolution technique<br>- Non-contact manufacturing process | - High cost<br>- Limited control over surface properties<br>- Compatible with a limited number of materials | - |
| Soft lithography [17] | Scaffolds with details in the nano- and micrometer scale that have been transferred onto a range of polymers with different properties | - Wide range of materials<br>- Easy and straightforward process<br>- Low cost | - Can lose some of the detail during the stamp/mold creation process | Fabrication of components for microfluidic systems and lab-on-a-chip devices. Like compression molding, should be used in combination with other methods. |
| Laser texturing [18] | Structuring/texturing of the material surface can be localized without affecting the surrounding areas and is in the region of nano- and micrometers | - Local excitation of certain areas of a material, with minimal damage to the surrounding areas<br>- Non-contact fabrication method<br>- Can be used with a wide range of materials | - Requires specialized and expensive equipment | Precise and controllable micro-/nano-patterning of scaffolds for nerve regeneration |
| Fiber mesh / Fiber bonding [19] | Fibrous scaffolds with large surface areas | - Creation of scaffolds with large surface areas<br>- Mesh structure allows the rapid diffusion of nutrients<br>- Mechanical stability | - Lacks structural stability<br>- Poor mechanical property control<br>- Limited applications in other polymers (except for PGA, PLLA) | - |
| **Electrohydrodynamic techniques** | | | | |
| Electrospraying [20] | Highly charged droplets are formed. Their charge prevents their coagulation and | - The size distribution of the droplets is usually narrow,no droplet agglomeration and | - Requires specialized equipment<br>- Difficult to control the droplet's size | Fabrication of carriers for drugs and therapeutic molecules for brain delivery. |
| | promotes their self-dispersion. | coagulation are occurring<br>- The motion of charged droplets can be easily controlled<br>- Higher deposition efficiency of charged spray compared to un-charged droplets.<br>- Single-step processing | - Can induce macromolecule degradation | |
| Electrospinning [21] | Continuous micro- and nano-scale fibres from a rich variety of materials. By blending different polymers nanofibers with internal morphology and secondary structures e.g porous, hollow, or core–sheath structure can be fabricated. Also, fibers can be organized into ordered arrays or hierarchical structures by modulating their stacking, arrangement and folding. | - Easy to use, simple, versatile, efficient and ideal for large scale production<br>- high surface area to volume ratio structures<br>- large number of inter-/intra fibrous pores (high porosity)<br>- fabrication of fibers from various types of raw materials (from natural and synthetic polymers to composites, consisting of organic and inorganic components) leads to unlimited applications<br>- the ability to control many factors, such as the fiber diameter, orientation, and composition | - The requirement for specialized equipment (although it is inexpensive)<br>- The use of organic solvents<br>- The limited control of pore structures<br>- The process depends on many variables | Synthetic nerve conduits to facilitate axonal guidance and to enhance nerve regeneration. |
| **Solid freeform fabrication (SFF) / Rapid prototyping (RP)** | | | | |
| Photolithography-based techniques [22] | Precise internal architectures and external geometries, which match those of human tissue (structures with ≥50 µm features) | - Good mechanical Strength<br>- high spatial resolution<br>- Easy to achieve small features<br>- High degree of fabrication accuracy<br>- low printing time<br>- Bio-resins can be incorporated to create bioinks | - resins with cytotoxic residuals may be used<br>- Often during the processing supporting structures are required<br>- Specialized equipment is required<br>- high intensity UV light may be used | Fabrication of complex 3D-tissue structure with high resolution for brain regeneration as long as biocompatible hydrogels are used. |
| Selective laser sintering (SLS)/ Selective laser melting (SLM) [23] | Fabrication of complex geometries with intricate and controllable internal architectures | - Good mechanical strength<br>- High accuracy<br>- Broad range of materials that can be used<br>- Easy to create layered 3D structures, as new layers of powder can be layered on top of the previous sintered layers<br>- no supports required<br>- many commercial machine providers are existing | - Elevated temperatures<br>- Local high energy input<br>- Uncontrolled porosity<br>- Laser beam diameter (~400µm) and powder particle size limit the dimension of the generated scaffolds<br>- Difficulty in building specially shaped scaffolds with sharp corners or clear boundaries | - |
| | | | - Restricted by material properties (the material has to be available as a powder and the powder must have suitable melting and welding behaviour) | |
| Microsphere sintering (sub-category of sintering) [24] | Microspheres are fused together to create a single macroscopic unit, with complex shapes and architectures | - Can load biological material into the water droplets<br>- Suitable for obtaining composite structures from polymeric and inorganic substances<br>- customization (patient-specific) of the scaffolds<br>- freedom from toxic solvents | - very expensive, because large quantities of raw materials are needed<br>- requires specialized equipment<br>- microspheres may stick together or may not being spread evenly in successive layers | Fabrication of macroporous, 3D shape–specific constructs, conductive to infiltration and with controlled release of bio-active molecules for nerve regeneration. |
| Fused deposition modelling [25] | scaffolds with honeycomb-like pattern, fully interconnected channel network, and controllable porosity and channel size | - Low cost<br>- Good mechanical strength<br>- Versatile pattern design<br>- does not require any solvent | - Elevated temperatures<br>- Small range of bulk materials<br>- thermal degradation and spatial resolution | New biocompatible and biodegradable filament materials must be formulated in order to use FDM for nerve regeneration applications. |
| 3D bio-printing [26, 27] | Precise layering of cells, biologic scaffolds, and growth factors in order to create bioidentical tissue for a variety of uses. | - Broad range of materials and conditions<br>- Incorporation of cells and macromolecules<br>- accurate reproduction of tissue<br>- potential of an industry-scale robotic tissue-fabrication line | - Slow processing<br>- Time consuming<br>- high cost and size<br>- Low mechanical Strength<br>- Low process resolution<br>- Lack of full automation<br>- Cartridge and nozzle whenever they used, can negatively affect cell viability | Construction of brain-like structures to serve as *in vitro* 3D models and custom-made platforms for personalized medicine. |

## 2.1. Conventional methods

### 2.1.1. Solvent casting / Particulate leaching

Solvent casting/particulate leaching (Fig. 2a) is a popular, traditional method used for porous scaffold fabrication and is based on a relatively simple technique. A homogeneous polymer solution that contains some type of porogen is cast into a mold and the solvent is allowed to evaporate. The resulting composite material contains the polymer and the porogens. For final scaffold fabrication, this composite material is submerged in a bath so that the porogens can dissolve in order to reveal a porous structure that can be used in various tissue engineering applications [9].

A variety of different materials can be used as porogens, with salt, sugar and wax being the most common candidates [28-30]. Through this fabrication method, it is possible to create scaffolds with controlled porosity, as the method allows for optimisation both through the use of porogens of different type, shape and size as well as through the concentration of porogen added to the polymeric solution [31]. Additionally, by utilising various types of pre-treatments (such as pre-fusing porogen particles or using more than one type of porogen), scientists have been able to control the porosity even further and improve pore interconnectivity [32].

#### 2.1.2. Phase separation

Phase separation (Fig. 2b) can be split into two distinct sub-categories depending on the method of fabrication followed: nonsolvent-induced phase separation (NIPS) and thermally induced phase separation (TIPS).

NIPS requires three components: a polymer, a solvent and a non-solvent. A homogeneous polymer solution that contains at least one solvent is first cast on a suitable support structure, exposed to the air for a short amount of time and then immersed in a bath containing the non-solvent solution. The exchange between the solvent and the non-solvent leads to the formation of two phases that are created due to the precipitation of the polymer: the polymer-rich and the polymer-lean phase. As the polymer-rich phase becomes more enriched in the non-solvent, the porous structure is created [10, 33, 34]. Used mostly for the preparation of membranes, NIPS-generated scaffolds usually have heterogeneous pore structures and, as such, have limited applications in tissue engineering.

On the other hand, TIPS (Fig. 2b) can occur with solutions of polymer and solvent that are homogeneous at elevated temperatures. Through the cooling of these homogeneous solutions, they tend to separate into polymer-lean and polymer-rich phases. The latter phase solidifies, creating the scaffold matrix, while the polymer-lean phase creates the pores due to the removal of the solvent. TIPS can be further split into two categories: solid-liquid (S-L) phase separation and liquid-liquid (L-L) phase separation. S-L phase separation uses the lowering of the temperature to induce solvent crystallisation, with the removal of these crystals leading to pore formation. L-L phase separation takes advantage of an upper critical temperature where both phases (polymer-rich and polymer-lean) co-exist and the temperature and concentration at which their de-mixing occurs leads to the formation of the porous structure [9, 11]. More specifically, the separation between the two phases can occur due to binodal de-mixing and/or spinodal decomposition. Binodal de-mixing is characterised by nucleation and growth, takes place in the metastable region (between the binodal and spinodal curves) and tends to create a porous structure with a poorly interconnected network. On the other hand, spinodal decomposition takes place within the spinodal curve (unstable region) and leads to the formation of a well interconnected network. As a result, the way in which the phase separation happens and the thermodynamic region in which it occurs are of critical importance in the determination of the morphology attained [35].

### 2.1.3. Self-assembly

Self-assembly (Fig. 2c) is the autonomous organisation of well-defined components into ordered structures without external instruction. It has been widely utilised for the fabrication of various nanofibers using biological molecules and is based on the presence of both noncovalent and weak covalent interactions (e.g. van der Waals, hydrophobic interactions, hydrogen bonds) [36]. Though the individual interactions are quite weak, the fact that there are a large number of them leads to the formation of the assembled scaffold/structures. The process of self-assembly is directly inspired by nature, where viral proteins are able to self-assemble to create the viral capsids and phospholipids – which are naturally amphiphilic molecules – are able to assemble into different types of higher order structures, including lipid bilayer membranes, vesicles, micelles or even tubules.

Two types of natural materials are widely used in the self-assembly fabrication process: collagen and elastin. Both these molecules are extremely abundant *in vivo* since they are components of all connective tissues and the extracellular matrix (ECM). These natural molecules have been used as the basis for the design of new materials that are collagen- and elastin-like, as well as paved the way for the *de novo* design of synthetic peptides [12].

A large number of the molecules used in this fabrication method are amphiphilic peptides and can contain a variety of features essential for the self-assembly process to take place. Such examples are: 1) long hydrophobic tails that aggregate in aqueous solutions, 2) consecutive cysteine residues to enable the formation of disulphide bonds, 3) glycine residues to provide flexibility to the peptide, 4) phosphorylated serine residues to enable interactions with calcium ions and/or 5) RGD ligands to enhance cellular adhesion. The RGD sequence is a well-known bioactive motif found in ECM proteins such as fibronectin [37] and, as part of the design strategy, incorporation of this motif into peptide and protein scaffolds is used to convey cell adhesion properties.

Amphiphilic peptides usually start to self-assemble based on the adjustment of certain parameters of the solution they are in. The main parameters are the light, the pH, the temperature, the salt ion concentration of the solution and the presence/absence of a reducing/oxidizing agent [11, 12, 38]. Self-assembling peptides can also be modified in order to become tissue-specific through the integration of different signalling peptide sequences into the peptides or even through the addition of growth factors specific for each cellular environment [39]. Understanding all the different characteristics and elements that can be incorporated into these peptides gives researchers the ability to precisely design and finetune the morphological features of the scaffold they desire. Changing the aminoacidic composition of the self-assembling peptides enables the creation of a variety of different structures, including vesicles, micelles, mono- and bi-layers, fibers and tapes [40], while also allowing the more precise mimicking of the extracellular environment. Furthermore, recent developments have proposed that novel computationally engineered self-assembling peptides can offer open-ended capabilities to future multifunctional tissue engineering scaffolds [41].

Self-assembling peptides usually form hydrogels with various types of nanofibers and nanoscale networks and as such, have poor mechanical stability, making it quite difficult to create stable 3D structures [11]. This can result in the fragmentation of the engineered nanofibers, which are more susceptible to endocytosis. Due to the high cost of synthesis for the biomaterials used in this type of fabrication method, the applications in tissue engineering and regenerative medicine are quite limited, though there are ongoing efforts to improve the fabricated scaffold properties.

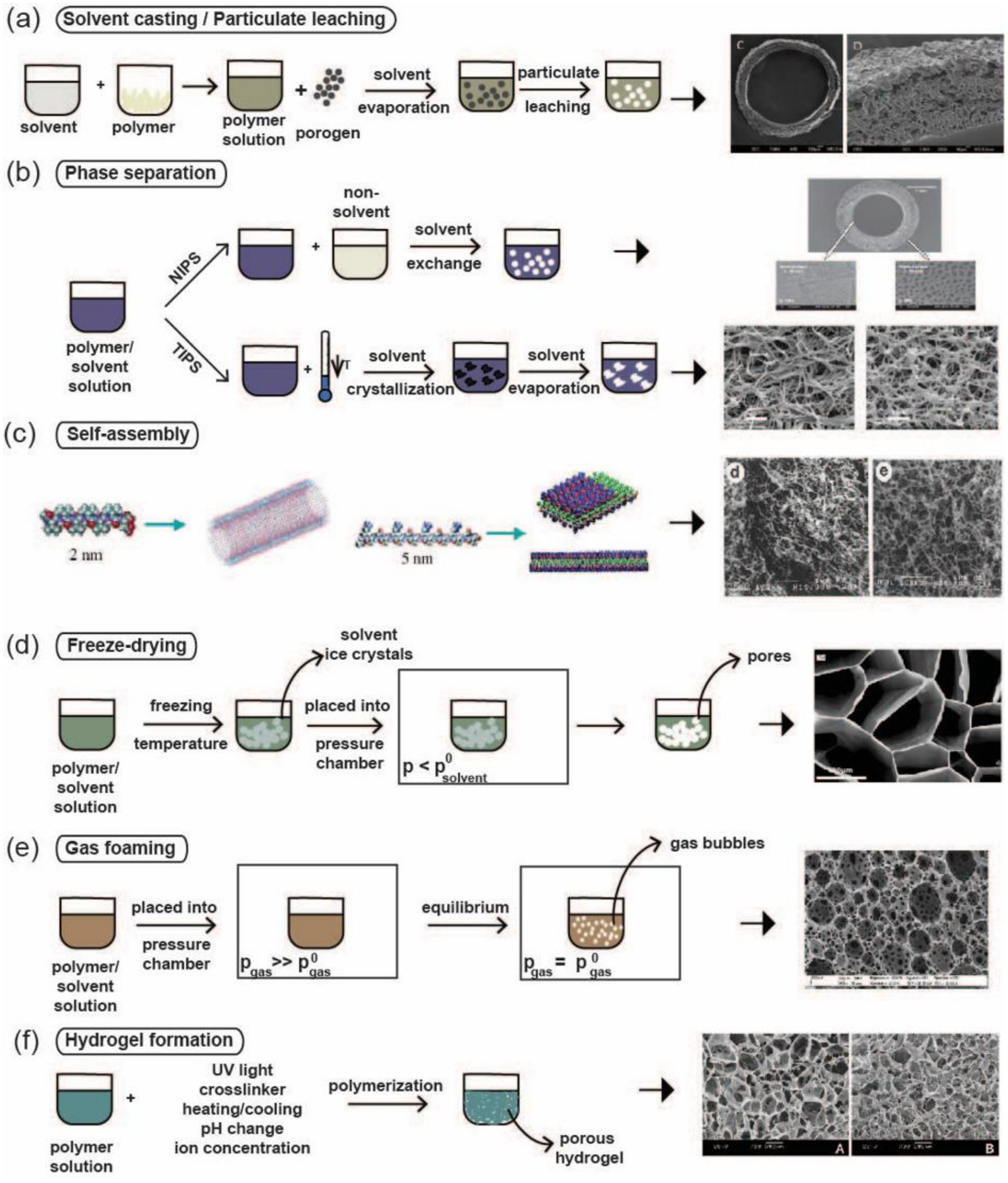

**Figure 2:** (a) Solvent casting/particulate leaching fabrication method. The polymer powder is dissolved in a solvent to create a polymer solution, to which a porogen is added. The polymer hardens through the evaporation of the solvent and the final network within the scaffold is created by leaching the porogen out of the structure [42], (b) Phase separation processes

depicting the differences between NIPS [43] and TIPS [44] (c) Example of the structure of a peptide amphiphile that is used in the self-assembly fabrication technique. The chemical structure and the molecular model of peptide amphiphiles are shown [45]. Fabrication processes of (d) Freeze drying [46] (e), gas foaming [47] and (f) hydrogel formation [48]. Scaffold images reprinted with permission from (a) Kokai et al, 2009 , (b) Oh et al, 2008 and Sun et al, 2012, (c) Zhao & Zhang, 2007 and Holmes et al [49], 2000 (Copyright (2000) National Academy of Sciences), (d) Stokols & Tuszynski, 2006, (e) Barbetta et al, 2010 and (f) Zhang et al, 2011.

#### 2.1.4. Freeze-drying

Freeze-drying (Fig. 2d) is a method that is used to produce highly porous scaffolds, without the need for a porogen. It is a relatively simple technique in the sense that it requires a polymer solution to be cooled down to a low temperature (-70$^{o}$C to -80$^{o}$C), a process which leads to the formation of ice crystals from the solvent. This results in the polymer aggregating in all the spaces left around the ice crystals. The latter are subsequently removed through the application of pressure at a level that is lower than the equilibrium vapor pressure of the frozen solvent, leading to the formation of dry scaffolds with interconnected pores through sublimation. Any residual water that was not in a frozen state is removed in a secondary drying process through desorption [11, 50]. Scaffold porosity can be controlled to an extent through the use of different polymeric material solutions as well as by controlling the freezing temperatures that are used during the freeze-drying process [50, 51]. Scaffolds that have been prepared through freeze-drying have been widely used in neural tissue engineering, as the good interconnectivity of the pores achieved through this fabrication method has been found to promote cellular infiltration and tissue growth when used *in situ* [50].

#### 2.1.5. Gas foaming

Gas foaming (Fig. 2e) is a technique that requires the formation of a gas in a polymer solution. The first way in which this can occur requires a molded polymer together with a gas-foaming agent, such as carbon dioxide ($CO_2$), nitrogen ($N_2$) or water ($H_2O$). The polymers are pressurized with the gas-foaming agents until they become saturated, in order to achieve nucleation – the formation of gas bubbles within the polymer. Through this process, the bubbles that are formed range from 100µm to 500µm, essentially forming pores of the same size within the polymer [47, 51]. Another way to induce gas foaming is to cause a reaction during the mixing process. Through the addition of two chemicals that, upon reaction, cause the release of a gas (e.g. $N_2$), foaming is rapidly induced, leading to the formation of a highly porous network [14]. However, despite the fact that it is a relatively easy technique that does not require the use of a solvent, it has one major drawback. Because of the way in which the bubbles are formed, most of the scaffolds fabricated through this process have poorly connected pores with a non-porous outer surface [52]. Such scaffolds are very difficult to use in tissue engineering, due to the necessity for both a porous outer surface that will facilitate cell infiltration and a well interconnected porous network that will allow the elongation of axons during the repair process. As such, its applications in neural tissue engineering are extremely limited.

#### 2.1.6. Hydrogel formation

Hydrogels are highly hydrophilic polymers that can incorporate very large amounts of water into their structures. The material that is used to create the hydrogel structures is a liquid solution of polymer chains of varying sizes, which also contains unreacted monomers or crosslinking compounds [53, 54]. For the formation of porous hydrogel structures, the components of this material need to be crosslinked. This process can occur in various ways, ranging from UV photopolymerization and ionic concentration-dependent crosslinking, to changes in the temperature or the pH of the material (Fig. 2f) [15, 55]. Depending on the type and the properties of the polymeric material, as well as the crosslinking method used, the hydrogels can acquire different properties and can range from being soft and flexible (mimicking soft tissues such as the brain) to rigid (mimicking cartilage or bone) or elastic (mimicking the skin). Essentially, the changes that can occur in the porosity and stiffness/flexibility of the resulting hydrogel play a crucial role in how hydrogels interact with cells [53].

The polymeric materials that are used for the formation of hydrogels can be of natural or synthetic origin. Naturally occurring molecules able to form hydrogels include collagen, gelatin, chitosan, agarose, hyaluronate and alginate, while synthetic molecules that are commonly used include poly(acrylic acid) and its derivatives (with one of the most well-known ones being poly(2-hydroxyethyl methacrylate) (PHEMA), poly(ethylene oxide) (PEO) and its copolymers and poly(vinyl alcohol) (PVA)) [53, 56]. The materials that will be chosen for any given application are always dependent on the specific requirements of the application and the potential adverse effects must always be taken into consideration. Natural biomaterials are more reminiscent of the extracellular matrix environment but can often have disadvantages such as high variability between batches, as well as high immunogenicity. On the other hand, synthetic biomaterials may be more consistent between batches and have very low immunogenicity but may not always be compatible with the tissue of interest [57].

## 2.2. Molding and texturing methods

### 2.2.1. Compression/Injection molding

Compression/Injection molding (Fig. 3a) is a fabrication process very similar to solvent casting/particulate leaching. With the particulate leaching component of the process being the same in all cases, the way in which they differ is related to the way in which the polymer is prepared. The process of melt molding requires three separate components: a mold, a polymer powder and a porogen. Similar to the methodology described above for solvent casting, a polymer solution is prepared, wherein the polymer is dissolved in a solvent and poured into a mold, in order to allow the solvent to evaporate and create the scaffold structure. In this fabrication method, there are two ways to prepare the polymer and porogen used for scaffold creation. In the first method, the polymer is melted and becomes mixed with the porogen through an extruder or a mixing machine, creating a polymer matrix with an even distribution of porogen. Then the molding can take place either through compression or injection. In the second method, the polymer powder becomes physically mixed with the porogen in solid form and added to the mold. The scaffold is created through the use of compression [9, 51]. The porosity of the sample can be controlled through the different chemical composition and size of the porogens and the method that is used to prepare the scaffold, as the use of the melting process in the first method nearly always leads to the porogen shrinking in size [9].

### 2.2.2. Soft lithography and photolithography

Soft lithography (Fig. 3c) is a family of techniques that are used for the fabrication and/or replication of structures through the use of molds, stamps and photomasks and it includes methods such as microcontact printing (μCP), replica molding (REM), microtransfer molding (μTM), micromolding in capillaries (MIMIC) and solvent-assisted micromolding (SAMIM) [58]. It was developed as a way to enable microfabrication at the nano- and micrometer scale, in order to partly replace photolithography as the main fabrication method. The use of the word "soft" refers to the materials that are used in these fabrication techniques, which are planar, flexible, curved and soft substrates, as opposed to the more rigid ones used in other lithographic techniques, such as polystyrene-based polymers [17]. Photolithography is routinely used for the manufacturing of microelectronic structures using a projection-printing system. It is the main tool used in the semiconductor industry and is used to produce all integrated circuits. The way in which this method works is by projecting the image of a reticle through a high numerical aperture lens system onto a thin film of photoresist that has been spin-coated on a wafer.

The application of this technique in biology and biotechnology comes with several limitations, due to factors such as its very high cost, the limited amount of control over the properties of the surface, as well as being compatible with a limited number of materials [16]. As a result, soft lithography techniques were developed in order to overcome these limitations. In each one of the techniques that make up the soft lithography technique family, an elastomeric stamp or mold is used to transfer a pattern to the substrate. Firstly, the pattern is designed, followed by the fabrication of the mask and the master. Once these are made, they are used to make the PDMS stamps/molds, which are then used in the printing/molding/embossing stage of each individual protocol [58].

### 2.2.3. Laser texturing

In order to create various structures on surfaces and tailor their morphology for a variety of different applications, there have been several approaches that have been developed, including photolithography and plasma treatment. One of the most recently developed techniques that belongs to the subtractive manufacturing approaches, is surface texturing through the use of ultrafast lasers (Fig. 3b) [59, 60]. This particular technique is based on the principle of laser ablation and has found a large number of applications due to its advantage to locally excite certain areas of a material, with minimal damage to the surrounding ones. This is due to the short duration of the pulse, which allows for a higher excitation threshold intensity to be attained, without increasing laser beam intensity. As a consequence, precise structuring/texturing on the material surface at both the nano- and the micrometer scales can be attained [18].

Additional advantages of the laser texturing method include the fact that it is non-contact, can be applied to a variety of different materials (including metals, ceramics and polymers) and can be used to texture materials on their surface, as well as etching deeper regions within the bulk of the material to create structures with complex geometries [61]. Materials that have been laser-patterned with structures such as grooves, pillars, micro- and nanocones [62] and channels have been widely used as cell culture platforms in order to investigate the effect of topographical cues on various cellular responses [63].

Another phenomenon that can occur as a result of laser irradiation is foaming. This can be induced superficially on the surfaces of biopolymers, such as gelatin, collagen and chitosan, through the application of single pulses in the nanosecond to femtosecond domains. More specifically, the single pulse that is delivered to the surface of the biopolymer creates a transient acoustic wave which eventually leads to nucleation and the creation of bubbles [64-66]. Through controlling the number, duration, wavelength and fluence of the pulses, the foaming process can be fine-tuned and can lead to the creation of differentially nanostructured biomaterial scaffolds [65, 67, 68]. As a consequence, this method provides us with a tool to create scaffolds with micrometer precision [69].

### 2.2.4. Fiber mesh / Fiber bonding

The fiber mesh fabrication technique (Fig. 3d) requires the formation of fibers made out of polymeric materials, which are then used in the same way as threads would be and are woven into three-dimensional patterns in order to create meshes with a range of pore sizes [70].

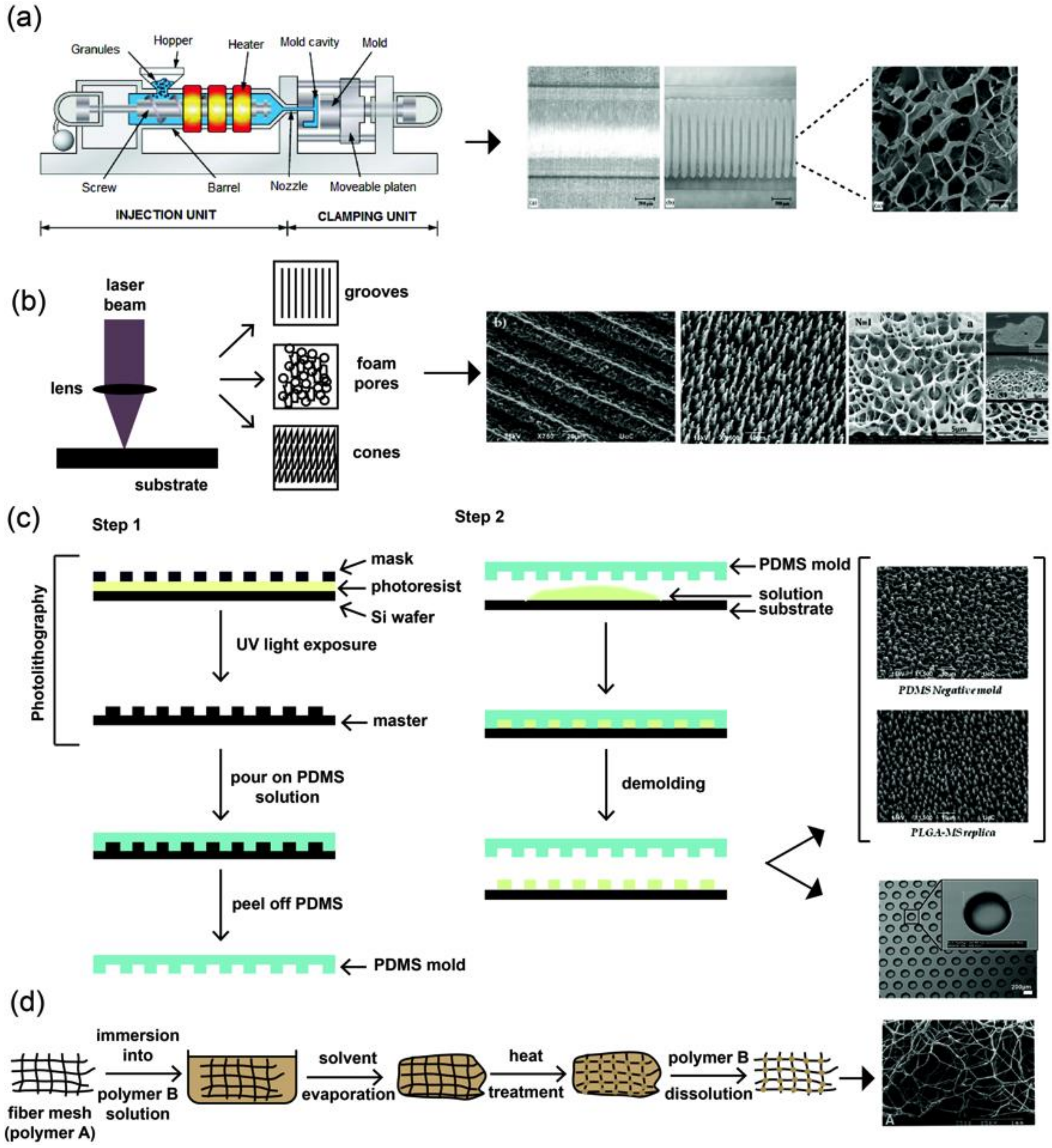

**Figure 3:** (a) Schematic representation of an injection molding machine [71], and SEM images of representative structures [72], (b) Laser texturing [68, 73], (c) Photolithography and soft lithography. Step 1: Photolithography and the process for the creation of the negative PDMS mold for use in soft lithography techniques [73, 74]. Step 2: Process for the transfer of the pattern of the negative PDMS mold onto another material, (d) Fiber bonding [19]. Reprinted with permission from (a) Svecko

et al, 2013 and Freier et al, 2005, (b) Babaliari et al, 2018, Daskalova et al, 2016 (c) Babaliari et al, 2018, Leijten et al, 2016 and (d) Mikos et al, 1993.

The scaffolds obtained via this method have quite a large surface area and allow the rapid diffusion of nutrients essential for growth and survival. However, since they are not stable enough for use *in vivo*, this technique was further developed into fiber bonding, where the scaffold created as part of the fiber mesh fabrication technique was placed into a second polymer/solvent solution. Once the solvent has been evaporated, the scaffold is heated to above the melting temperature of both polymers and after cooling, the second polymer is removed through selective dissolution. The result is a scaffold made out of one polymer that has been physically bonded at the intersections between the fibers [19, 75]. This fabrication method does not offer fine control over scaffold porosity and the scaffolds have limited use due to the types of polymers that can be used and the use of a solvent during the fabrication process.

### 2.3. Electrohydrodynamic techniques

Electrohydrodynamic methods of fabrication are based on the electrostatic attraction of a liquid which, while coming out of a nozzle, is subjected to an electric field and subsequently collected on a plate. The typical setup (Fig. 4a) for these fabrication methods consists of three primary components: i) a high voltage power supply (usually in the kV range), ii) a syringe with a metallic needle (spinneret) and iii) a grounded collector. In a typical electrohydrodynamic process, a viscoelastic solution is placed in the syringe and as a small droplet is forced out of the tip, it becomes charged by applying an electrical potential difference between the droplet and the plate. The electrical field is created by connecting the positive electrode of the high-voltage power supply to the metallic needle and the negative electrode to the grounded conductive collector. As the voltage increases, the electrostatic repulsion starts to overcome the surface tension of the fluid and the pendant droplet deforms into a conical droplet, known as the Taylor cone. When the electrostatic repulsion finally overcomes the surface tension, a fine, charged jet of solution is ejected from the tip of the needle towards the grounded collector [76]. On the way to the collector, the solvent is evaporated and nano-/microstructures are obtained once the process is complete.

One critical parameter of the process, which affects the morphology of the resulting structure is the concentration of the solution (Fig. 4c) [77]. At lower solution concentrations, spherical particles are observed. As the concentration increases, fibers also begin to be formed, resulting in bead-on-string morphologies (i.e. electrospun fibers with particulates). Above the critical solution concentration, continuous, uniform fibers can be generated. Finally, when the solution concentration is too high, helix-shaped microribbons are formed instead of smooth fibers. Hence, two different electrohydrodynamic approaches have been developed: a) electrospraying and b) electrospinning. Electrospinning and electrospraying are two highly versatile and scalable electrohydrodynamic methods for fabrication of nano- and micro-scale fibers and particles, respectively.

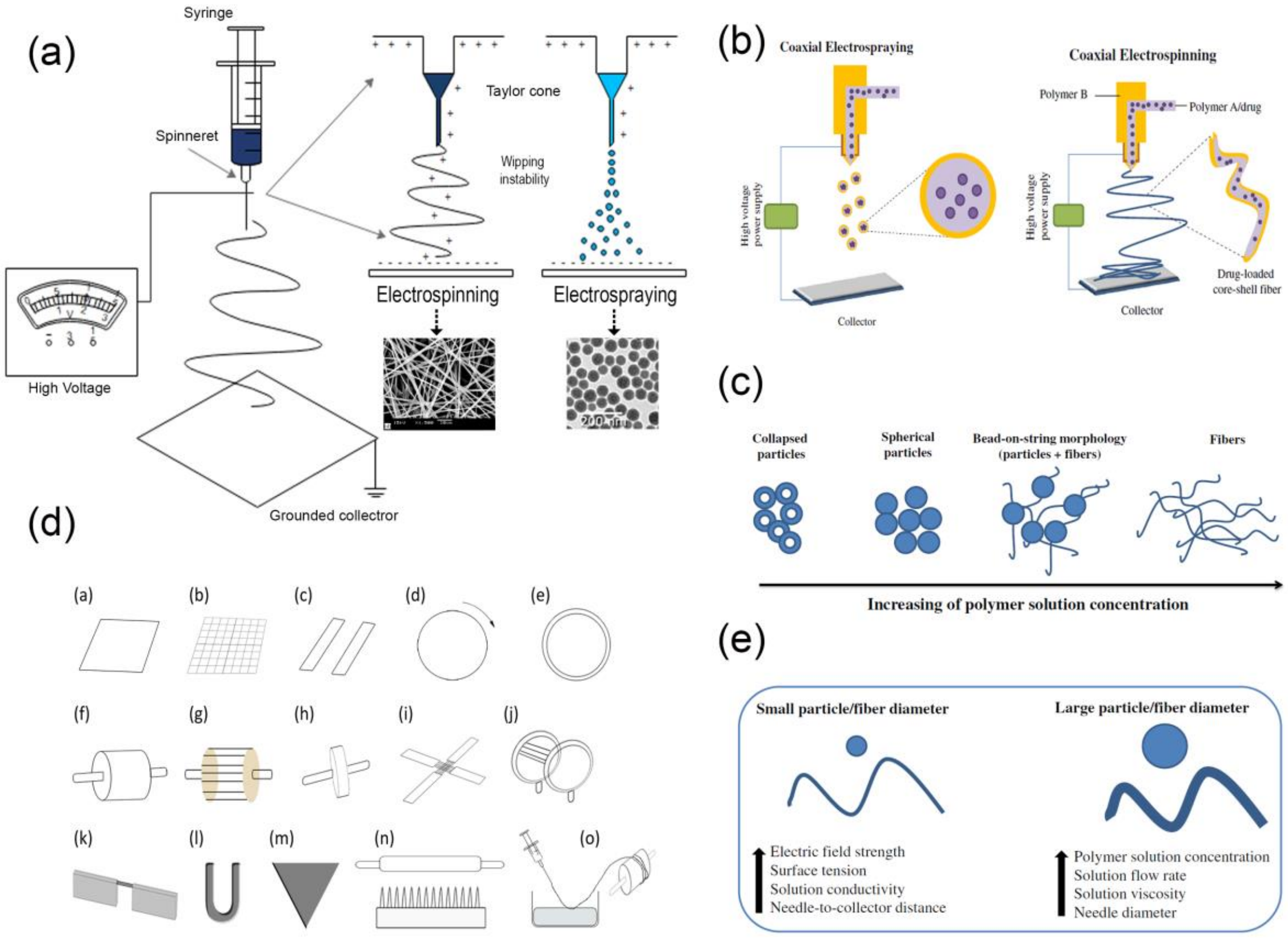

**Figure 4:** Electrohydrodynamic techniques. (a) Schematic representation of an electrospinning/electrospraying setup, (b) coaxial nozzles, (c) resulting electrospun structures depending on the concentration of the solution, (d) different types of collectors used in electrospinning and (e) Factors affecting the diameter of particles/fiber created by electrohydrodynamic methods. SEM images in (a) have been reprinted with permission from Yang et al, 2005 [78], Eaton et al, 2017 [79], images in (b),(c) and (e) have been reprinted with permission from Nikolaou et al, 2018 [21].

### 2.3.1. Electrospraying

Electrospraying (Fig. 4a) is a process of liquid atomization by electrical forces. As mentioned above, its main difference from electrospinning is the concentration of the solution. In electrospraying, the low solution concentration destabilizes the jet due to varicose instability and hence, fine droplets are formed. These highly charged droplets self-disperse in space, thereby preventing droplet agglomeration and coagulation. Furthermore, the solvent evaporation over time promotes the contraction and solidification of the droplets resulting in solid particles deposited on the grounded collector.

During the electrospraying process, the jet deforms and disrupts into droplets due mainly to electrical force as no additional mechanical energy is needed for liquid atomization. The charge and size of the droplet can easily be controlled to some extent by adjusting the flow rate and voltage applied to the nozzle. Consequently, electrospraying has some advantages over conventional mechanical spraying systems, where droplets are charged by induction: i) it produces droplets of a smaller size, ii) their size distribution is usually narrow with low standard deviation, iii) the charged droplets are self-dispersing and no droplet agglomeration and coagulation occurs, iv) the motion of charged droplets can easily be controlled by electric fields such as deflection or focusing and v) the deposition efficiency of the charged spray is much higher than for uncharged droplets [80].

For small-scale processes, electrospraying with a single capillary is used, however, for traditional industrial spraying approaches, multi-nozzle or slit-nozzle systems have been established. An alternative coaxial electrospraying approach has been developed for the formation of multilayered particles for biomedical applications (Fig. 4b). Coaxial electrospraying is based on conventional microencapsulation/nanoencapsulation processes and modifies the single-axial electrospray process by using a coaxial capillary needle to deliver two liquids independently. The key component in the setup is a coaxial nozzle that consists of an outer needle and an inner needle and two immiscible liquids, one in each needle (Fig. 4b). Two pumps control the flow rates in the respective needles creating two-layer droplets using an electric field.

### 2.3.2. Electrospinning

Electrospinning (Fig. 4a) is the method of drawing thin fibers out from a viscoelastic fluid using electrostatic forces. During this process, as the charged jet is ejected towards the collector, it is forced to a whipping or bending motion due to the unevenly distributed charges and the high concentration of the solution. As a result, after the elongation of the jet and the rapid evaporation of the solvent, a solid, thin fiber is deposited on the grounded collector. The final structure is usually a randomly oriented, non-woven mat. However, electrospinning offers the possibility of the production of fibers with different orientation and hierarchical structures. To obtain aligned orientation in electrospinning, the type of the collector plays a crucial role. In the past decades, to achieve aligned electrospun fibers, several special designed collectors have been used, such as rotating drums, metal frames or two conductive substrates separated by an insulating gap (Fig. 4d) [81]. The shape of the collector also influences the resulting nanofiber structure. The deposited patterns include shapes such as circles, triangles, squares, crosses, rectangles and more, and can be made out of different materials such as copper, aluminium, gold, wood etc. [81]. Additionally, a liquid system with a coagulation bath can be used instead of a collector or can be combined with a rotating mandrel, in order to obtain a continuous yarn made from electrospun fibers. In general, the necessary fiber alignment can be obtained by mechanical, magnetic or electrostatic means [82, 83]. In the mechanical approach [84-86], it is usually the use of a rotating mandrel that aligns the fibers along the direction of the rotation. Alignment can also be achieved by magnetic forces [87]. In this case, a small number of magnetic nanoparticles are incorporated in the polymer solution to magnetize it and the solution can be electrospun into nanofibers in the presence of a magnetic field. In this way, the magnetized fibers are stretched into essentially parallel fibers, across a gap, to generate a uniaxially aligned array between the two magnets. In the electrostatic approach, the collector typically consists of a pair of electrodes with an insulating gap between them which forces the fibers to be stretched and align themselves perpendicular to each edge of the gap [88].

In recent years, hierarchical nanofiber structures, including core-shell, hollow and porous structures, can be obtained with electrospinning. Typically, the nanofibers produced using electrospinning have a solid structure, however two strategies have been explored for generating porous nanofibers: i) selectively removing one of the components from the fibers and ii) inducing polymer−solvent phase separation by rapidly cooling the fibers prior to complete solidification [82]. Both approaches have been used and can be extended to obtain porous nanofibers from a variety of polymers. A coaxial electrospinning set up (Fig. 4b) can be used to create tubular fibers. Electrospinning of two immiscible solutions through a coaxial spinneret can generate core−sheath and hollow nanofibers with controllable dimensions by selective removal of the core after electrospinning. By adjusting the experimental parameters, the wall thickness and the inner diameter of the tubular nanofibers can be tuned from tens to several hundred nanometers.

Electrospun and electrosprayed products possess several structural advantages such as high surface-to-volume ratio, porosity, tailored morphology and sub-micron and nanoscale size. Furthermore, the size and the final structure can be controlled by manipulating critical factors such as the solution, as well as instrumental and ambient parameters. The solution parameters include the molecular weight, the type and concentration of the polymer, the solvent used and the solution properties (ie. pH, conductivity, viscosity and surface tension). The instrumental parameters include the applied electrical potential, the flow rate of the solution, the distance between the tip of the needle and the collector and occasionally the nature of the collector material and the shape. Finally, the ambient conditions in the process chamber, such as the temperature, humidity and air velocity, can collectively determine the rate of evaporation of the solvent from the electrospun or electrosprayed product and affect the characteristics of the final structure. Regarding the size of the particle/fiber diameter, a general rule that applies is that it is decreased when the electric field strength, the surface tension, the solution conductivity and the needle-to-collector distance are increased. On the contrary, any increase in the solution concentration, viscosity, flow rate and needle diameter results in particles and fibers with larger diameters (Fig. 4e) [21].

Apart from the particle/fiber size and structure, which can be fine-tuned by adjusting the electrospraying/electrospinning parameters, post-processing of the final products may be required to produce functionalized structures for several applications [89]. The most frequent functionalization strategies that can facilitate a better cell response for biomedical applications are surface treatment by chemical and physical means and the coating or binding of bioactive molecules. Methods for surface modification include plasma treatment, poly(dopamine) treatment, layer-by-layer technique, coating with blends and emulsions, grafting of natural polymers etc. [89].

Besides the independent use of nano/microparticles or nano/microfibers, electrospraying and electrospinning have also been combined together for the generation of hybrid particle/fiber composite materials [90, 91]. For the generation of these hybrid structures, a simultaneous electrospinning and electrospraying process can be performed using the same collector or the electrospraying process can take place on the same collector after the completion of the electrospinning procedure [91]. Both of the aforementioned processes result in the deposition of electrosprayed particles onto the surfaces of the electrospun fibers, generating advanced composite materials useful in the field of biomedical research for tissue engineering.

### 2.4. Solid freeform fabrication / Rapid prototyping

Rapid prototyping or Additive manufacturing is a group of techniques which are used to quickly fabricate physical models and prototypes using three-dimensional computer-aided design (CAD) data (Fig. 5a). The first step in rapid prototyping is the creation of a 3D computer model using software, which builds spatial image models. The 3D images are corrected or modified and are subsequently cut into sequences of layers by a slicer software. These are then used during the second step of the process, to generate complex objects layer-by-layer via the solidification of melts, layer photo-polymerization or the bonding of particles using either laser beam-induced sintering or special binders. The final step is related to post-printing procedures such as curing, sintering or final finishing. Irrespective of the technology used for the creation of a 3D model, all the methods in additive manufacturing are based on the same principle: the laying down of material in a layer-upon-layer fashion in order to create a whole object. In other words, the successive addition of 2D layers of material continues until a complete 3D object is fabricated.

A variety of additive manufacturing techniques have been developed with the most common methods being Selective laser sintering and melting (SLS and SLM), stereolithography, fused deposition modeling and 3D plotting/printing. All these bottom-up approaches can be used with various forms of materials such as liquids, solids or powders and offer the ability to create highly organized three-dimensional structures with architectures that cannot be attained via traditional manufacturing processes. They additionally address issues such as internal porosities, lack of residual stress and interlocking shapes without connection [92]. However, the most important advantage of additive manufacturing is the capability of the approach to produce customized structures which can serve as surgical prototypes, prostheses or scaffolds in biomedicine.

### 2.4.1. Photolithography-based techniques

In photolithography-based techniques, the solidification of liquid photosensitive polymers takes place by exposure to light in order to build three-dimensional models. These techniques include Stereolithography (SLA), Digital Light Processing (DLP), Continuous Liquid Interface Production (CLIP) and Two-Photon and Multiphoton polymerization (2PP/MPP) [22]. Both SLA and DLP work by selectively exposing liquid resin to a light source, with their difference being in the light source: a laser in the case of SLA and a projector in the case of DLP. During the process of SLA (Fig. 5b), a built platform is lowered in a vat filled with a liquid photopolymer resin [93]. UV or visible light is focused on the resin-platform interface in a precise pattern and solidifies the resin to create the first layer. Once this layer is solidified, the platform is lowered for the polymerization of the subsequent layer and this process is repeated until the whole 3D structure is complete. Although stereolithography originally used UV light for photocuring, the visible light has been explored as an alternative to the harmful for live cells UV radiation, in order for stereolithography to be used for bioprinting. In DLP, just like in SLA, a resin tank with a transparent bottom and a built platform that descends into a resin tank are used and the objects are build upside down, layer by layer. The difference here is that a digital screen is used as a projector and flashes an image of a layer across the entire platform, curing all points simultaneously. This makes DLP much faster than SLA. DLP has been used for the fabrication of 3D constructs with high spatial resolution, excellent structural stability and reliable biocompatibility which mimic the complex architecture of biological tissues [94-96]. CLIP, also known as Digital Light Synthesis (DLS), is based on SLA but the process doesn't pause after each layer. Instead, the resin continuously flows and the objects are built upside-down. Due to this continuous process, the parts have smother surfaces and are printed much faster. 2PP differs from other laser-based additive manufacturing techniques because it applies the principle of two-photon absorption for the generation of micro- and nanostructures inside a polymerizable solution [97]. This can be achieved by the use of a tightly focused, femtosecond pulsed laser beam (with wavelengths in the near-infrared range), which penetrates the solution and confines the polymerization to the focal point instead of the entire area. Thus, 2PP does not operate in a layer-by-layer fashion and as a result has virtually no geometrical restrictions when producing a structure. Compared to other conventional additive manufacturing techniques, 2PP provides higher resolution 3D scaffolds [98-100] and is therefore able to mimic the extracellular matrix to a greater extent which is beneficial for cell attachment and proliferation [101, 102].

### 2.4.2. Selective laser sintering / Selective laser melting

SLS and SLM (Fig. 5b) belong to the powder-based additive manufacturing techniques because they induce consolidation of a powdered material using high-energy light sources [23]. The powders which can be used must be materials that can be melted by lasers. During the process, a layer of powder is deposited and melted using a laser which creates a precise pattern dictated by CAD data. Next, a new layer of powder is deposited on top and melted and the process is repeated as many times as necessary to complete the whole structure. SLS and SLM are very similar methods. The main difference between them is that in the former, the powder does not fully melt, but it is heated to a degree where it can be fused together on a molecular level. As such, sintering allows for controlled porosity of the material. SLM uses the laser to achieve a full melt of the powder into a homogeneous liquid, creating stronger structures with fewer or no voids as long as a single metal powder is used.

A special case of sintering is microsphere sintering [24], where microspheres of polymer are used instead of powder. Prefabricated polymeric microspheres are poured into a mold and heated to a specific temperature (usually above the glass transition temperature of the polymeric matrix) for several hours. This causes the melting of the surface layer of the microspheres, which are fused with adjacent microspheres, creating a three-dimensional porous scaffold with excellent mechanical properties. Besides heat sintering, which is the most used method for microsphere sintering, other techniques can be used such as solvent-based sintering, and subcritical $CO_2$ sintering.

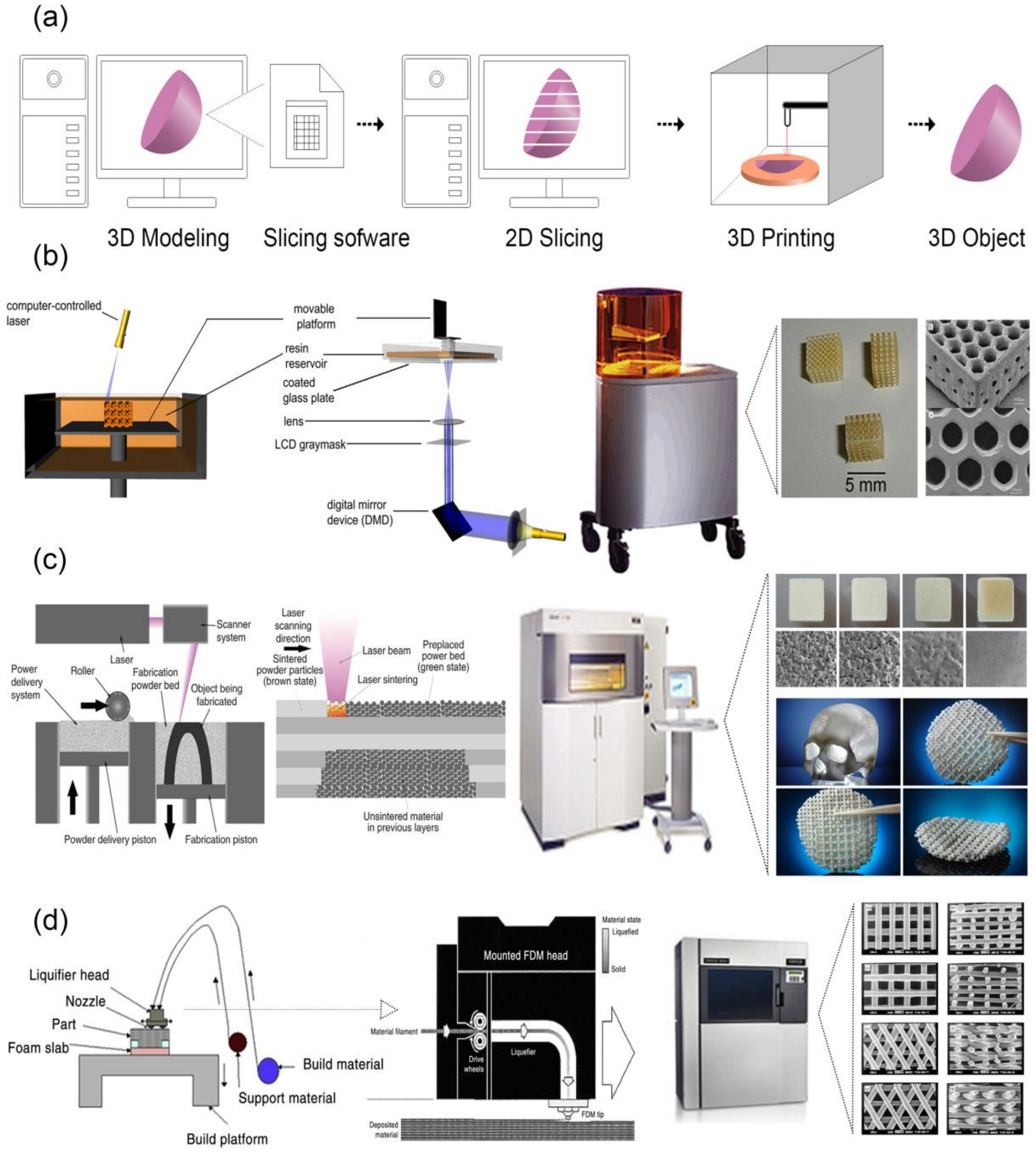

**Figure 5:** Principles and methods of Rapid prototyping/Solid free form fabrication. (a) Schematic representation of the Rapid prototyping process. A 3D computer model is created, sliced into layers and printed in a layer-by-layer fashion. Schematic representation of the printing processes of (b) Stereolithography as a representative example of the photolithography-based techniques, (c) Selective Laser Sintering and (d) Fused Deposition modeling. For each method we also see its corresponding equipment and representative examples of fabricated scaffolds. Images were adapted with permission from (b) Melchels et al, 2010 [93], (c) Molitch-Hou et al, 2018 [103] and Gayer et al, 2019 [104], (d) Masood et al, 2014 [105] and Zein et al, 2002 [25] for SEM images. (Scale bar in (c) 500 µm x/y/z).

### 2.4.3. Fused deposition modelling

Fused deposition modeling (Fig. 5c) belongs to the material extrusion additive manufacturing techniques and uses a small temperature-controlled extruder to force out a thermoplastic filament material [105, 106]. The polymer filaments are deposited onto a platform in semi-molten condition, in a shape according to a computer-designed model and in a layer-by-layer process. Once a layer is completed, the base platform is lowered and the next layer is deposited. Fused deposition modeling offers the capability of creating highly porous scaffolds with a fully interconnected channel network, as well as controllable porosity and channel size. Due to the nature of the materials processed in FDM (e.g thermosensitive plastics and polymers), this method has been widely used in medicine for the fabrication of customized patient-specific medical devices, such as implants, prostheses, anatomical models and surgical guides, as well as customized platforms for personalized medicine.

### 2.4.4. 3D bioprinting

Three-dimensional printing includes several flexible techniques for the direct manufacturing of complex shapes with high resolution, as well as for the processing of highly customized medical products combined with image reconstitution techniques. As an additive manufacturing technique, 3D printing is based on the deposition of biomaterials in a layer-by-layer manner. In most cases, a computer-controlled three-axis mechanical platform drives the movements of the print-head in the required algorithm and shape. Bioprinting differs from conventional 3D printing in that a bioink with living cells and growth factors is used to create tissue-like structures that imitate natural tissues [107]. Bioink usually consists of a carrier material where living cells have been enveloped; however, it can be composed only of cells or the cells can be loaded later on. The carrier material in most cases is a polymeric gel that supports cell attachment and function but also provides protection to the cells during the printing process.

The most commonly used methods of bioprinting for tissue engineering applications are the inkjet-based, the extrusion-based, the stereolithography-based, and the laser-assisted printing (Fig. 6) [26, 108].

In inkjet-based printing (Fig. 6a), droplets of biomaterials are selectively placed on the platform in a layer-upon-layer fashion for the creation of a complete structure. Droplets can be formed either by thermal, piezoelectric or electrostatic forces. The main limitation of the procedure is the inherent inability of the print-head to provide continuous flow and the low cell densities in the ink (only bioinks with viscosities lower than 10 mPa are printable via inkjet). Thermal inkjet printing is not common in tissue engineering due to the loss of activity of the macromolecules from the very high temperatures used (potentially higher than 200°C).

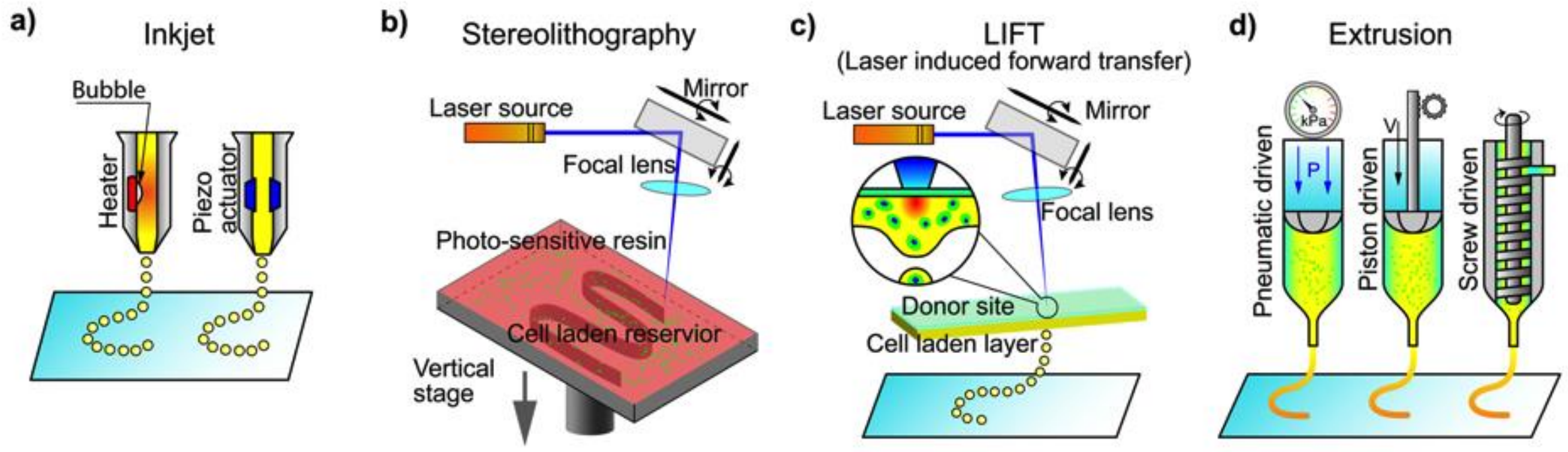

**Figure 6:** Illustration of 3D printing processes: Inject-based 3D printing (a), Stereolithography-based 3D printing (b) laser-assisted 3D printing (c) and Extrusion-based 3D printing (d). Image reprinted with permission from Jiang et al, 2019 [108].

In stereolithography-based bioprinting (Fig. 6b), visible light can be used as an alternative to the harmful UV light, for the curing of biocompatible hydrogels and biopolymer blends. Two modifications have been described: the 'top-down' approach, which is similar to conventional stereolithography but with the platform being lowered into a tank filled with bioink (this usually consists of a pre-polymer solution with living cells), and the 'bottoms-up' approach, where the prepolymer solution with the cells is pipetted into the container one layer at a time from the bottom to the top [109]. This 'bottom-up' setup is suitable for cell encapsulation applications. The advantages of stereolithography-based bioprinting are high resolution and velocity, the high cell concentrations that can be used and the absence of problems due to nozzle clogging. The main disadvantages are the high cost of the devices and the cytotoxicity of the lights and photo-initiators.

Laser-assisted bioprinting (Fig. 6c) is a non-contact, nozzle-free technique, which allows high resolution deposition of biomaterial in a solid or liquid state [110]. It is based on the Laser-induced forward transfer technique and the typical setup includes the laser and two parallel laser-transparent glass slides: the donor and the receiver. The printed cells are spread onto a donor slide, embedded in a biological polymer or suspended in culture medium or hydrogel. The receiving slide is usually coated with a hydrogel in order to enhance cell adhesion and proliferation. The hydrogel on the receiver slide also offers protection and minimizes the cell damage upon impact. A laser pulse induces the propulsion of cells from the donor slide toward the collector slide in a computer-controlled manner. In some cases, an absorbing layer is interposed between the donor slide and the bioink to avoid the direct interaction of the laser with the living cells in the bioink. This technique was previously used for the printing/patterning of a variety of biomolecules, such as proteins and DNA for engineering of biosensors, diagnostics or cell culture platforms [111-113].

In extrusion-based printing (Fig. 6d), biomaterials are extruded out of the print-head due to mechanical or pneumatic pressure. This method allows the incorporation of cells and biomolecules because it does not involve any heating processes. When compared to inkjet bioprinting, extrusion-based bioprinting offers higher cell densities but lower speed and resolution (bioinks with viscosity in the range of $30^6$-$10^7$ mPa are printable with extrusion-based printing).

## 3. Applications in neural tissue engineering

As mentioned above, the vertebrate nervous system (Fig. 7) is subdivided into the CNS and the PNS. The main components of the CNS are the brain and the spinal cord, but it includes also the optic, olfactory and auditory systems due to the fact that they are connected directly to brain tissue without intermediate nerve fibers. The PNS consists of the cranial nerves that branch out from the brain, the spinal nerves that branch out from the spinal cord, as well as sensory nerve cell bodies (dorsal root ganglia) and their processes.

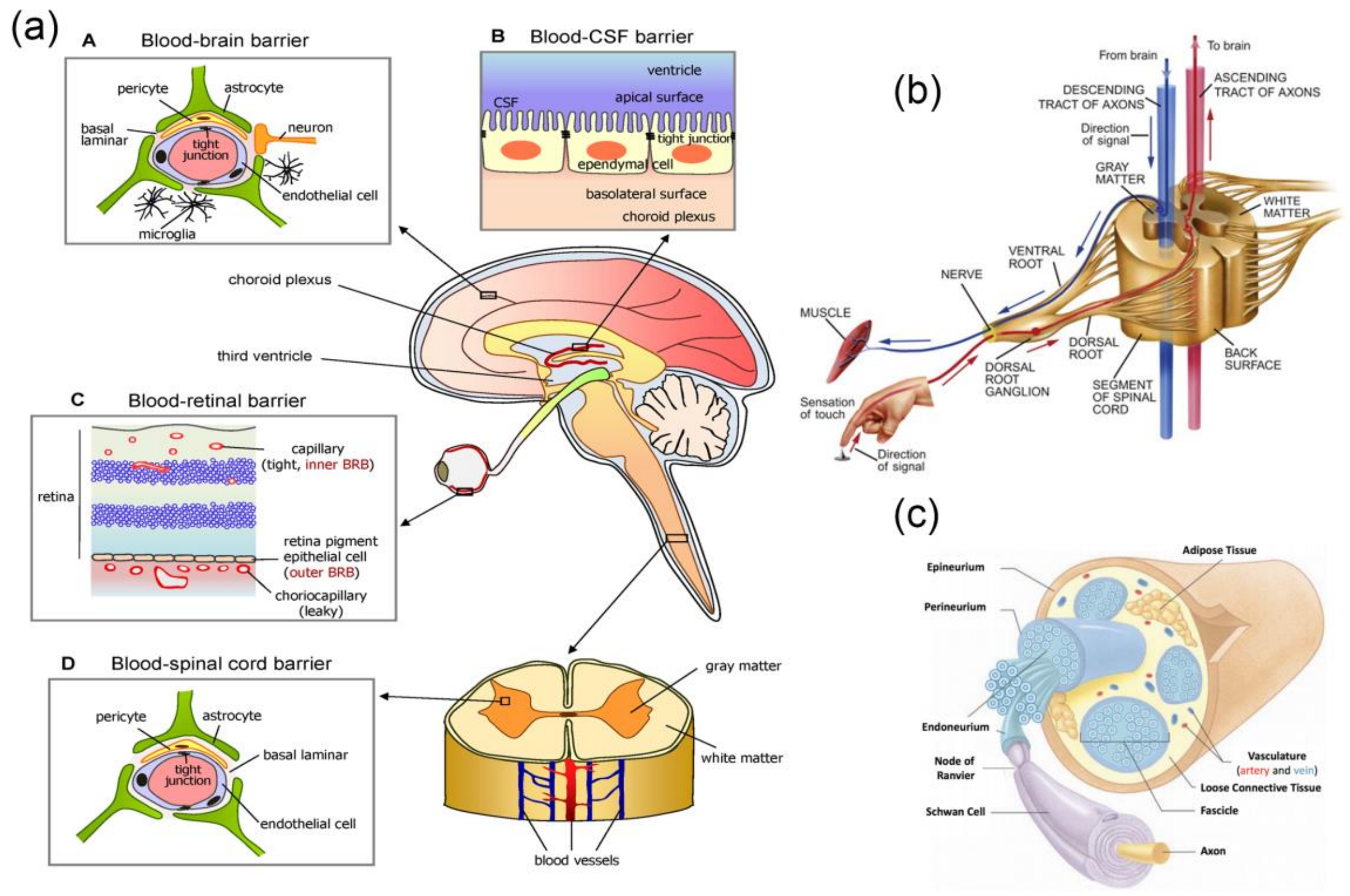

**Figure 7**: Schematic representation of the major components of the human nervous system. (a) anatomy of the brain and the blood brain barrier [114], (b) cross section of the spinal cord [115] and (c) anatomy of the peripheral nerve [116]. Images reprinted with permission from (a) Choi et al, 2008, (b) McDonald et al, 2014 and (c) Kuliasha et al, 2018.

Neural tissue in both CNS and PNS is composed of neurons and glial cells. Neurons are specialized cells that can receive and transmit chemical or electrical signals and consist of a cell body (soma) and its extensions (axons and dendrites). Clusters of sensory nerve somata are located just outside the spinal column and are known as ganglia. Dendrites transmit electrical signals to the neuron cell body and the axon conducts impulses away from it. Glial cells provide support and protective functions for the neurons and include Schwann cells in the PNS and astrocytes and oligodendrocytes in the CNS.

Although neurons and glial cells are found in both divisions of the nervous system, a large variation concerning the basic cell types and tissue organization can be observed between the two. One of the major differences between the CNS and PNS lies in their potential for regeneration. Peripheral nerves can spontaneously regenerate if the injury is small, therefore suturing together the ends of the damaged nerves is sufficient to repair the defect. For larger injuries, transplantation using an autograft (tissue harvested from elsewhere in the body) is the gold standard. Thus, tissue engineering applications for peripheral nerve injury treatment aim firstly to facilitate nerve regeneration and guidance and, secondly, to prevent fibrous tissue formation that impedes the regenerating nerve. In contrast, the CNS in adults shows little to no regeneration after physical or chemical damage due to several limiting factors. The most important of these factors are the intrinsic inability of neurons for growth, the glial scar formation which creates physical barriers and the presence of inhibitory molecules around the lesion area. Therefore, regeneration strategies for the CNS focus on transplantation of neural stem cells or differentiated neural cells in combination with the incorporation of growth factors and glial cells, which can create a permissive environment to promote neuronal differentiation. Consequently, the CNS and PNS responses to injury are distinctively different and, therefore, any therapeutic approach must be specialized accordingly. In the following, we present an overview on how the biofabrication methods presented above have been applied to address both PNS and CNS injuries.

### 3.1. Peripheral nerve injury applications

As mentioned above, the PNS consists of two types of cells: neurons and glial cells. The main function of the neurons of the PNS is the connection of the CNS with sensory and motor targets. In order to achieve this, the body of each nerve cell is located close to the spinal cord or the brain, with a long axon extending as far as required. This axon is protected by the glial cells of the PNS, the Schwann cells, which wrap around the axons and form sheaths containing a protein called myelin. The presence of these sheaths is responsible for the insulation and enhancement of the signal that is transduced through the axon. The morphology of the peripheral nerves is complex as they form cable-like bundles and are surrounded by support tissue, creating anatomically defined trunks (Fig. 7c). Individual axons are covered by the myelin sheath and both are covered by oriented collagen fibers, which create the endoneurium. A group of axons are bundled together with many layers of flattened cells and collagen creating a fascicle, which is wrapped around with a delicate sheath and is known as the perineurium. Finally, many fascicles are bundled together with blood vessels and fatty tissue and are wrapped around with a third protective sheath of connective tissue, a structure which is known as the epineurium. During the process of peripheral nerve injury (PNI), parts of the axons that are connected to a peripheral nerve become severed and degrade in a process called Wallerian degradation. In this case, the Schwann cells change to a pro-regenerative phenotype and secrete factors that assist axon regeneration, while the neurons themselves focus on the process of axon regeneration by initiating protein synthesis.

PNI treatments differ depending on the length of the gap that is created during the injury process. When the gap is small, the two ends of the severed axon can be surgically re-attached. When the gap is significantly longer, healthy nerve segments from elsewhere in the body are used to reconnect the two nerve edges in a surgical process that is called an autograft. In order to avoid the use of autografts, there have been ongoing efforts to create implantable scaffolds which can be placed in the gap that occurs after nerve injury and can facilitate the regeneration of severed axons by providing support and guidance. The majority of these scaffolds are in the form of conduits that serve as guidance channels. A variety of fabrication methods have been used to create scaffolds for peripheral nerve regeneration and these will be analyzed below.

Yannas et al. were the first to report that porous biodegradable collagen-glycosaminoglycan scaffolds, produced by a freeze drying process, were able to induce regeneration of myelinated and unmyelinated axons over large distances (15 mm) between the severed ends of the adult rat sciatic nerve [117].

In 2001, Miller et al. used both compression molding and solvent casting in order to produce microgrooved substrates made of poly-D,L-lactic acid (PDLA) (Fig. 8a), with the solvent casting method producing more stable substrates that exhibited lower degradation rates when compared to the ones created through compression molding. The adsorption of laminin into the grooves, as well as the optimization of groove width were found to be significant in promoting Schwann cell alignment, with grooves between 10-20μm showing optimal results [118].

Porogen leaching was used by Kokai et al. to produce nerve guides made out of polycaprolactone (PCL) [42]. They were able to deposit a NaCl:PCL solution onto a mandrel, which, after drying and porogen leaching, created a porous nerve guide structure with 80% porosity and a pore size in the range of 10-38μm. Even though they did not directly test these scaffolds with cell cultures, they assessed their ability to be fouled as little as possible with lysozyme while also maintaining glucose permeability, which is required for cell growth within the scaffold structure.

In 2009, Chiono et al. were able to produce non-porous nerve guides made out of PCL through the use of the melt extrusion technique [119]. When tested *in vivo* for the repair of severed peroneal and median nerves in Wistar rats, they were found to be effective in promoting repair of small and medium nerve gaps (0.5-1.5cm), but were ineffective when used for the bridging of longer nerve gaps (4.5cm). As a result, further efforts were able to produce porous PCL scaffolds through the use of particulate leaching of a PCL/PEO (poly-ethylene oxide) solution, where the PEO component was selectively dissolved by immersing into water in order to create the porous structure [13]. The obtained scaffold was then filled with a gelatin/genipin (GL/GP) solution, which upon freeze-drying forms a sponge-like structure. Cell adhesion and survival of PC12 and NOBEC cells were assessed on the GL/GP structure alone, while several NOBEC cells were also found in the gelatin of the porous nerve guides.

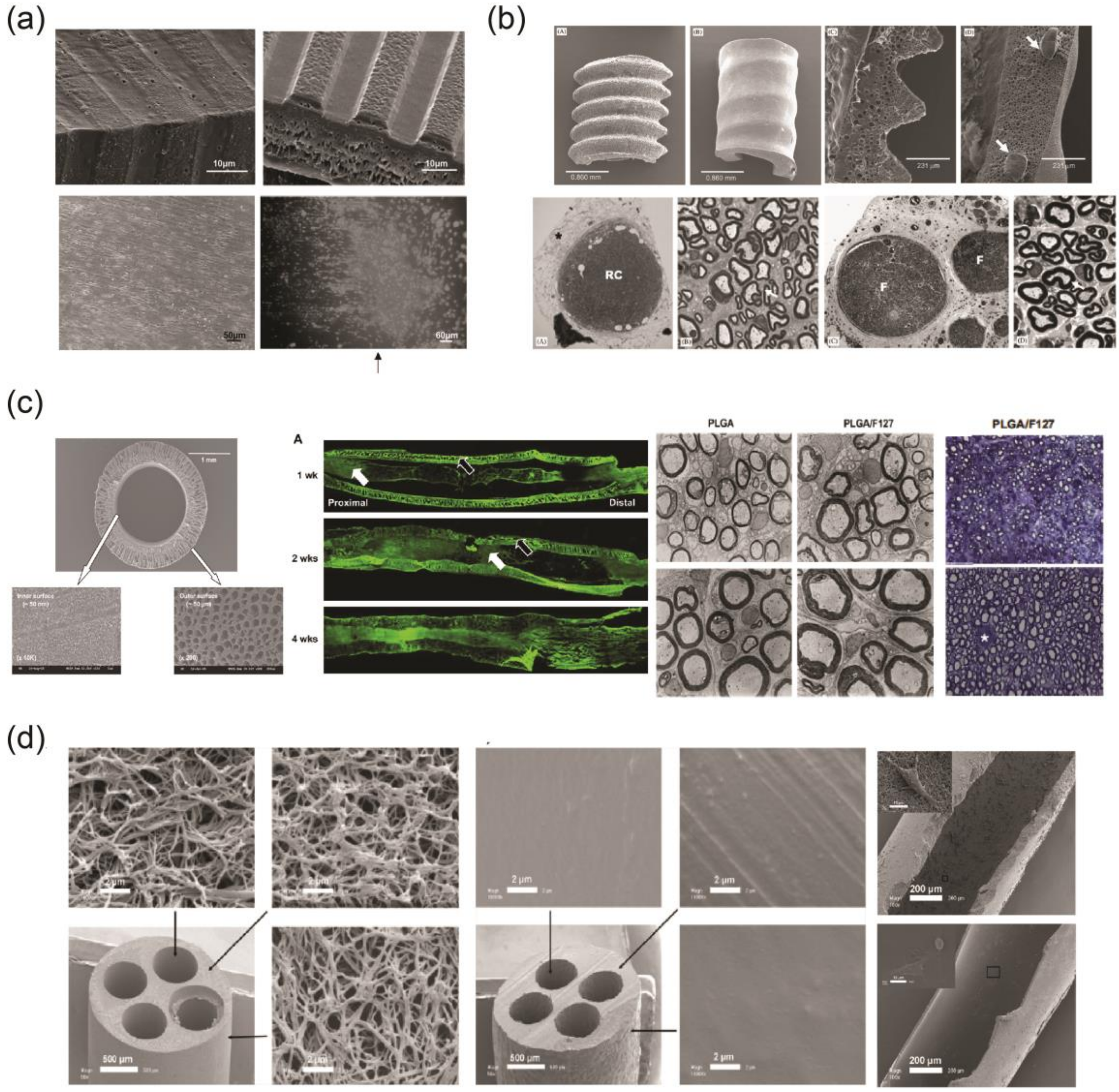

**Figure 8:** Fabrication of scaffolds for peripheral nerve regeneration applications through the use of conventional, porous and molding methods. (a) Solvent cast, molded, microgrooved PDLA substrates were able to induce alignment in Schwann cell cultures (Miller et al, 2001) (b) effect of reinforced PHEMA-MMA hydrogel tubes in the regeneration of a rat sciatic nerve injury model [120] (c) PLGA/Pluronic F127 scaffold tubes were able to induce nerve regeneration in a rat sciatic nerve injury model [43] and (d) PLLA nanofibrous scaffolds were created without the use of electrospinning and showed good adherence of PC12 cells and fibroblasts [44]. Reprinted with permission from (a) Miller et al 2001, (b) Katayama et al, 2006, (c) Oh et al, 2008, (d) Sun et al, 2012.

Belkas et al. tested the capacity for peripheral nerve regeneration through the use of hydrogel nerve tubes [121]. These tubes were created using poly(2-hydroxyethyl methacrylate-co-methyl methacrylate) (PHEMA-MMA), which is prepared by using 2-hydroxyethyl methacrylate (HEMA), methyl methacrylate (MMA) as the material to be polymerized and ethylene dimethacrylate (EDMA) as a crosslinking agent, as well as a redox initiating system composed of ammonium persulfate (APS) and sodium metabisulfite (SMBS). The final polymer blend produced was placed into custom built disposable molds [122]. The hydrogel nerve tubes that were created through this process were measured to have an inner diameter of 1.3 mm, an outer diameter of 1.8 mm and were 12 mm long. The nerve tubes were implanted into adult male Lewis rats, which had undergone an 8-mm segment nerve excision in the sciatic nerve and the nerve regeneration process was followed over a 16-week post-implantation period. The results were compared to the ones obtained when using the standard autograft technique, with nerve graft segments being obtained from isogenic donor rats. It is shown that even though the difference between the hydrogel nerve tubes and the autografts was quite significant at 4-weeks post-op, with no axonal regeneration being seen in the PHEMA-MMA tubes, this was eliminated by 8-weeks post-op and both categories exhibited similar regeneration rates. At 16-weeks post-op, 60% of tubes exhibited the same regenerative effects with the autografts, while the remaining 40% were significantly deteriorated due to the collapsing of the nerve tubes.

A follow-up study conducted in 2006 by Katayama and co-workers showed that the researchers were able to reinforce the hydrogel nerve tubes through the use of PCL coil structures within the mold (Fig. 8b) [120]. The “reinforced” nerve tubes were implanted in the same way as described above and were able to significantly improve nerve regeneration when compared to the gold standard of autografts.

In another *in vivo* study, Oh and co-workers used the NIPS method in order to produce nerve guides made of PLGA (Fig. 8c) [43]. Their fabrication technique included the immersion of an alginate hydrogel rod into a poly-lactic-co-glycolic acid (PLGA)/tetraglycol or PLGA/Pluronic F127/tetraglycol solution that were precipitated onto the alginate rod through the diffusion of water from the latter to the PLGA solutions. After washing with additional water, the resulting scaffolds were tubes made of PLGA and PLGA/Pluronic F127 (an additive that was used in order to make the PLGA more hydrophilic) respectively, with an inner diameter of ~1.5mm and a porous wall thickness of ~0.4mm. Further characterization of the tubular scaffolds was able to show that the PLGA/Pluronic F127 nerve guides had an asymmetric porous structure, with variably sized pores (~50nm on the inner tube surface and ~50μm on the outer tube surface). The *in vivo* tests were performed in Sprague-Dawley rats and involved the severing of a 10mm segment of the sciatic nerve and subsequent implantation of 12mm scaffold tubes that were placed between the two severed nerve stumps. This work indicated that the addition of Pluronic F127 played an important role in the maintenance of the tubular structure of the scaffold, while also deterring the formation of fibrous scar tissue, thus resulting in better nerve regeneration behavior. This was estimated at 0.35mm/day, with the 10-mm gap being bridged within the first 4 weeks, while at 12 and 24 weeks, the healed rat sciatic nerve also displayed a larger axon diameter as well as a thicker myelin sheath.

NIPS, in combination with microprinting, was also used to create nerve guides made of poly-lactic acid (PLA) [123]. More specifically, a PLA/dioxane solution was cast onto a negative polydimethylsiloxane (PDMS) microgrooved mold and the resulting samples were placed into varying concentrations of the non-solvent ethyl alcohol (20%, 40% and 95%) for 24h. This process resulted in the formation of both symmetric and asymmetric porous structures, while scaffolds with the same porous structures but lacking the outer microgrooved pattern were also created. Cell alignment was tested using the BCRC-60046 glioma cell line, with the findings showing that the microgrooved pattern was able to induce alignment. The *in vivo* tests were along the same vein as the ones described above, with the asymmetrically porous microgrooved scaffolds showing increased myelination rates at 4 and 6 weeks, compared to the symmetrically porous ones.

Sun et al. employed a combination of injection molding and TIPS in order to produce nanofibrous scaffolds without the use of electrospinning techniques (Fig. 8d) [44]. Through the use of a mold with varying numbers of acupuncture needles and a poly-L-lactic acid/tetrahydrofurane (PLLA/THF) solution, they were able to create single- and multiple-channel scaffolds, by injecting the solution into the molds and inducing TIPS at -80$^{o}$C for 12h. Control solid-walled scaffolds were also fabricated through the use of a solvent casting method, where a PLLA/DCM solution was placed into the same molds and placed into a fume hood in order to allow the evaporation of the solvent. By using different polymer concentrations, the length of the fibers forming the scaffold as well as its porosity could be adjusted accordingly. PC12 cells and fibroblasts that were seeded on the scaffolds showed better adherence to the nanofibrous ones compared to the solid-walled ones, making the former good candidates for further optimization in order to facilitate the orientation of regenerating tissues.

Electrospinning can produce highly aligned fibers of different diameters and, therefore, is ideal for the manufacture of nerve conduits. As a result, many scientists have used this technique for the creation of artificial grafts to replace peripheral nerve autografts (Fig. 9). In 2004, Bini et al. created hollow tubes of PLGA electrospun nanofibers which were collected with a rotating teflon tube mandrel [124]. The tubes were examined in the rat sciatic nerve model with a 10 mm gap and the results showed good nerve regeneration, high tube stability and no signs of inflammation. In a similar work, empty tubes made of biodegradable polymers (a blend of PLGA/PCL) were used for the repair of 10mm defects in rat sciatic nerve model (Fig. 9a) [125]. Also, hollow tubes of electrospun fibers of a copolymer of PCL and ethyl ethylene phosphate (PCL/EEP) were rolled up and wrapped with a thin film of PCL/EEP to create conduits for peripheral nerve regeneration (Fig. 9b) [126]. For better results, scientists encapsulated human glial cell-derived neurotrophic factor (GDNF) into the electrospun fibers. The protein was randomly dispersed throughout the polymer matrix in aggregate form and released in a sustained manner for up to two months. Such conduits were tested in a rat sciatic nerve defect model and showed great regeneration in all groups and sufficient electrophysiological recovery, compared to control animals.

In order to create conduits with adequate molecular and structural functionalities, Madduri et al. developed silk fibroin (SF) conduits loaded with GDNF and nerve growth factor (NGF) (Fig. 9c) [127]. In a first step, silk fibroin films with GDNF and NGF were created through the air-drying method. Then, aligned and randomly oriented nanofibers were collected on the SF films, using two parallel electrodes or flat plates as collectors respectively. SF films, decorated with nanofibers, were rolled around a teflon-coated steel mandrel and the ends glued together to form conduits. Finally, the tubes were sprayed with PLGA to prevent release of the neurotrophic factors to the exterior of the conduit. Nerve conduits were seeded with Neuro-2A and PC12 cell lines, DRGs and spinal cord explants. The results showed good cell response and sustained release of neurotrophic factors for 4 weeks.

Fibers of polysialic acid, a bioreactive molecule with a decisive role in peripheral nerve regeneration, were also fabricated via electrospinning [128], in combination with PEO to improve the mechanical properties. Substrates showed good cell viability and directed cell proliferation along the fibers *in vitro* using immortalized Schwann cells. 'Semi-solid" 3D-cylindrical constructs made of PCL fibers were constructed using a two-pole air gap electrospinning system (Fig. 9d) [129]. The collector in this experimental setup consisted of two vertical piers grounded to a common voltage with an additional set of horizontal piers from each vertical pier, projecting inwards with respect to the upright piers. Fibers were collected in the gap between the terminal ends of these projecting piers. The gap was adjustable from about 1 to 6 inches, giving the ability to create conduits with distinct material properties and intrafiber spacing, independent from the size of individual fibers. This 3D construct was coated with electrospun fibers of PGA/PLA copolymer and used for the reconstruction of 10 mm lesions in sciatic nerve in rats. The results showed dense, parallel arrays of myelinated and non-myelinated axons in the lesion site and functional blood vessels scattered throughout the implant.

As hollow tubes represent a relatively early evolutionary stage in nerve guide design, the trend in bioengineering is heading toward the creation of advanced nerve guidance channels (NGCs) with intraluminal structures, using sophisticated bio-fabrication methods (Fig. 10). These NGCs aim to direct the sprouting of axons, to provide growth factors and to inhibit scar formation at the site of injury. In this regard, Jeffries and Wang, in 2013, used electrospun PCL and PLGA fibers to fabricate conduits resembling a decellularized nerve [130]. Parallelly-aligned PCL fibers were wrapped around channel templates and randomly aligned PLGA fibers were used to cover the PCL fibers. Then, the composite was rolled up to create a conduit and the templates were removed to leave behind empty channels. To achieve this structure, a new type of collector was used: a parallel plate, open collector with combs attached to the long sides of the collector to hold the templates in place (Fig. 10a). The use of Schwannoma cells showed cell infiltration of the structure *in vitro*, supporting the idea of directing neurite outgrowth during regeneration.

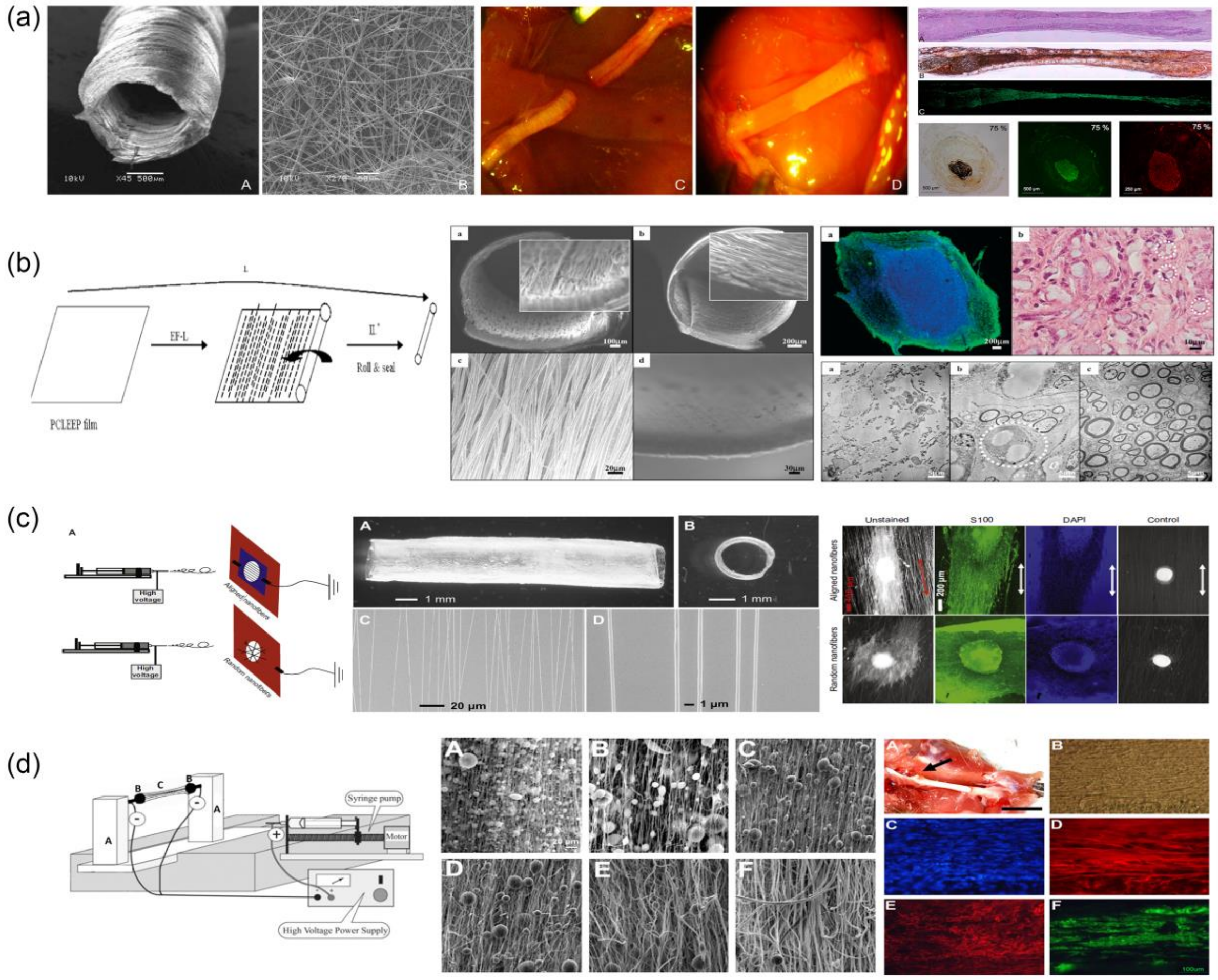

**Figure 9**: Fabrication of hollow conduits via electrospinning for peripheral nerve regeneration. (a) hollow tubes of PLGA electrospun nanofibers used to bridge a 10-mm gap in a rat sciatic nerve [124], (b) hollow tubes made of biodegradable blend of PLGA/PCL used for the repair of 10mm defects in rat sciatic nerve [125], (c) a thin film of PCLEEP PCL/EEP were roll up to create empty conduits for peripheral nerve regeneration [126], (d) silk fibroin conduits loaded with GDNF and NGF supported the growth of neural cells [127] and (e) "Semi-solid", 3D-cylindrical constructs made of PCL fibres used for the reconstruction of 10 mm lesions in rat sciatic nerve [129]. Images adapted with permission from (a) Panseri et al, 2004, (b) Chew et al, 2007, (c) Madduri et al, 2010 and (d) Jha et al, 2011.

From previously published results, it became obvious that the creation of conduits resembling peripheral nerves required the fabrication of different structures such as fibers and pores and a combination of materials with different mechanical properties. Therefore, scientists started searching for new materials and composites for potential applications in conduit manufacturing. Random and aligned nanofibers have been fabricated from poly(3-hydroxybutyrate-co-3-hydroxyvalerate (PHBV)/collagen which support proliferation and direct orientation of PC12 cells *in vitro* [131]. 3D scaffolds comprised of aligned electrospun PCL nanofibers, functionalized with laminin coating or embedded in 3D hyaluronic acid hydrogel, controlled neurite direction and enhanced neurite outgrowth of the neuroblastoma SH-SY5Y cell line [132]. Alignment of PCL fibers was achieved by collecting electrospun nanofibers with a grounded rotating drum. A poly(glycerol sebacate)-PMMA co-polymer was electrospun into fibers, collected on a flat plate and then coated with gelatin in order to be used for the culture of PC12 cells [133]. In this work, authors claimed that such 4D scaffolds can act as a temporary ECM to provide essential signalling cues to the cells that come into direct contact with it.

More recently, in 2018, advanced conduits with aligned conductive fibers were fabricated by Jing et al. [134]. PLGA nanofibers were electrospun and collected with a rotating cylinder to create a parallelly-aligned fibrous mesh. This fibrous mesh was rolled up and used to fill up an empty conduit which was created by dipping a mandrel in an emulsion of a co-polymer of PLGA and polypyrrole (PP) several times. Both materials (PLGA and PP) were non-cytotoxic to PC12 cells and were able to promote cell proliferation and differentiation. The aligned orientation of the fibers also provided strong guidance for neurites. In addition, the conduits displayed conductivity and degradability as well as oriented inner structure. Conduit effectiveness was proven by their ability to promote regeneration of injured sciatic nerves in a rat model.

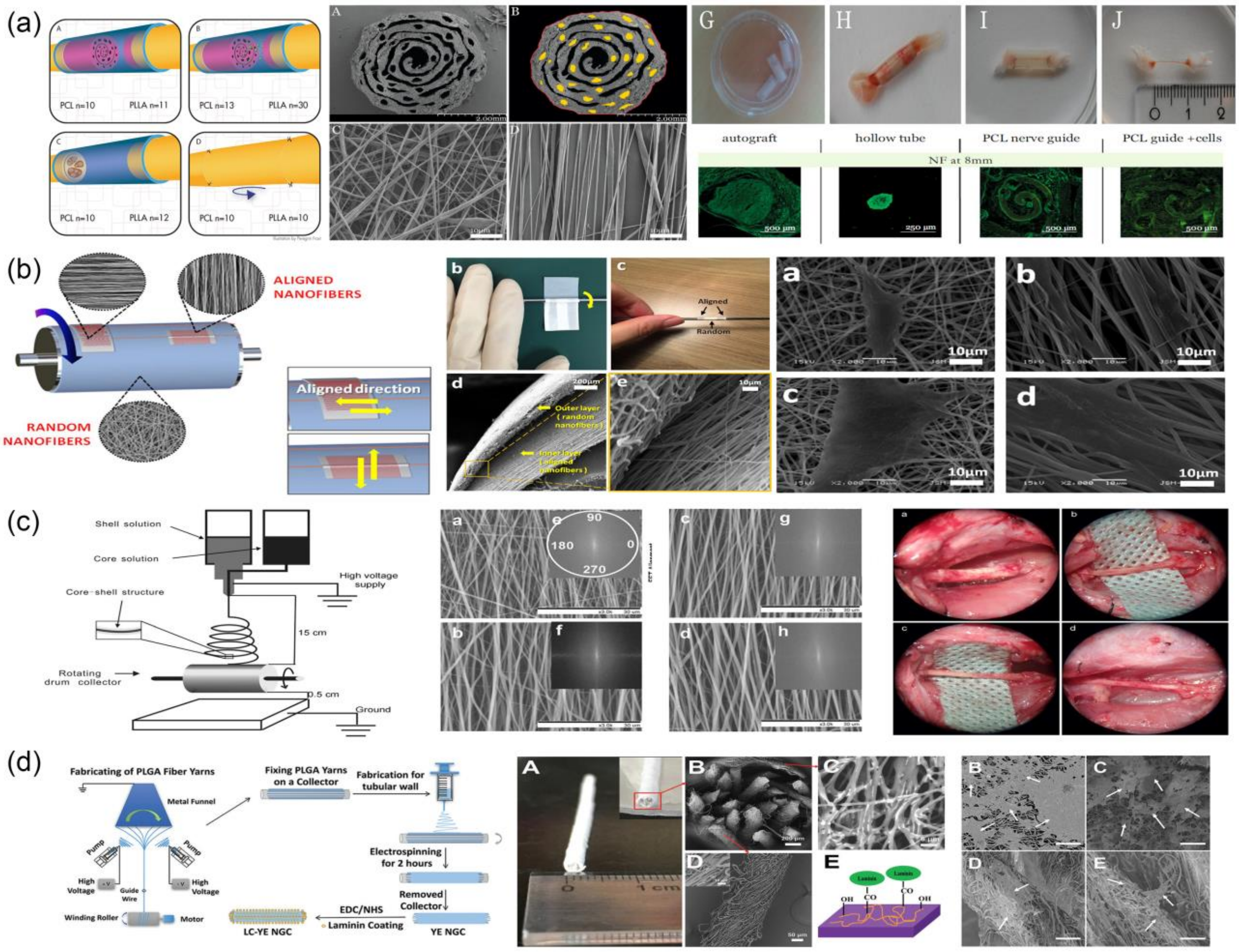

**Figure 10**: Fabrication of advanced nerve guidance channels (NGC) via electrospinning for peripheral nerve regeneration. (a) PCL/PLLA conduits seeded with autologous stem cells for repair of sciatic nerve defects in rats [135], (b) 3D bi-functionalized nerve conduits made of PLGA and PU supported the attachment and growth of neural cells [136], (c) core-shell PLGA nanofibrous NGCs loaded with NGF implanted across a 15-mm defect in the sciatic nerve [137] and (d) laminin-coated and yarn-encapsulated PLGA NGCs supported proliferation and migration of glia cells [138]. Images reprinted with permission from (a) Frost et al, 2018, (b) Kim et al, 2016, (c) Zhang et al, 2014 and (d) Wu et al, 2017.

Similar results were obtained from a different research group using autologous stem cells and PCL/PLLA conduits for repair of sciatic nerve defects in rats (Fig. 10a) [135]. For the creation of an aligned fibrous mesh of either PCL or PLLA, a U-shaped collector was used with sutures as templates to create empty channels. After electrospinning, the fibrous mesh was rolled up to form conduits and the sutures were carefully removed using tweezers. *In vivo* experiments in rats showed that the PLLA nerve guides supported axonal regeneration, although the results were inferior compared to autologous nerve grafts. The PLLA nerve guides combined microfibers for guidance and pores for cellular infiltration and could therefore also act as vehicles for co-transplanted cells.

In 2016, Kim et al. reported the use of a new electrospinning set up for the fabrication of 3D bi-functionalized nerve conduits made of PLGA and polyurethane (PU) [136]. This new setup included a rotating wheel as a collector, which was equipped with a single angled U-shape copper wire (Fig. 10b). The copper wire was set up horizontally on the collector and cellophane tapes were set over the fixed copper wire, along the vertical and horizontal axis. Vertically aligned nanofibers were collected on the horizontal cellophane tape, while horizontally aligned nanofibers were collected on the vertical cellophane tape. In addition, randomly aligned nanofibers were collected on the same mat in the gap between the U-shape copper wire. This setup allowed the fabrication of a mat comprised of aligned, in different directions, nanofibers and randomly oriented nanofibers in a single step. For the creation of a conduit, the mat was rolled up and covered with randomly aligned electrospun nanofibers. This new setup lends itself well to mass production, as it allows the fabrication of more than fifty mats at once. NGCs supported the attachment and growth of PC12 and S42 cells, thus proving that they can be used to assist peripheral nerve regeneration.

In a study by Zhang et al., core-shell PLGA nanofibrous NGCs loaded with NGF were fabricated via coaxial electrospinning (Fig. 10c) [137]. NGCs were seeded with PC12 cells and the results demonstrated that they presented a sustained release pattern and remained biologically active for over 60 days. Moreover, they supported peripheral nerve regeneration in rats, when implanted across a 15-mm defect in the sciatic nerve. Emulsion co-axial electrospinning was also used for the fabrication of scaffolds with PLLA/PCL nanofibers to form the shell and bovine serum albumin (BSA)/NGF to serve as the core [139]. The sustained release of NGF was sufficient to support differentiation of PC12 cells *in vitro*. Additionally, conduits of PLLA-PCL/silk fibroin nanofibers loaded with monosialoganglioside (GM1) and NGF in the core were constructed using the same method [140]. Although both conduits supported growth and differentiation of PC12 cells, the PLLA-PCL/silk fibroin conduit showed a synergistic effect between GM1 and NGF. The latter was also tested *in vivo* in rats and showed good performance in sciatic nerve regeneration. In a follow up study, PLLA-PCL/silk fibroin nanofibers loaded with B5 were again created using emulsion electrospinning [141]. These nanofibers supported Schwann cell proliferation and the *in vitro* release of B5 was characterized as sustained. A laminin-coated and yarn-encapsulated PLGA NGC (LC-YE-PLGA NGC) was fabricated using a double-nozzle electrospinning system (Fig. 10d) [138]. The conduits consisted of PLGA yarns with axial alignments in the lumen and PLGA fibers as the external tube wall, while both layers were covalently linked with laminin. LC-YE-PLGA NGCs supported proliferation and migration of SW cells *in vitro*.

A different approach to fabricate conduits for nerve guidance was proposed by Valmikinathan et al. [142]. They constructed PLGA microsphere-based spiral scaffolds using the method of microsphere sintering. Initially, PLGA microparticles were synthesized using traditional oil-in-water (O/W) emulsion techniques. In the next step, the microspheres were laid on an aluminium foil and sintered in an oven at 80$^{o}$C for 1 h. The microsphere sheet was then cut into thin strips, which were rolled into a spiral structure to form novel spiral scaffolds. In some cases, the strips were covered with PLGA electrospun fibers. These novel scaffolds were found to support the growth and proliferation of rat Schwann cells *in vitro*.

Alternatively, NGCs with more precise dimensions and complex internal architectures resembling natural structures in the nervous system can be fabricated by adapting Rapid Prototyping processes. Yamada et al. optimized the FDM technique to fabricate 3D-microstructures using biodegradable polylactides (PLA, poly-glycolic acid – PGA and PLGA) in order to establish a method for the creation of biodegradable implantable devices on demand for clinical applications [143]. PLA scaffolds were found to be biocompatible and supported the growth of PC12 cells. In 2009, Cui et al. formed double-layer polyurethane (PU)–collagen nerve conduits for peripheral nerve regeneration [144]. They used a modification of the FDM technique, known as double-nozzle low-temperature deposition manufacturing (DLDM). The DLDM system was equipped with two nozzles which were fed with the two different polymers: PU and collagen. CAD files directed the deposition of the inner circle (made out of collagen) first from the first nozzle and then the outer circle (made out of PU) from the second nozzle. When this procedure was finished, the platform was lowered for a fixed amount of time and the process was repeated until the 3D-structure completed. With this technique, scientists were able to simultaneously build a 3D-conduit comprised of an inner layer of aligned fibers of the natural polymer collagen and an outer porous layer made out of the synthetic polymer PU, thus combining the biocompatibility of the collagen and the mechanical stability of the PU.

Radulescu et al. in 2007 used ink-jet technology to build cylindrical conduits of PLA/PCL copolymer able to support a cell-regulated NGF-delivery system [145]. Their conduits were fabricated using standard inkjet micro-dispensing technology with piezoelectric devices and a stainless-steel mandrel, connected to a temperature control system, as a printing substrate. Using the same inkjet technology, Hu et al. were able to manufacture printed bio-conduits [146]. Their bio-ink was composed of cryopolymerized gelatin methacryloyl (cryoGelMA) gel cellularized with adipose-derived stem cells (ASCs). These conduits supported the survival and proliferation of the seeded ASCs *in vitro* and the re-innervation across a 10mm sciatic nerve gap in rat models *in vivo* (Fig. 11a).

Electrohydrodynamic Jet 3D Printing (EHD-jetting) was used for the fabrication of PCL-based three-dimensional porous NGCs [147]. The method belongs to the inkjet-based 3D-printing category, where the material droplets are formed using electrostatic forces. With the use of EHD-jetting, 3D porous scaffolds of PCL were printed and then rolled into tubes. The tube ends were heat-sealed together to form conduits. These scaffolds supported PC12 proliferation and differentiation *in vitro.*

Several extrusion-based printing methods have been used for the fabrication of conduits for peripheral nerve regeneration. Owens et al. in 2013 were able to bioprint for the first time 3D conduits composed only of mouse bone marrow stem cells (BMSCs) and Schwann cells supported by agarose rods [148]. The scaffolds were used in a rat sciatic nerve model and showed a good regeneration rate which was comparable to the one achieved by empty collagen conduits. More recently, scientists have used Schwann cells encapsulated in composite hydrogels of alginate, fibrin, hyaluronic acid, and/or RGD peptides as bio-ink to bioprint scaffolds for nerve regeneration [149]. These results show that the bioprinted scaffolds can promote the alignment of Schwann cells inside scaffolds and thus provide haptotactic cues to direct axon extension and regeneration.

In 2015, Johnson et al. fabricated anatomical nerve regeneration pathways via a microextrusion printing principle. For the pathway manufacture, they used 3D models, which were reverse engineered from rat anatomies by 3D-scanning [150]. The resulting structure was bifurcating pathways of silicone tubes (Fig. 11b) decorated with microgrooves and filled with hydrogel drops containing gradients of neurotrophic factors (GDNF and NGF). Alginate and methacrylate gelatin hydrogels were used as printing materials, while a custom microextrusion based 3D-printing system with a pneumatic dispensing system was used for the fabrication. With these 3D pathways, they achieved successful regeneration of complex nerve injuries *in vivo* in rats.

The same year, Evangelista et al. built single-lumen and multi-lumen conduits using a stereolithography-based 3D printing method and poly(ethylene glycol) diacrylate (PEGDA) as the main material [151]. UV light at 325nm was used for the crosslinking of the PEGDA hydrogel and the conduits were tested *in vivo* for sciatic nerve regeneration in rats. Among the fabricated scaffolds, the single-lumen conduits were superior in nerve regeneration compared to the multi-lumen ones and the number of regenerating axons was similar to that of a normal nerve.

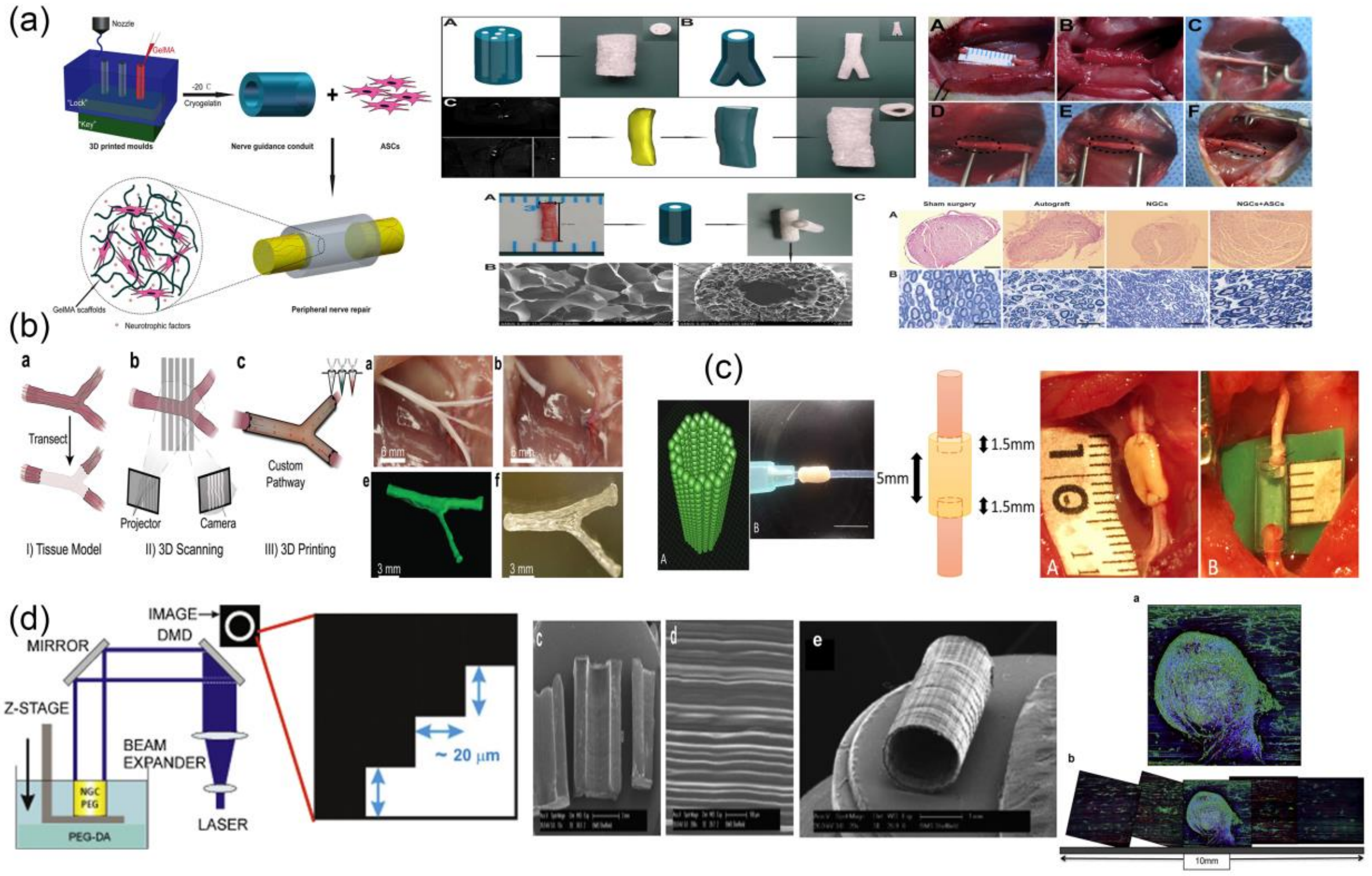

**Figure 11**: Nerve conduits for peripheral nerve regeneration fabricated using 3D printing. (a) bio-conduits composed of cryoGelMA and adipose-derived stem cells supported the re-innervation across a 10mm sciatic nerve gap in rat [146], (b) bifurcating silicone tubes decorated with microgrooves and filled with hydrogels with GDNF and NGF induced nerve regeneration in rats [150], (c) 3D tube-like conduits composed solely of cell spheroids used to repair a 3mm gap in a rat sciatic nerve [152] and (d) 3D channels made of photocurable PEGDA resin implanted in mouse[153]. Images adapted with permission from (a) Hu et al, 2016, (b) Johnson et al, 2015, (c) Yurie et al, 2017 and (d) Pateman et al, 2015.

Zhu et al. created a printable nano-bioink composed of gelatin methacrylamide (GelMA), neural stem cells (PC12) and bioactive graphene nanoplatelets that can promote neural cell differentiation [154]. They used this ink in a stereolithography-based 3D-bioprinting technique to fabricate a neural construct for potential neural tissue regeneration. The cells encapsulated in the hydrogel presented good cell viability at the low GelMA concentration, good differentiation and neurite elongation in a culture period of 2 weeks.

Laser-based microstereolithography was introduced by Pateman et al. in 2015 for the fabrication of NGCs [153]. The microstereolithography setup (Fig. 11d) was equipped with a 405 nm laser for the photocuring of PEGDA resin. They fabricated 3D channels for Schwann cells and dorsal root ganglion explant cultures *in vitro,* and conduits for *in vivo* implantation in the thy-1-YFP-H mouse model. Although the fabricated scaffolds supported the attachment and growth of neural cells *in vitro*, the *in vivo* results highlighted the need for intraluminal structures into conduits for better nerve guidance and regeneration.

In contrast to all natural and synthetic materials that have been used for conduits fabrication, Yurie and associates in 2017, created 3D tube-like conduits composed solely of homogeneous multicellular spheroids of fibroblast cells, using a 3D bio-printer (Fig. 11c) [152]. To assemble the conduits, cell spheroids were robotically placed into an array of skewers using a microneedle-based method known as “the Kenzan method” for bioprinting. In this method, which was invented by professor Koich Nakayama at Saga University, cell spheroids are placed in fine needle arrays and allowed to merge with adjacent spheroids until the final structure is firm enough not to require needles [155]. In the above-mentioned 3D tube-like conduits created by Yurie and associates, the Kenzan needles were removed one week after printing. The conduits were cultured further until the desired function and strength of the tissue was achieved. These scaffold-free Bio 3D-Conduits were tested for their ability to bridge a gap of 3 mm in a rat sciatic nerve. The results confirmed strong nerve regeneration, which was superior compared to the one achieved by a hollow silicone tube.

As a conclusion, it should be emphasized that the ideal conduit, according to the FDA, must be biocompatible and bioresorbable and should support neurite extension within a microenvironment that promotes peripheral nerve regeneration while minimizing the interactions between the myofibroblasts and axon growth. Based on these criteria, the FDA has approved several conduits for use in humans [156, 157], such as Neuragen®, AxoGuard™, Neuroflex™, NeuroMatrix™, SaluTunnel™, NeuraWrap™, NeuroMend™, Surgis® Nerve Cuff, Neurotube® and Neurolac®. However, as they are recommended only for small gaps (up to 3cm) and their biological performance is inferior compared to autografts, the need for new materials and for the development of alternative conduits for peripheral nerve regeneration remains imperative.

### 3.2. CNS – Traumatic Brain Injury and Neurodegenerative Disorders

The central nervous system (CNS) consists of the brain and the spinal cord (SC) and they are both characterized by the same types of glial cells as well as the presence of the blood-brain barrier (BBB) (Fig. 7a). There are four different glial cell types that exist within the CNS, which are quite different to the glial cells in the PNS, both morphologically, as well as functionally. Each type of glial cell in the CNS plays a distinct role: oligodendrocytes are responsible for the myelination of axons, astrocytes are more general supportive glial cells and play important roles in the traumatic injury response, as well as maintaining the BBB, ependymal cells have epithelial characteristics and form a continuous sheet that lines the ventricles, choroid plexuses and the central canal of the spinal cord, while microglia are scavenger-like cells whose function is similar to that of macrophages and who also play an important role in injury response.

When it comes to the neurons, the brain has some unique characteristics that make it differ greatly from the SC, which is the main reason that, even though they are both a part of the CNS, they are treated differently in response to injury. One such main difference is the fact that the brain contains a number of different neuronal cell types, depending on the area they are located in and the function they carry out. The brain has a distinct structure with three main compartments: the cerebrum, the brainstem and the cerebellum, each one of which is split into further smaller sub-compartments. Depending on their function, nerve cells in the brain can either be extended and branched, covering large areas and creating synapses with other neurons, or they can be much shorter in size and transmit the information they receive in a more localized fashion. There are a vast number of pathologies that can occur in the brain, ranging from injuries to disorders and diseases, and depending on the area in the brain that each one occurs, the patient in question displays characteristic symptoms.

In the CNS, injury occurs in two phases: the primary, or acute, phase and the secondary phase, with the latter being the one in which the injury spreads and causes further damage and cell death. During this phase, a number of different growth inhibitory molecules become released, such as certain proteoglycans and other molecules that are exposed due to the breakdown of the myelin sheaths and the surrounding supportive cells, all of which block the axon regeneration process. On the other hand, neurodegenerative disorders are characterized by the selective destruction of neurons in distinct areas of the brain. One such example is Parkinson's disease, where the neurons that produce the neurotransmitter dopamine are selectively destroyed in the regions of the basal ganglia. Additionally, the astrocytes in this region die, while the amount of microglia increases. In the human brain, there are 5 different pathways that are connected to the basal ganglia, including the motor and limbic circuits, so, as a result, all five are affected in Parkinson's disease, leading to characteristic symptoms in normal functions such as movement, attention and learning. Another example is Huntington's disease, which is caused by a genetic alteration in the huntingtin (HTT) gene, leading to the production of a mutated protein that is toxic to certain cell types. Early signs of damage are particularly evident in the striatum, but, during disease progression, the damage spreads to other areas of the brain. HTT has been found to be involved in a number of different pathways: it is essential for embryonic development, acts as an anti-apoptotic factor, controls the production of brain derived neurotrophic factor (BDNF), facilitates vesicular transport, controls neuronal gene transcription and is also involved in the machinery that controls synaptic transmission [158].

In order to repair any damage caused by injury or a degenerative disorder, the brain mounts an inflammatory response which differs from the one that occurs in other organs in the body and which becomes even more complicated when the trauma results in a compromised BBB. When implanted materials are used to facilitate the repair process, further injury is usually caused, but the effect of the implanted material far outweighs the negatives that may be caused during the implantation. Due to the fact that the inflammatory response in the brain aims to protect it from further damage, scientists are actively investigating which structures and which materials can facilitate the switching of the inflammatory response towards the reparative phenotype [159]. Because the CNS is a more complex system than the PNS, it means that even though the purpose of any implanted scaffold/structure is to guide the axon regeneration process, the CNS requires an additional component: the need for the implanted material to change the environment from inhibitory to pro-repair [3].

Considering the fact that current strategies for brain tissue regeneration are still insufficient, a variety of fabrication methods have been used in order to create scaffolds capable of assisting in the repair process after traumatic brain injury or neurodegenerative disorders and these will be analyzed in the following paragraphs.

In 2000, Holmes et al. investigated the formation of scaffolds through the design of self-assembling peptide sequences inspired by nature (Fig. 12a) [49]. In order to mimic the RGD motif found in a large number of extracellular matrix (ECM) proteins, they created two self-assembling peptides, RAD16-I (with a “RADA” repeated motif) and RAD16-II (with a “RARADADA” repeated motif). Both sequences are able to assemble into strings, tapes or sheets when added into an environment containing millimolar concentrations of monovalent salts (e.g. NaCl, KCl). The scaffolds created through this fabrication process were tested for their affinity to support neurite outgrowth in both cultured (rat PC12 and human SY5Y neuroblastoma cells) and primary cells (mouse cerebellar granule and hippocampal neurons, rat hippocampal neurons), with the neurites extending between 0.4-0.5mm in the former and 0.1-0.3mm in the latter. The scaffold was also shown to support the formation of active synapses.

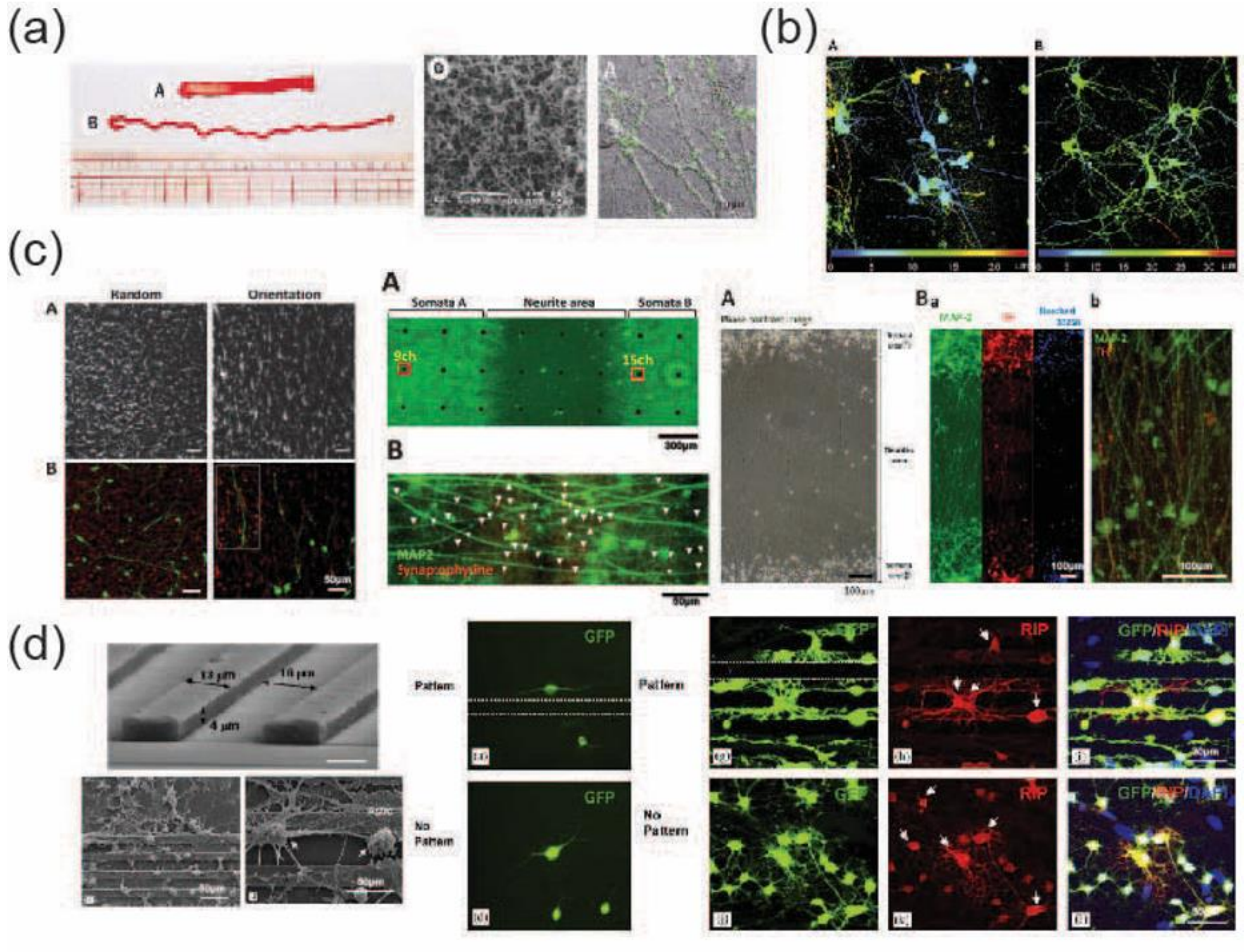

**Figure 12:** Use of conventional, porous and molding methods for the creation of scaffolds used for central nervous system regeneration. (a) Design of two novel self-assembling peptides (RAD16-I and RAD16-II) which mimicked the RDG motif found in ECM proteins exhibited strong affinity for neural cells, promoting neurite outgrowth [49], (b) entrapment of hippocampal neuron cells in 3D collagen hydrogels [160], (c) design of an oriented collagen fiber culture model showed oriented neurite extension [161] and (d) induction of differentiation of AHPCs due to the synergistic effect of patterned microgrooves, laminin and glial cell presence [137]. Images reprinted with permission from (a) Holmes et al, 2000 (Copyright (2000) National Academy of Sciences), (b) Xu et al, 2009, (c) Odawara et al, 2013 and (d) Recknor et al, 2006.

The RAD16-I self-assembling peptide scaffold was also used to entrap migrating potential neuroprogenitor cells from postnatal hippocampal slices [162]. After growing hippocampal slices on RAD16-I scaffolds that were 100-500µm thick for ~1 week, glial and neuroprogenitor cells were able to migrate into the scaffold. These were then easily collected and used to initiate new cultures, which were mitotically active for at least 2-3 days after collection. The affinity of this scaffold for the migrating cells makes it a useful tool for the enrichment of neural populations, so that they can then be used for further studies (such as characterizing neural cells or even harvesting them to investigate their potential in cell-based therapies). The same peptide was used in another study by Ellis-Behnke et al. where they attempted to investigate optic tract regeneration after completely severing the retina from the brachium of the superior colliculus and blocking the reinnervation with the injection of saline into the wound [163]. The creation of a scaffold made from the RAD16-I peptide injected into the site of injury was found to allow the elongation of axons through the gap created by the injury, while in adult hamsters, the researchers also noted a return of functional vision. From these results, they postulated that the scaffold could potentially enhance cell migration and thus, promote the elongation of the axons through the lesion site. This particular study was the first-time researchers were able to demonstrate healing in an *in vivo* setup, paving the way for the further development of such biomedical technology tools.

In an attempt to create more physiologically relevant 3D culture models to assess neural cell-cell and cell-material interactions, a type I collagen hydrogel was used to create a 3D structure that could entrap neuronal cells (Fig. 12b) [160]. Culture of rat embryonic hippocampal neurons either in entrapment 3D collagen hydrogel scaffolds, 3D sandwich hydrogel scaffolds (where the cells were seeded at the interface between two separate collagen hydrogels) or on 2D collagen coverslips showed that the cells were able to maintain their viability, metabolic function as well as their electrophysiological activity, regardless of the type of culture they were in. However, they found that the 3D entrapment model resembled the real *in vivo* conditions more closely, due to the fact that the cells were evenly distributed within the matrix at the beginning of the experimental process and the neurites were able to extend within the pore network of the hydrogel structure, while both the 3D sandwich and 2D coverslip methods only exhibited neurite outgrowth on the same plane that the cells had been seeded during the experimental setup. In a similar study, Frampton et al. entrapped a variety of cell types within an alginate 3D construct and assessed the effect of the entrapment and the material on the viability of the cells [55]. Using primary astroglial, microglial and hippocampal neuron cells that were entrapped in an alginate hydrogel, it was shown that all cells remained viable and metabolically active, while neuronal cells were able to create functional synaptic elements which displayed electrical activity.

In 2013, Odawara et al. took the complexity for the creation of 3D culture models one step further, by creating a cell culturing platform that allowed the control of the neurite extension direction, while maintaining the neural cell somata in a limited space, thus mimicking the organization that can be found in the brain (Fig. 12c) [161]. More specifically, the fabrication process that was followed was the following: (1) creation of PDMS sheets that contained microchambers of a specific volume and placement onto a non-treated polystyrene (PS) tissue culture dish, (2) coating of the chambers with poly-D-lysine (PDL) and addition of the cell suspension, (3) incubation at 37°C and 5% $CO_2$ for 30min and subsequent removal of the PDMS sheet, (4) placement of the culture dish at a 45° angle, addition of liquid rat tail collagen solution and incubation at 37°C to achieve gelation of the collagen with simultaneous collagen fiber orientation/alignment. The cells embedded within this collagen matrix were either rat embryonic hippocampal neurons or human iPSC-derived neurons and they were cultured for certain time points in order to observe the elongation of the neurites. The researchers found that the oriented collagen fibers were able to promote the outgrowth of functional and electrically active neurite extensions in the same direction as the fibers.

Using a rather different approach, Recknor and coworkers used solvent casting to create micropatterned substrates using polystyrene (PS) [164, 165]. Reactive ion etching and photolithographic techniques were used to create the micropatterned surface on the silicon wafer, which was then transferred to the PS substrates using solvent casting. A PS/toluene solution was spin- or gravity-cast onto the micropatterned surface to create substrates between 50 and 70μm thick. After the substrates dried for 24h, they were removed from the micropatterned surface by soaking in $dH_2O$. These substrates were subsequently used to assess their effects on astroglia and neural progenitor cells. Initial experiments with the former were able to show that the single cue of the grooved micropatterned substrate required the additional presence of adsorbed laminin in order to induce directionality in the astrocytes [164]. When the same substrates were used with adult hippocampal progenitor cells (AHPCs) in co-culture with astroglial cells, the synergistic effect between the microgrooved substrate, the adsorbed laminin and the presence of the glial cells induced a much higher degree of differentiation in AHPCs, when compared to AHPCs grown on the same substrates without the presence of the astroglial cells (Fig. 12d) [165].

In 2006, Ao and coworkers were able to create multimicrotubule chitosan conduits through a combination of novel molds and thermally-induced phase separation (TIPS) [166]. The diameter of the microtubules was controllable through the use of different cooling temperatures and polymer concentrations. When Neuro2a cells were grown on these conduits, they were able to grow within the microtubules, as well as respond to the addition of the differentiation agent retinoic acid, thus making the conduits attractive candidates for nerve tissue engineering applications.

Möllers et al. produced collagen scaffolds using a directed freeze-drying process that leads to the formation of parallel oriented pores [167]. These scaffolds were then tested in cell culture of different CNS cell populations: rat olfactory ensheathing cells (OECs), rat astrocytes and the human neuroblastoma SH-SY5Y cell line, with the glial cells (OECs and astrocytes) displaying excellent cell attachment, proliferation and migration, while the neuronal cells also showed directed axonal growth.

Electrospun fibrous scaffolds are very popular for applications in CNS repair. In 2004, Yang et al. reported the fabrication of a 3D scaffold with random oriented electrospun PLLA fibers and interconnected pores that supported neural stem cell differentiation [168]. The nanofibrous structure was created using electrospinning with a flat collector and was designed to resemble the natural extracellular matrix structure in the human body. In a later study, the same group reported the fabrication of highly aligned and randomly oriented PLLA micro- and nano-fibers using electrospinning for neural tissue engineering applications (Fig. 13a) [78]. For the aligned fiber collection, a rotating disk was used, whereas a flat aluminum plate was used for collecting random fibers. The fiber diameter was dictated by the polymer concentration and was proportionally increased. These scaffolds were seeded with neonatal mouse cerebellum C17.2 cells and the results showed that the direction of NSC elongation and neurite outgrowth was parallel to the direction of PLLA fibers for the aligned scaffolds. Moreover, aligned nanofibers induced longer neurite outgrowth than fibers with diameters on the microscale implying a connection between fiber diameter and neurite outgrowth.

Nikkola et al. created diclofenac sodium (DS)-releasing PCL nanofibers using electrospinning for drug delivery in the brain [169]. This highly porous nanofiber scaffold exhibited a fast drug release rate, with about 45% of DS being released within the first 24 hours. Additionally, for nerve regeneration and guidance applications, randomly or aligned nanofibers were created using the co-polymer PCL/gelatin [170]. Electrospinning with a flat or rotating wheel was used for the formation of nanofibers and the ratio of PCL/gelatin 70:30 was proven to be the most suitable for the adhesion and proliferation of C17.2 neural stem cells (Fig. 13b). 3D scaffolds of electrospun PCL nanofibers were created by Horne et al. to be used for the repair of damaged brain tissue [171]. PDL nanofibers were deposited on a rotating aluminum mandrel and were functionalized by immobilization of BDNF. These scaffolds were seeded with mouse embryonic neural stem cells (NSCs) and the results showed that they enhanced proliferation and directed cell fate toward neuronal and oligodendrocyte cells, which are essential for neural tissue repair.

In order to shed more light on the effect of the fiber's characteristics on cell response, He and coworkers created PLLA nanofibers, with diameters ranging from 200-900 nm. They conducted cell culture experiments with neonatal mouse cerebellum C17.2 cells on and showed that cell viability and proliferation was best on 500 nm fibers and reduced on smaller or larger fibers (Fig. 13c) [172]. These findings showed that the connection between neurite outgrowth and fiber diameter is more complex than it was initially thought. Fon et al. also used electrospinning with a rotating aluminium mandrel to create aligned PCL nanofibers, encapsulated with a BDNF-mimetic ligand [173]. These nanofibers were implanted in rat brains to investigate their potential in the regeneration of damaged tissue (Fig. 13d). The results showed the migration of neuroblasts towards the implant and infiltration of nanofibers by these cells. There is also evidence of infiltrated neuroblast cell survival and maturation. Parallel studies reported the use of 3D electrospun PLA nanofibers which release lactate for the regeneration of tissue within the CNS [174]. Scaffold implantation into rat brain cavities induced vascularization, neurogenesis and survival of newly generated neurons into normal brain circuits (Fig. 13e). At the same time, Zanden et al. created electrospun polyether-based PU nanofibers treated with plasma [175]. They showed that their scaffolds promoted adherence, proliferation and differentiation of human embryonic stem cells (hESCs) and rat postnatal neural stem cells (Fig. 13f).

Working toward the development of an injectable hydrogel for brain tissue regeneration, in 2015, Rivet et al. created a hybrid scaffold comprised of electrospun PLLA and PLLA/fibronectin fibers embedded in agarose/methylcellulose hydrogel [176]. To do that, first the fibers were electrospun on a thin film of polyvinyl alcohol (PVA) on a glass slide (Fig. 13g). Then, the film with the fibers attached was removed from the glass and cut into short segments. In the next step, the film of PVA was dissolved in $dH_2O$ and the segments were separated by centrifugation. Finally, the segments with the fibers were added into a tube which was filled with hydrogel and they were dispersed within the gel by repeatedly passing the mixture through a syringe. The hybrid scaffolds were injected into the rat striatum and the results showed the migration and infiltration of macrophages/microglia and resident astrocytes toward the implant. These findings indicate that neural cells were able to locate the fibers and utilize their cues for migration into the hybrid matrix.

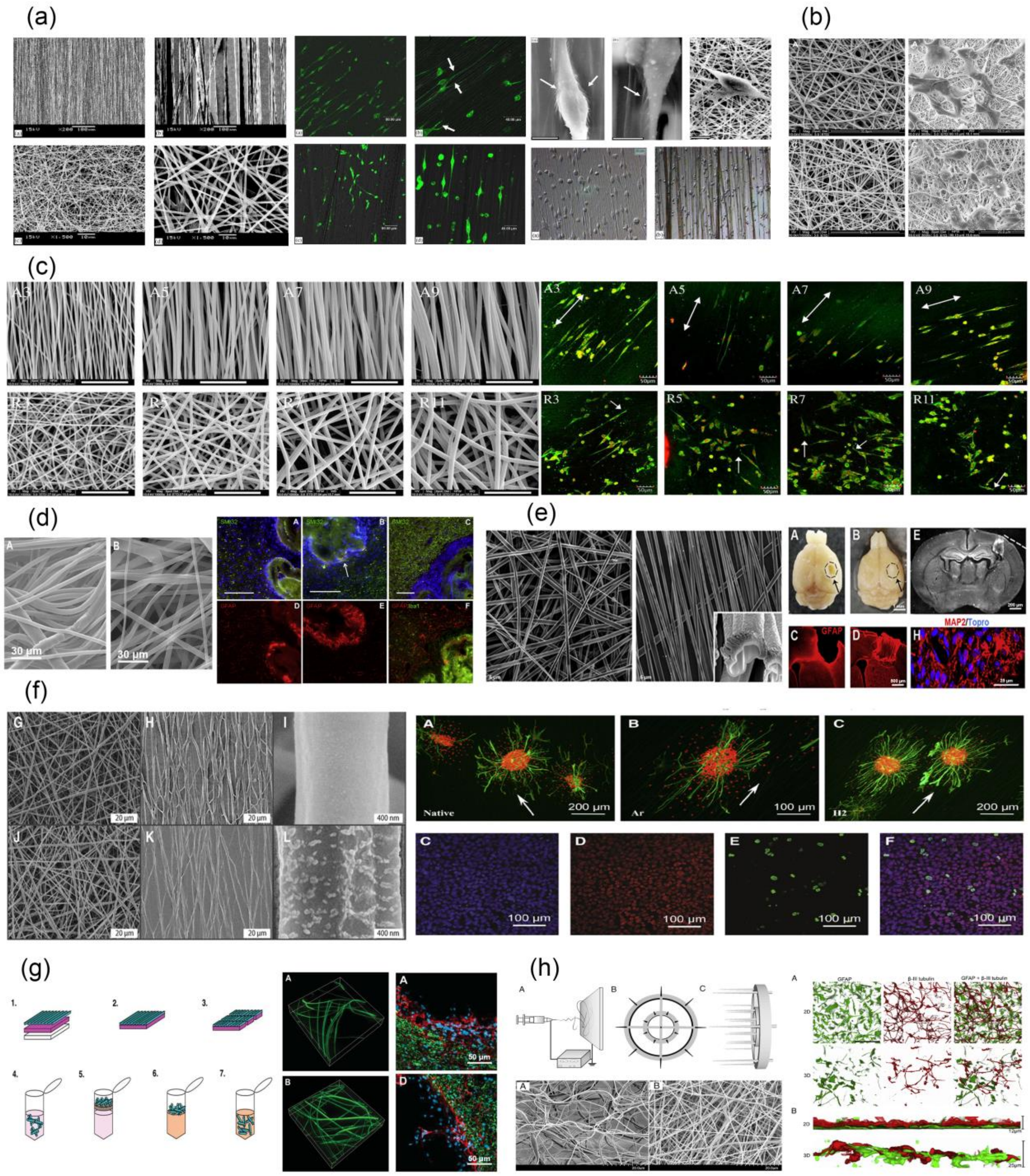

**Figure 13**: Use of electrospinning for the creation of fibrous scaffolds for central nervous system regeneration. (a) electrospun PLLA micro- and nano-fibers for neural stem differentiation [168], (b) aligned nanofibers made of PCL/gelatin co-polymer support adhesion and proliferation of neural cells [170], (c) PLLA nanofibers direct neurite outgrowth of [172], (d) aligned PCL nanofibers encapsulated with a non-peptide ligand were implanted in rat brain [173], (e) lactate-releasing 3D electrospun PLA nanofibers implanted in rat brain to support neurogenesis [174], (f) electrospun polyether-based PU nanofibers with surface modification [175], (g) hybrid

scaffolds comprised of electrospun PLLA/fibronectin fibers and hydrogels injected into the rat brain [176] and (h) fabrication of uncompressed low-density electrospun PCL fiber scaffolds using a custom-made rotating collector for 3D cell assay with human neural progenitor cells [177]. Images adapted with permission from (a) Yang et al, 2005, (b) Ghasemi-Mobarakeh et al, 2008, (c) He et al, 2010, (d) Fon et al, 2014, (e) Alvarez et al, 2014, (f) Zanden et al, 2014, (g) Rivet et al, 2015 and (h) Jakobsson et al, 2017.

More recently, Jakobsson et al. used a custom-made rotating collector (Fig. 13h) for the fabrication of uncompressed low-density electrospun fiber scaffolds made out of PCL [177]. The collector was designed to hold two circular arrays of stainless-steel needles, which were connected to a common grounded base. The result of electrospinning using this collector was a highly porous low-density fiber scaffold with maintained interconnecting pores. These fibrous scaffolds were seeded with cells from a human neural progenitor cell line originating from a 7-week old forebrain. Confocal images revealed that the cells in 2D cultures created a dense layer of glia beneath the layer of neurons, while in 3D cultures, the glial and neuronal cells were found to intermingle, creating networks spanning a few μm in depth. These findings indicated that this 3D cell assay will most likely provide more physiologically relevant results compared to conventional 2D cultures.

3D printing has also been used in applications for CNS regeneration. Zhang et al. in 2007 created three-dimensional hydrogel structures, composed of gelatin and gelatin/hyaluronan in order to investigate their effects in the reparation of injury in the central nervous system [178]. They used a custom-made cell assembly machine for the deposition of the polymers in predefined 3D patterns. The 3D patterns consisted of square grids which were designed using a software package. These 3D constructs were implanted into cortical defects of rat brains for the evaluation of their ability to improve tissue reconstruction. The results proved that both constructs were biocompatible, with the gelatin/hyaluronan scaffolds showing better contiguity with the surrounding tissue. In a similar study, a bio-ink composed of thermo-responsive polyurethane, which can form a hydrogel at 37$^{o}$C without a crosslinker, and NSCs were used to bioprint 3D scaffolds [179]. For the scaffold fabrication, a self-developed FDM equipment was used (Fig. 14a). NSCs, which were embedded into PU before gelation, exhibited excellent proliferation and differentiation *in vitro*. Another point worth mentioning in this report is the use of zebrafish as neural injury model. PU dispersion and NSCs were injected in adult zebrafish and reduced the mortality after traumatic brain injury to 37%.

Taking the use of hydrogels for brain regeneration one step further, Gu et al. incorporated cells, thus creating a bio-ink. They  used a 3D-Bioplotter® system to bioprint functional 3D mini-tissues made of bio-ink composed of hydrogel and neural stem cells (hNSCs) [180]. For the formation of the hydrogel solution, agarose in several concentration was dissolved by heating, and alginate and carboxymethyl-chitosan were added subsequently in that order. The hydrogel solutions were cooled afterwards to room temperature and combined with hNSCs for the direct-write bioprinting (Fig. 14b). Finally, the samples were extrusion-printed into a cubic construct. The constructs supported cell cultivation for several weeks, indicating the beneficial mechanical properties of the hydrogel. Importantly, hNSCs were able to differentiate into functional neurons and neuroglia, which supports the idea of the potential use of the 3D constructs for investigation of the human neural regeneration, function and disease. Lee et al. combined stereolithography-based 3D printing and electrospinning techniques to fabricate a novel 3D biomimetic neural scaffold [181]. Aligned PCL and PCL/gelatin microfibers were collected on a rotating mandrel and placed at the bottom of a petri dish. Then, printable hydrogel inks, composed of PEG and PEGDA covered the electrospun fibers. Scaffolds were seeded with NSCs and primary cortical neurons and the results showed increased cell adhesion and proliferation. Also, the fibers increased the neurite length and guided the neurite extensions of primary cortical neurons along them.

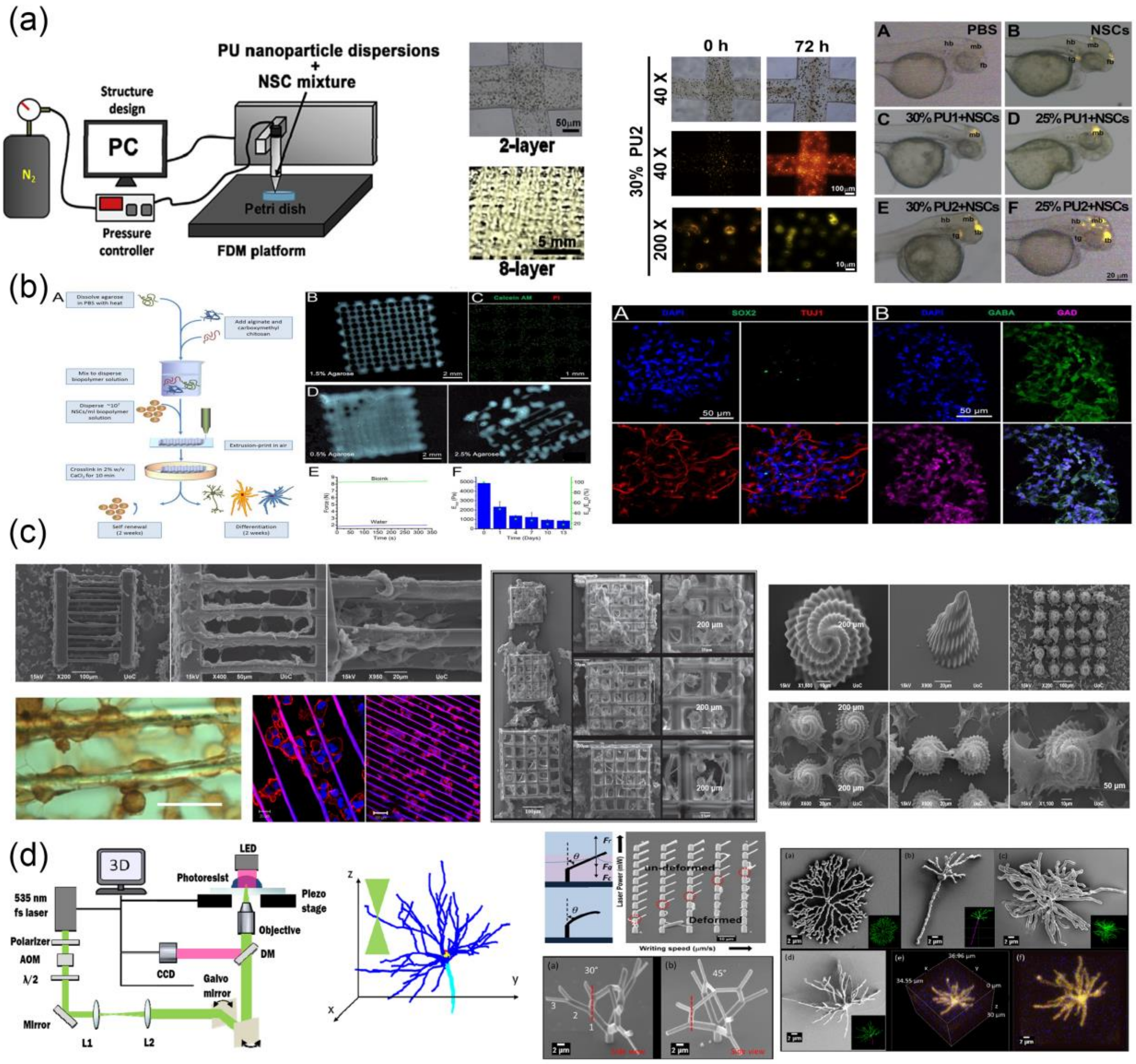

**Figure 14**: 3D printed scaffolds for CNS applications. (a) use of FDM technique for bioprinting 3D scaffolds composed of PU and NSCs to support neural cell differentiation [179], (b) direct-write printing of bioink composed of hydrogel and hNSCs into porous 3D scaffolds to study the human brain development and function [182], (c) PLA-based 3D structures written via DLW can be used as test-bed for neural tissue engineering applications [183] and (d) zirconium-based 3D biomimetic neuron structures fabricated with DLW method [184]. Images adapted with permission from (a) Hsieh et al, 2015, (b) Gu et al, 2016, (c) Melissinaki et al, 2011 and (d) Yu et al, 2018.

A different approach for the fabrication of 3D scaffolds for brain tissue regeneration was introduced by Melissinaki et al in 2011. They used direct laser writing (DLW) (a method that belongs to Stereolithography-based 3D printing) for the fabrication of 3D scaffolds made out of PLA, for neural tissue applications (Fig. 14c) [183]. A Ti:Sapphire femtosecond laser was employed for the polymerization of the 4-star arm PLA in a layer-by-layer fashion with the last layer printed on the surface of the coverslip. Several 3D structures were created such as cross-hatched structures, woodpile structures and sea-shell structures, demonstrating the suitability of the material for DLW. The minimum feature size achieved with their optical setup was approximately 800 nm and the structures were constructed on a timescale of 10–30 min. Neuronal cell lines (NG108-15 and PC12) were cultured on the fabricated scaffolds and the results showed good neurocompatibility, suggesting that this material can readily be used as scaffolds for neuronal tissue engineering applications. DLW was also used by Yu and coworkers in 2018 to fabricate 3D biomimetic scaffolds resembling the multiple branches structures of neurons [184]. Their set up was equipped with a femtosecond laser, operating at 535nm and a zirconium-based hybrid organic–inorganic photoresist was used as scaffold material (Fig. 14d). Moreover, in their study they included a branching model to adjust the mechanical properties of the scaffold and to eliminate the capillary force. Using this branching model, they were able to control the laser power and the writing speed in order to tune the branch dimensions which satisfied the balance between capillary force and elastic restoring force. The final result was the creation of biomimetic neuron structures with different branch angles, branch lengths and branch diameters imitating neuron cells from different parts of the brain in rats and rabbits. This work paved the way for the use of DLW towards applications in engineered neural networks.

### 3.3. CNS – Spinal cord injury (SCI)

The spinal cord (SC) is part of the central nervous system and, as such, contains neurons and CNS glial cells (oligodendrocytes, astrocytes and microglia). However, its architecture is more reminiscent of the PNS, as the spinal cord is characterized by long cables of myelinated axons that are bundled together (Fig. 7b). An additional characteristic of the SC over the PNS is the fact that the neurons are able to connect to each other and create rather complex signalling loops.

Injury in the SC leads to a very different response when compared to injury in the PNS. As mentioned earlier, PNI leads to activation of Schwann cells into a pro-regenerative phenotype that aims to promote axon regeneration. In contrast, SCI activates inhibitory signals to block this process. The response to SCI occurs in two major phases: the primary or acute phase and the secondary phase. In the primary phase, the SC becomes injured, swells, disrupts the blood flow and the neural cells become uncontrollably activated due to a disruption in the ionic concentrations within the cells, which leads to a loss of local neural function. In the secondary phase, the zone of cell death expands, due to the disruption of the local vasculature, and the bulk release of neurotransmitters during the primary phase leads to the production of free radicals which further enhances the level of apoptotic cell death. During the second phase, the inhibitory phenotype of SCI comes into play. The astrocytes and microglia that are present at the site of injury take on a reactive phenotype and lead to the formation of a glial scar containing growth-inhibitory factors.

Strategies to treat SCI are limited, mainly due to the fact that the acute phase occurs rapidly after the injury, making it impossible to interfere with. With the steroid methylprednisolone being the only drug approved by the FDA for SCI, there is an imperative need to discover alternative approaches for neuroprotection and regeneration. Such alternative treatments are targeted at either limiting the extent of injury caused during the secondary phase or bridging the gap caused during injury to facilitate axon regeneration. Both types of treatments rely heavily on the development of novel biomaterials and the field of tissue engineering. The biomaterials can either be used as drug delivery systems for targeted administration at the site of injury or even act as bridges offering guidance cues for axon regeneration and elongation, while also blocking the formation of glial scars. Such applications for spinal cord injury treatment will be further analyzed below.

Despite the fact that DRGs lie within the PNS, the regrowth of sensory axons of DRGs into the spinal cord after crush lesions makes them a particularly useful experimental model to study axon regeneration within the CNS [185]. This is because the sensory neurons of DRGs are pseudo-unipolar which means that they have an axon with two branches (a peripheral and a central branch) that act as a single axon. The peripheral branch is easily accessible to manipulations without disrupting the spinal cord, while the regeneration of the central branch into the SC offers the ability to study the inhibitory factors within the CNS. Therefore, many studies use the DRG model to test materials and scaffolds for treatment of SCI.

The ability of a self-assembling peptide scaffold to promote cell differentiation was tested by assessing the differentiation of neural progenitor cells (NPCs) [186]. The peptide amphiphile (PA) that was designed in this study contained an IKVAV sequence (isoleucine-lysine-valine-alanine-valine), an epitope that is found in laminin and is known to promote neurite sprouting and growth. A control PA containing the sequence EQS (glutamic acid-glutamine-serine) (EQS-PA) was also created and found to be biocompatible but non-bioactive. Self-assembly of this peptide amphiphile (IKVAV-PA) was achieved through adding the aqueous peptide amphiphile solution to suspensions of NPCs in media at a 1:1 ratio, with the resulting structure being a hydrogel scaffold with incorporated dissociated NPCs or clusters of NPCs known as neurospheres. Cells remained viable throughout the process and neurite outgrowth was observed at a much faster rate than in cells grown on PDL- or laminin-coated substrates. The PA nanofibers were also able to elicit the same response when in a 2D culture system. Additionally, the peptide amphiphile scaffold self-assembly process was also triggered *in vivo* by injecting the PA solutions directly into the tissues (such as the spinal cord), where the aqueous environment led to the formation of a solid scaffold in the area of injection and which were found to be well tolerated by the animals. This was further investigated in 2008, where the IKVAV-PA self-assembling peptide was used in an *in vivo* study in order to test its properties and assess its therapeutic potential in spinal cord injury (Fig. 15a) [187]. Using a clip compression model of SCI in mice, researchers investigated the effect of the IKVAV-PA in the recovery process. They injected IKVAV-PA,

IKVAV alone, EQS-PA or glucose or gave the mice a mock injection and studied them for a number of weeks thereafter. Behavioral improvement was more pronounced and astrogliosis, as well as apoptotic cell death, were reduced in the IKVAV-PA group. All these results showed that the IKVAV-PA self-assembling peptide was able to promote partial recovery of SCI, however, further optimization is required in order to improve its mechanical and biological properties and achieve more significant results.

In 2005, Yang et al. created single and multiple lumen nerve conduits and bridges with NGF, for application in PNS and SCI respectively (Fig. 15b). They used a combination of molding, gas foaming and particulate leaching fabrication techniques [188]. The first step involved the creation of microspheres using a cryogenic double emulsion technique, where a protein solution containing NGF was added to a poly(lactic acid)/dichloromethane (PLA/DCM) solution and frozen in liquid nitrogen, followed by the addition of two different PVA solutions which resulted in microsphere formation. These NGF-loaded microspheres were subsequently mixed with NaCl and added to a mold containing different numbers of stainless-steel pins to create the lumens. This was then placed under pressure for 12-16h, followed by a leaching time of 4h in water in order to remove the porogen (NaCl). The NGF that was released from the conduits was found to be bioactive and promoted rat DRG neurite extension. The resulting nerve conduits were also implanted subcutaneously in order to assess the stability of the structure as a whole. After 13 days, the implants had become invaded by cells from the surrounding tissue, while the channels remained intact. This characteristic, combined with the controlled delivery of NGF, makes these nerve conduits attractive to use in tissue engineering applications to promote nerve regeneration.

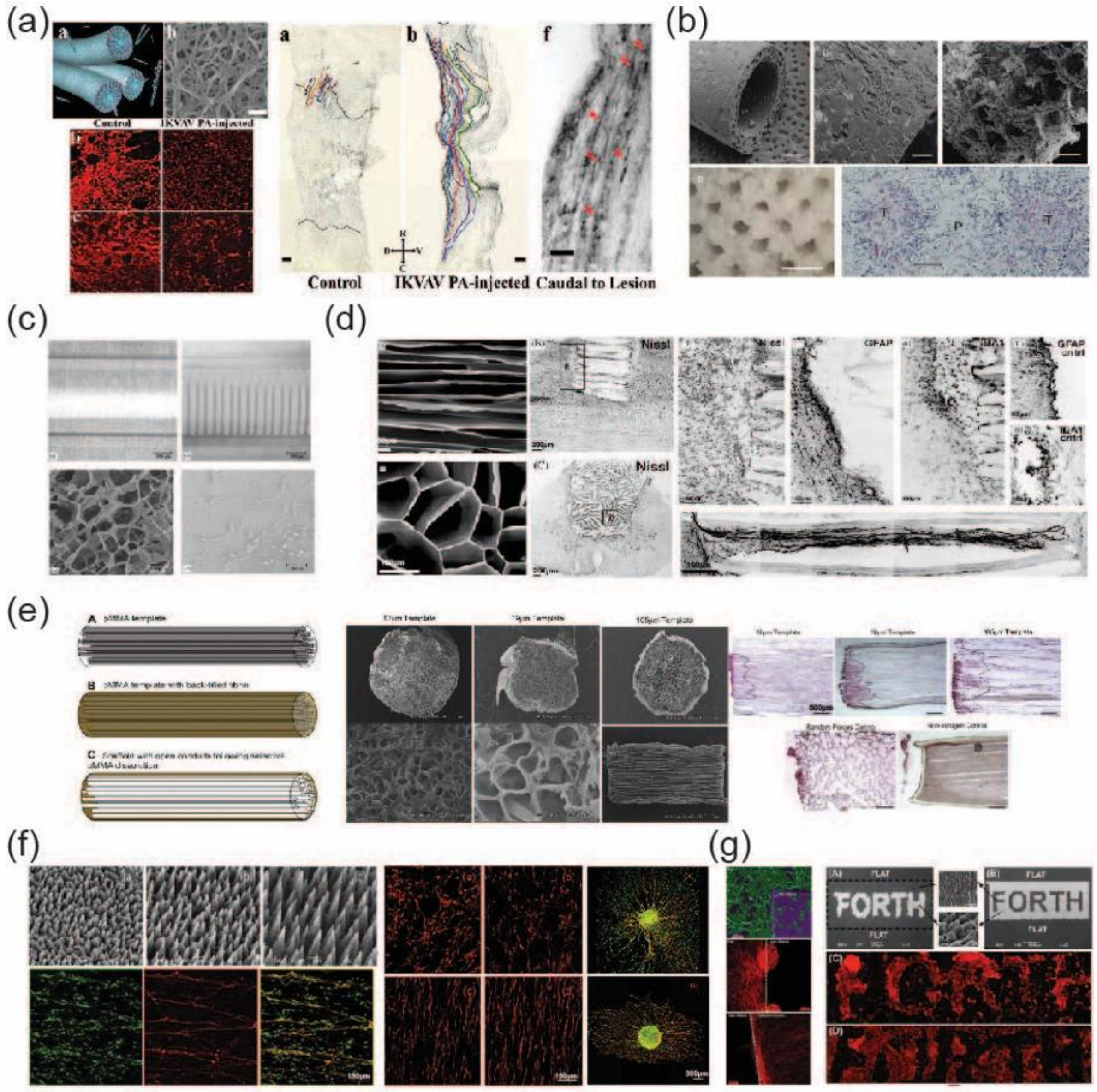

**Figure 15:** Conventional, porous and molding methods for the creation of tissue engineering scaffolds for spinal cord regeneration following injury. (a) Use of a self-assembling peptide amphiphile for the creation of a scaffold that was able to promote neurite outgrowth in a spinal cord injury model [187], (b) NGF-loaded single and multiple lumen nerve conduits were fabricated using a combination of methods and remained intact for 13 days after implantation *in vivo* [188], (c) Chitin and chitosan hydrogel tubes were produced through injection molding and were shown to induce neurite extension in chick DRG models [72], (d) Freeze-dried agarose scaffolds were able to guide injured axon growth in *in vivo* models [46], (e) Combined use of methods to create fibrin scaffolds showed that scaffold porosity could play a more important role than the geometrical dimensions of the conduits [189], (f) Discontinuous and anisotropic microconical structures created through femtosecond laser texturing were able to control the directional outgrowth of Schwann cells, sympathetic neurons and DRG explants [190]

and (g) Creation of nano-rippled and microgrooved hierarchical patterned surfaces showed that the combination of the two morphologies supports Schwann cell outgrowth [63]. Images adapted with permission from (a) Tysseling-Mattiace et al, 2008 (Copyright (2008) Society for Neuroscience, (b) Yang et al, 2005, (c) Freier et al, 2005, (d) Stokols & Tuszynski, 2004, (e) Scott et al, 2011, (f) Simitzi et al, 2015 and (g) Yiannakou et al, 2017.

Another application saw the creation of chitin and chitosan hydrogel tubes through the use of injection molding [72]. Freier and coworkers produced wet and dry chitin tubes, chitin tubes reinforced with PLGA coils, as well as chitosan tubes (produced through the deacetylation of the chitin tubes) by injecting an N-acetylated chitosan solution into a cylindrical mold containing a cylindrical core (Fig. 15c). The samples were allowed to gelate for 24h and removed from the molds in order to be used for experiments. The honeycomb-like structure that resulted from the formation of the hydrogel completely collapsed when the samples were air-dried. The compatibility of the chitin and chitosan hydrogels and films was tested through the use of primary chick dorsal root ganglion (DRG) neurons and the analysis of their adhesion and neurite extension on the various types of scaffolds produced. DRG neuron adherence and neurite extension was found to occur on all surfaces, though chitosan films displayed greater compatibility, with a higher number of adhered cells as well as longer neurite extensions. Due to the fact that chitin tubes are mechanically strong, biocompatible and biodegradable, as well as the ease of chemical modification due to chitin properties, makes them ideal candidates for neural tissue engineering.

Stokols & Tuszynski fabricated agarose scaffolds using freeze-drying in order to assess their ability to facilitate recovery after spinal cord injury (Fig. 15d) [46]. More specifically, they created a porous agarose scaffold by injecting a melted agarose solution into a Teflon mold with individual grooves. After cooling, the mold was placed in Styrofoam in a way which only left the bottom surface of the Teflon mold exposed and this was subsequently placed on a block of dry ice that was partly submerged in liquid nitrogen, in order to create a uniaxial thermal gradient. After removing the liquid nitrogen through the application of a vacuum, the mold was frozen for 45min and lyophilized overnight. The next day, the agarose scaffolds were removed from the mold and cut before being used. Additionally, some of the scaffolds were also rehydrated so that the growth factor BDNF could be added. The scaffolds were then implanted into lesions caused in the spinal cord of Fischer 344 rats. After one month *in vivo*, the scaffolds were found to be stable and biocompatible, while also supporting and guiding the growth of the injured axons. This was found to be highly correlated with the presence of host cells within the channels.

Hyaluronic acid (HA) hydrogels were created in order to assess their effect on neurite elongation in a spinal cord regeneration model [191]. Thiolated hyaluronan was crosslinked with PEGDA to create the hydrogels and these were tested in both *in vitro* and *in vivo* experimental setups. As a first step, the assessment of the HA hydrogels was carried out in chick DRG explant cultures, with polymerization of the hydrogels occurring around the DRG in the culture plate wells. The HA hydrogels were found to promote neurite extension, with neurites being at least 50% longer than the ones grown on control fibrin hydrogel matrices, while they were also able to support sustained culturing of the DRG explants for 8 days, as opposed to the fibrin ones, where the neurites collapsed after <60 hours. When the HA hydrogels were further tested in *in vivo* rat models through implantation into the gap caused by a T9-10 laminectomy and complete transection of the spinal cord at that point, the researchers were only able to notice large amounts of fibrosis at the site of injury, with the HA hydrogel most possibly becoming degraded soon after the implantation process. A follow-up work from the same group in 2016 was able to show that the HA hydrogel implant was unable to support axonal or neuronal regeneration after a T10 hemisection SCI in rats, but it did seem to exhibit a neuroprotective effect at the site of the injury signified by the decreased presence of inflammatory cells at the site and a reduction in scar tissue [192].

In 2008, Wong et al. used particulate leaching in order to create scaffolds for SCI treatment [193]. The scaffolds were by adding PCL/acetone solution into 5 different mold shapes that had been prepacked with salt crystals. After the solvent was evaporated, the scaffolds were then soaked in 70% ethanol in order to remove all the salt and create the randomly interconnected pores within the PCL. The constructed scaffolds were then implanted into the spinal cord of Sprague-Dawley rats in a 2mm gap that was created by the removal of tissue at the T8 vertebra. From the 5 different scaffold designs, the most effective ones were found to be the open-path ones, indicating that the surface area of the implant is not the most important characteristic. The macro-architecture and targeted contact guidance cues associated with the open-path scaffolds seem to play a much more significant role when it comes to neural tissue regeneration.

Injection molding in combination with TIPS was used in order to create multiple channel conduits made of PLGA. As described in He et al. in 2009, a PLGA/dioxane solution with or without sodium bicarbonate was injected into a mold containing 7 or 16 wires and frozen in order to create the scaffolds [194]. The phase separation in the PLGA/dioxane solution created micropores of ~17µm +/- 5µm, while the addition of sodium bicarbonate and the use of particulate leaching led to the further creation of macropores in the region of 125-200µm. Culturing of mesenchymal stem cells (MSCs) and co-culturing of MSCs and Schwann cells showed good affinity of the fabricated scaffold for these cells, with both cell types showing good proliferation rates and acceptable infiltration rates into the innermost regions of the conduits. The scaffolds were also tested *in vivo* and inserted into the spinal cords of Sprague-Dawley rats, where they became integrated into the spinal tissue and could not be removed without causing damage to the spinal cord. However, histological analysis was able to show that the number of macrophages recruited to the site was low, indicating good biocompatibility.

Scott et al. used a combination of methods in order to create scaffolds with highly aligned conduits made of fibrin (Fig. 15e) [189]. Firstly, they created monofilament strands of different diameters made out of poly(methyl methacrylate) (PMMA) or cellulose acetate (CA) through either melt extrusion or melt spinning. In the first case, a melt extruder with a spinneret containing 64 holes was loaded with the material and extruded through a temperature-controlled process with an extrusion speed of 200 or 800m/min. During the melt spinning, a spinneret with one hole was loaded with the material and extruded at a standard temperature at a rate of 27.5, 80 or 96.5m/min. These strands were then aligned, tightly bundled and placed into a Teflon tube in order to act as a sacrificial template for the creation of the fibrin scaffolds. A fibrinogen solution was added to the tube and allowed to polymerize over at least 16h in a thrombin working solution. Finally, the resulting tubes were placed in acetone baths to dissolve the PMMA or CA filaments, leading to the formation of the fibrin scaffolds. In order to test the affinity of these scaffolds for neural tissue regeneration, the researchers used a chick DRG model, where they cultured an explanted DRG in an environment containing the scaffolds. Initial studies showed that infiltration of the smaller conduit diameter scaffolds by Schwann cells was impossible due to their size and as such, they focused their study on measuring axon migration. They showed that there were no significant differences in the length of the axons extending from the DRG explant over the different conduit diameters that they were able to test as part of their study. This would indicate that the porosity of the scaffold that is achieved through this particular fabrication process could play a more important role than the geometrical dimension of the conduits.

While the aforementioned approaches involved the directional growth of neurites in continuous anisotropic scaffold structures, Simitzi et al. were the first to show the effectiveness of a discontinuous but anisotropic topography in the directional outgrowth of neural cells [190]. Using femtosecond laser texturing, the researchers were able to create discontinuous and anisotropic microconical substrates exhibiting three different geometrical characteristics, coined as low-, medium- and high-roughness respectively (Fig. 15f). All three surfaces were able to support Schwann cell and superior cervical ganglia sympathetic neurons growth, with oriented axonal extension being observed for the medium- and high-roughness substrates. Notably, although DRG explants cultured on the low-roughness substrates displayed an isotropic outgrowth pattern, in the medium- and high-roughness substrates, axonal outgrowth and Schwann cell migration occurred in an oriented fashion, parallel to the directionality of the underlying microcone texture (Fig. 15f). Besides this, laser texturing has been also reported as a means to create arbitrary patterns for neuronal cultures [63]. Through varying the laser irradiation conditions, different textures can be formed, ranging from 'simple' nanoripples to more hierarchical ones, comprising of microgrooves decorated with nanoripples (Fig. 15g). It was shown that although Schwann cells were able to grow well and orient on the micro-/nano- patterned grooves, the nanoripples inhibited cell outgrowth. As a consequence, arbitrary patterns of neurons can be formed via proper combination of the different micro- and nano- textures. The simplicity of the laser texturing technique makes it particularly attractive in order to facilitate the orientation of regenerating tissues, as well as enable spatial control of neuron cell growth with desired shapes and patterns.

In contrast to the study by Scott et al., a more robust analysis of the effect of fiber alignment and neurite extension was made by Corey et al in 2007. They proposed longitudinally aligned electrospun PLLA nanofibers for nerve guidance [195]. Highly aligned fibers were fabricated via electrospinning with a rotating wheel as a collector, where the rotation spin of the wheel determined the degree of fiber alignment. The wheel was rotated at 250 rpm, 110 rpm and 30 rpm to produce fibers of high, intermediate and low alignment respectively (Fig. 16a). Neurites from DRG explants cultured on these electrospun PLLA fibers showed robust guidance and neurites which they grew on highly aligned fiber had the longest length. Also, Schwann cells grown on fibers assumed a very narrow morphology compared to those on the surrounding coverslip. With this work, Corey et al. demonstrated for the first time that neurite extension from DRGs follow the direction of the aligned electrospun PLLA fibers of the scaffold. Moreover, they showed the connection between fiber alignment and neurite extension: greater fiber alignment resulted in more elongated extensions. Two more similar works were published in this past year. In the first one, Chow et al. created polydioxanone (PDS) fibers and cultured DRG on top of electrospun fibers with adherent astrocytes [196]. The length of neurites extending from the DRG was significantly increased by the presence of astrocytes. In the second, Schnell et al. cultured DRG explants on PCL and PCL/collagen electrospun nanofibers confirming the ability of electrospun fibers to guide the extension of neurites from DRGs (Fig. 16b) [197].

Several subsequent studies [172, 198-201] using DRGs from different species, different cell types, as well as different materials, all corroborate the aforementioned connection between fiber alignment and neurite orientation and length *in vitro*. Hurtado et al. fabricated films of electrospun PLA fibers and conduits for *in vivo* experiments [202]. For the conduits, two films of the same type were placed back-to-back and rolled into conduits in such a way as to create a middle insert within the conduit lumen (Fig. 16c). Conduits made of randomly oriented and aligned PLA nanofibers were transplanted in rat spinal cord complete transections where 3 mm gaps had been created. Remarkably, after 4 weeks, they observed robust CNS tissue growth, supported by migrating astrocytes.

In a similar study, Liu et al. fabricated conduits of electrospun collagen nanofibers [203]. Again, films of randomly oriented and aligned collagen fibers were created with electrospinning and spiral-shaped nerve guide conduits were subsequently prepared by rolling the collagen film into tubes of 4–5 layers. Conduits were transplanted to cover 2.5mm gaps in rat spinal cords. Although *in vitro* experiments with astrocytes and DRG explants showed good neurite extension and orientation, *in vivo* conduit transplantation exhibited little neurite sprouting around the lesion site and no astrocyte accumulation. In 2012, Downing et al. created patches of PLLA electrospun nanofibers that release Rolipram, a neuroprotective agent [204]. PLLA fibers were collected in a rotating mandrel and coated with an alginate hydrogel embedded with Rolipram (16d). When this scaffold was transplanted into hemisections of spinal cord in rats, a significant improvement in motor function was exhibited by the group, which received electrospun fibers that released a low dose of Rolipram compared to other groups, including the group with the high dose of Rolipram.

3D printing also offers the ability to fabricate complex structures resembling the natural microenvironment of the spinal cord. Recently, Koffler and co-workers introduced the use of a new technique, known as microscale continuous projection printing method (μCPP) for regenerative medicine applications in the spinal cord [205]. The 3D printer was equipped with a UV light source (365 nm) for photopolymerization of polyethylene glycol-gelatin methacrylate (PEG-GelMA). Cross sections of the T3 in the rat spinal cord were used to create digital images of gray and white matter. The gray matter was designed as a solid block and channels of 200 μm in diameter were incorporated into the white matter to provide linear guidance for the axonal regeneration. Scaffolds were loaded with NSCs and tested for their ability to support axon regeneration in complete transection at the T3 of the spinal cord. The findings showed evidence of injured host axons regeneration into the scaffolds and synapse formation with the implanted NSCs. Also implanted NSC axons were extended out of the scaffold and into the host spinal cord below the injury.

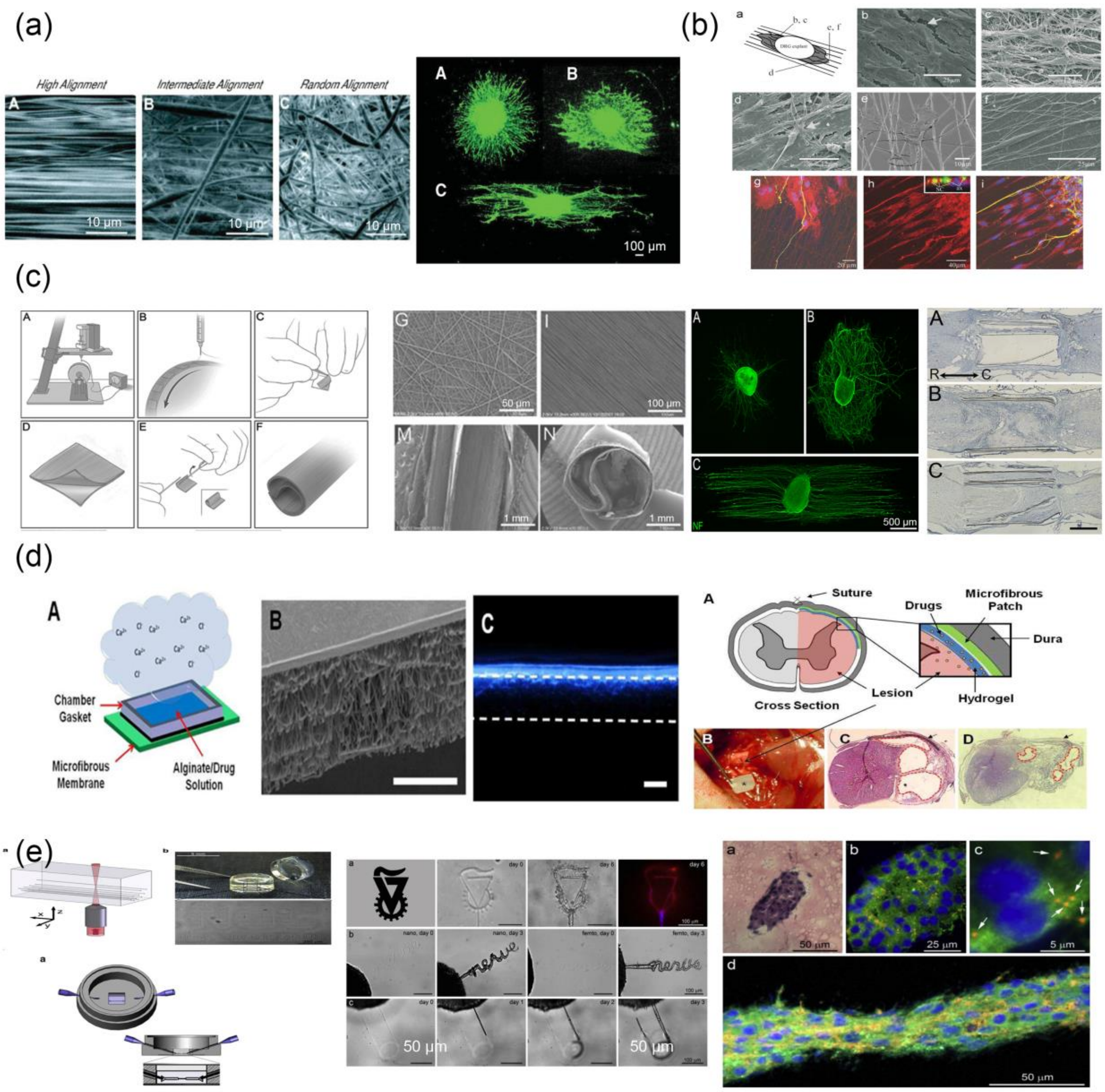

**Figure 16**: Different methods for fabricating scaffolds to support spinal cord regeneration after injury. (a) aligned electrospun nanofibers which support and direct the growth of DRGs [195], (b) aligned electrospun nanofibers made of biodegradable PCL/collagen blends for nerve regeneration [197], (c) fabrication of conduits from electrospun PLA fibers that support CNS tissue growth [202], (d) drug-eluting microfibrous patches for local delivery of Rolipram in spinal cord repair [204] and (e) patterned microchannels created by laser photoablation, guide the directional growth of neurites from dorsal root ganglia [206]. Images adapted with permission from

(a) Corey et al, 2007, (b) Schnell et al, 2007, (c) Hurtado et al, 2011, (d) Downing et al, 2012 and (e) Nadir et al, 2009.

In another approach, a stereolithography-based 3D printing method was used for the creation of guidance microchannels in transparent hydrogels to promote the directional growth of neurites from dorsal root ganglia *in vitro* (Fig. 16e) [206]. Specifically, nanosecond and femtosecond laser pulses were used for the photocuring of PEGylated fibrinogen hydrogels. The resulting structure consisted of photopatterned microchannels able to guide the directional growth of neurites from DRG.

At this point, it is worth mentioning that, although hundreds of millions of people worldwide suffer from neurological injuries and/or disorders that affect the CNS, repairing the nervous system remains one of the greatest challenges currently faced by regenerative medicine. Therefore, there is growing interest for the discovery of novel therapeutic approaches to promote regeneration of the CNS. These approaches should be combinatorial, incorporating cell-based delivery strategies, biomolecule delivery strategies and scaffold-based therapeutic methods. Current approaches and clinical trials are reviewed elsewhere [207-209].

### 3.4. Lab-on-a-chip devices

To date, the study of neurological diseases relies heavily on animal and human cellular models, however this combination has proven to provide insufficient information. Challenges in the field of translational research, ethical concerns and economic implications dictate the need and intensify the search for microphysiological neural systems (MPNS) as models for human neurological diseases, disorders and injuries. Bioengineering can offer alternative scaffold-free and scaffold-based MPNS that exhibit structural and/or functional aspects of the neuronal tissue. Some approaches to create MPNS will be reviewed below.

A successful example of MPNS are the chip-based systems, especially when combined with self-organized neurospheroids. These chip-based systems are 2D or 3D micropatterned platforms that incorporate scaffolding, mechanical, biochemical and topographical cues to recreate the physiological architecture and functions of the brain and the nervous system. Soft lithography is a common method used for microfabrication and therefore has been widely applied in chip manufacturing.

In 2012, Booth and Kim introduced a microfluidic BBB (µBBB) device which mimics the natural Blood Brain Barrier [210]. This µBBB system consisted of four patterned PDMS layers, two embedded electrode glass layers and a sandwiched polycarbonate membrane, all sequentially bonded to create a fully integrated device (Fig. 17a). For the fabrication of the PDMS layers, pre-polymer PDMS was first spin-coated and cured to create sheets and the sheets were subsequently patterned using lasers to create channels. The assembled device had two perpendicularly-crossing channels for dynamic flows and a porous membrane at the intersection of the flow channels for cell culture. In order to form a dual-layer BBB on the chip, a co-culture was performed by seeding C8D1A astrocytes in the first channel on the one side of the porous membrane and then b.End3 endothelial cells in the second channel on the other side of the membrane. Both cell lines exhibited high viability and expressed characteristic cellular markers. b.End3 cells formed tight junctions with high transepithelial/transendothelial electrical resistance (TEER), an index of high junction integrity. These results indicated that the fabricated µBBB could be a valid option for pre-clinical studies. In their follow-up study in 2014, the authors used the same device to co-culture b.End3 cells with glial cells (C6 cell line) in the two channels coated with collagen IV/fibronectin and poly-lysine, respectively [211]. They performed permeability analysis of neuroactive drugs and the results showed good correlation of their model under dynamic conditions with the *in vivo* conditions.

Park et al. manufactured a 3D brain-on-a-chip for the culture of neurospheroids as an *in vitro* model of Alzheimer's disease [212]. The chip had a top flat chamber, a bottom layer with concave microwell arrays and an interstitial space (Fig. 17b). The top flat chamber and the backbone of the bottom chamber containing 50 cylindrical wells were both fabricated using a standard soft lithography procedure. To fabricate the concave microwell array, the microwells of the backbone PDMS chamber were entirely filled with PDMS prepolymer and curing agent. The excess of the prepolymer removed using a glass slide and by applying slight pressure to the soft PDMS microwell plate. Next, half of the PDMS prepolymer in each microwell was removed by wiping. The remaining half of the PDMS prepolymer formed a concave meniscus in each well through surface tension. The prepolymer in the wells was polymerized by thermal curing, forming the final concave structure. Finally, the top and the bottom layer were bonded by treating with oxygen plasma for 20s creating an interstitial space for culture media flow. An osmotic micropump system was attached to maintain a slow interstitial level of flow. For the formation of neurospheroids inside the microwells, primary cortical neurons were isolated and seeded inside the chip using a micropipette. Cells were left to sink and become trapped inside the microwells, while the un-trapped cells were washed away by applying constant flow. Finally, the neural progenitor cells were cultured for 10 days to allow neurospheroid formation. Using this platform, the authors showed that neurospheroid growth was enhanced under flow and that the treatment with β-amyloids was more destructive in comparison to the one occurring in the culture under static conditions.

A 3D human BBB on a chip was developed as an *in vitro* model for neurovascular inflammation (Fig. 17c) [213]. First, a single square-shaped microchannel was created on PDMS mounted on a standard glass microscope slide, using soft lithography. Within this microchannel, a cylindrical collagen gel was formed using the viscous fingering method, which is the process of the unstable displacement of a more viscous fluid by a less viscous fluid. The entire process took about 30 seconds and resulted in the creation of a well-defined lumen with a diameter of approximately 600-800 μm protruding all the way through the 2cm long channel of the microfluidic chip. For the 3D cell culture, human astroglia cells were incorporated into the collagen gel.

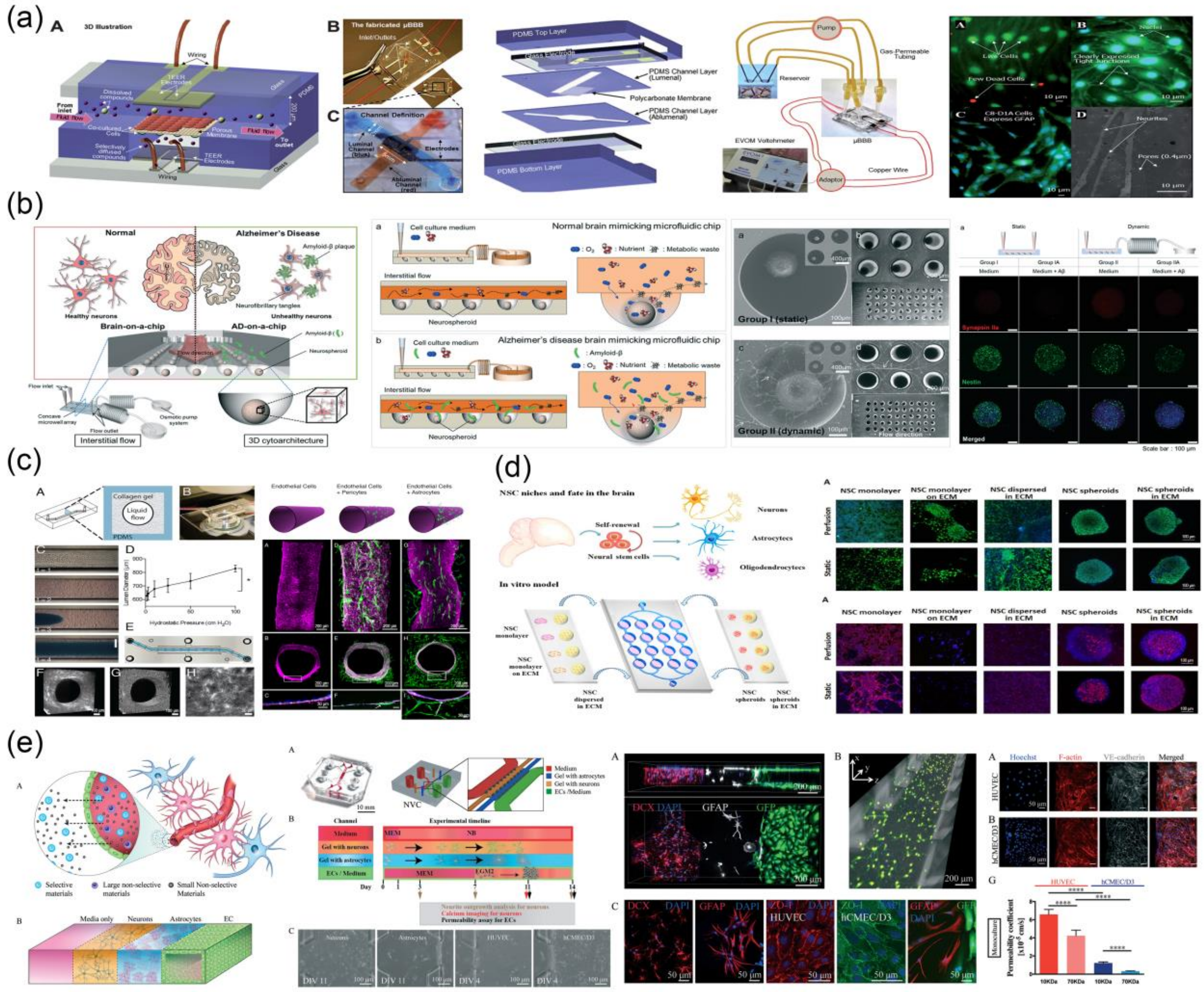

**Figure 17**: Use of soft lithography for the fabrication of lab-on-a-chip devices for neuronal applications. (a) a microfluidic device mimicking the natural BBB [210] (b) a brain-on-a-chip as *in vitro* model of Alzheimer's disease [212], (c) a 3D human BBB on a chip as *in vitro* model for neurovascular inflammation [213], (d) a microfluidic chip for studying neural stem cell fate determination [214] and (e) a 3D neurovascular microfluidic system as a model for BBB [215]. Images adapted with permission from (a) Booth et al, 2012, (b) Park et al, 2014, (c) Herland et al, 2016, (d) Wang et al, 2017 and (e) Adriani et al, 2017.

In 2017, Wang et al. fabricated a microfluidic chip for application in studies of neural stem cell fate determination [214]. The entire chip was made of PDMS and had two PDMS-coated glass slides as the cover (Fig. 17d). The lower layer, which had an array of microchannels and microwells, was constructed using the standard soft lithography technique. For the creation of the cover, PDMS was spin-coated and solidified on the two clear glass slides which served as substrates for cell adherence.

Primary cortical neurons were seeded on the two top glass slides under different conditions and were assembled on to the lower chamber of the chip using stainless steel clamps. Dynamic and static cultures were performed to evaluate the efficiency of the culture on the viability, self-renewal, proliferation and differentiation of NSCs into neurons, astrocytes or oligodendrocytes.

In another application using soft lithography, Adriani and associates created a 3D neurovascular microfluidic system as a model for the BBB [215]. The microfluidic device had a single layer made of PDMS with 4 different channels: two for 3D hydrogels embedded with cells and two fluidic channels for culture media (Fig. 17e). Collagen hydrogel mixed with primary astrocytes was first injected in device and left to polymerize for 30 min. Next, the neuron hydrogel was injected in the lateral fluidic channel, close to the astrocyte hydrogel and polymerized for 30 min. Appropriate media were added into the fluidic channels on either side of the hydrogels. Following 7 days in culture, the fluidic channel adjacent to the astrocyte hydrogel was seeded with HUVEC or hCMEC/D3 endothelial cells and the culture continued for an additional 7 days. Immunostainings showed that each cell type in the co-culture system exhibited a cell-type specific morphology and expressed characteristic cellular markers. Also, permeability tests proved the selectivity of the endothelial cell monolayer. All the above results were indicative of the functionality of the 3D microfluidic system.

A microfluidic device able to support the structuring of a multi-nodal neural network *in vitro* was presented by Van de Wijdeven and colleagues [216]. This microfluidic chip consisted of multiple cell compartments (4-nodal chip or 6-nodal chip), interconnected by funnel-shaped micrometer sized tunnels. The cell compartments were almost entirely open on top, whereas the areas where the compartments connected to the tunnels were enclosed within the microfluidic chip. The mold for the chip was formed in a two-step photolithography process, employing an SU8 series photoresist and the chip was fabricated by pouring PDMS into the mold. The authors showed that by culturing DRGs in this microfluidic chip, they were able to structure a neural network *in vitro* where neuronal aggregates were formed in the cell compartments, with their axons entering the tunnels, while some of them had even crossed to the opposite compartment.

Johnson and co-workers in 2016 printed a 3D biomimetic system reconstituting the critical function of glial cell-axon interfaces in order to study viral infections within the nervous system [217]. They used a micro-extrusion custom-made printer to manufacture a platform comprising of three different parts: a substrate of parallel microchannels (made out of PCL), a sealant layer and a top tri-chamber (Fig. 18a). Both of the latter parts were made out of silicone and had the same shape. After the chip assembly, three different compartments for cell isolation were created. Printing was conducted directly on poly-L-ornithine and laminin-coated petri dishes. Following printing, the different compartments were functionalized with primary hippocampal neurons, SCG neurons, Schwann cells and epithelial cells. Neurons were seeded in the first chamber and were left to grow axons that penetrated each of the individual chambers after 10–14 days. Schwann cells and epithelial cells were added later on in the other two compartments, which contained robust axonal networks, spreading out from the neurons in the first compartment. After the establishment of a complete network, viral particles were added only in the middle (second) chamber. Results showed that although Schwann cells were refractory to axon-to-cell infection, they appeared to participate in the infection response through axonal interaction.

Yi et al. created the glioblastoma-lab-on-a-chip, which represents a patient-specific tumor-on-a-chip model, useful for the identification of the most effective treatment or for testing new drug therapies [218]. An in-house 3D printer system based on SFF technology was equipped with a multi-head deposition system and used to deposit the inks on a sterilized surface-modified slide glass. Three different types of inks were loaded into different slots and deposited sequentially in specified directions to recapitulate the biochemical and biophysical cues of the glioblastoma microenvironment. The outer layer of the chip was printed out of modified silicone, the intermediate layer was made out of a mixture of brain decellularized ECM (BdECM) and vascular endothelial cells (HUVEC) and the core was made out of BdECM mixed with patient-derived glioblastoma cells. Tumor cells on the glioblastoma-lab-on-a-chip showed the expected treatment resistance with the patients from whom the cells had been isolated. Moreover, this ex-vivo GBM model was capable of identifying the best treatment with the help of personal bioinformatics analysis.

Very recently, Samper et al. fabricated a 3D-printed microfluidic biosensing platform for the *in vivo* recording in real time of the neurochemical changes in the human brain [219]. The system was based on the microdialysis technique for the detection of neurotransmitters and other molecules in interstitial tissue fluid and it could easily monitor biomarkers such as glutamate, glucose and lactate inside the brain, using electrochemical biosensors. The microfluidic platform consisted of three needle electrodes which were modified to work as biosensors and secured on a microfluidic channel, through which the dialysate flows (Fig. 18b). For the simulation experiments, the microfluidic channel used was a 3D-printed hard chip, which was replaced by a soft chip made of PDMS for the clinical monitoring. Several methodologies were combined for the fabrication of this device. For the creation of the biosensor, two layers of different hydrogels were added onto the electrode surface: Firstly, a protective poly(m-phenylenediamine) (poly(mPD)) film was electropolymerized onto the working electrode surface. To do that, the needle tip was placed into the mPD solution and held at 0.7 V for 20 minutes. A second layer of hydrogel was added later to immobilize enzymes onto the electrode surface. For the PDMS microfluidic channel, soft lithography was used to create a hollow tube. Once the PDMS had been cured, holes were pierced for the inlet/outlet tubing and for the sensors. For the hard microfluidic channel fabrication, two different 3D printers were used: an SLA-based and a polyjet 3D printer. Both printers fabricated chips with the same shape and approximately the same size but with slightly different microfluidic characteristics due to the different technology they use. The 3D printing technology was also used for the fabrication of the electrode holders. This microfluidic device was firstly used for the *in vitro* simulation of spreading depolarization (SD) events that often occurred in many pathophysiological conditions or trauma. More specifically, the levels of glutamate, glucose and lactate were monitored under physiological conditions and after an SD-like event. In the next step, the chip was used to monitor, for 4 days continuously, a patient with an intracerebral hemorrhage (ICH). Using this microfluidic device, it was possible to detect transient metabolite concentration changes that took place in as short as 8 seconds, as well as a dynamic glutamate change in a living human brain in real time for the first time.

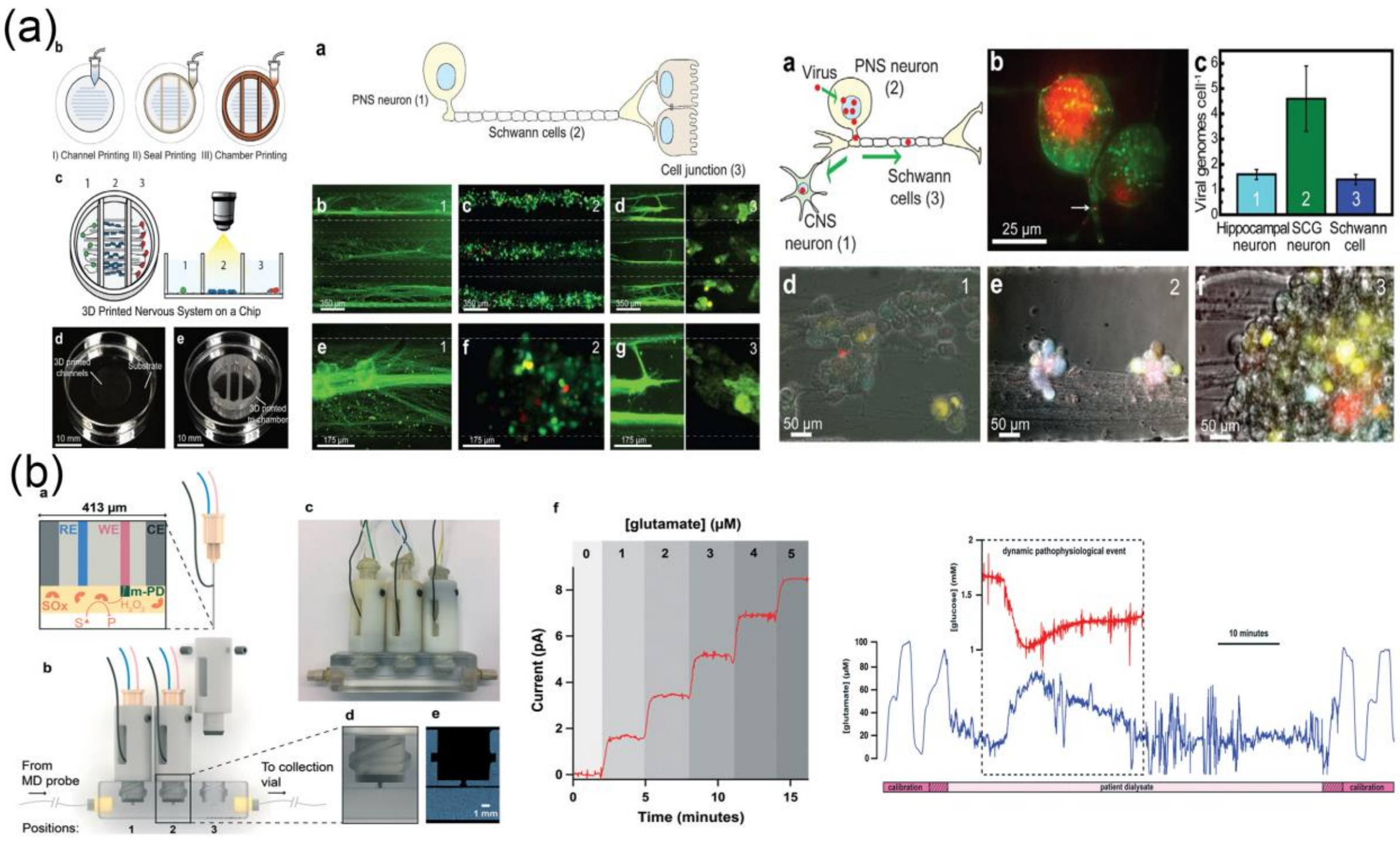

**Figure 18**: 3D printed of lab-on-a-chip devices for neuroscience applications. (a) micro-extrusion based 3D printing of an *in vitro* platform for the study of viral infection in the nervous system [217] (b) use of SLA-based and inkjet-based 3D printers for the manufacturing of a microfluidic biosensing platform for online monitoring of the chemical changes in the human brain [219]. Images adapted with permission from (a) Johnson et al, 2016 and (b) Samper et al, 2019.

## 4. Challenges and future outlook

Neural tissue engineering, a sub-field of tissue engineering, focuses on the formation of grafts scaffolds and systems that can promote nerve regeneration and the repair of damages caused to nervous tissue of both the CNS and PNS.

It is well known today that cellular behavior – both structurally and functionally – is different when seeded on 2D surfaces versus a 3D environment, which more closely mimics the one *in vivo*. Thus, the ultimate goal is the fabrication of structures that can mimic the natural environment constituting the ECM, which is a complex micro/nano environment that can mechanically support cell adhesion and migration and can provide topo-chemical signals via specific receptors, thereby determining cellular responses. The synthesis of novel and the utilization of natural biomaterials, as well as the fabrication of biomimetic scaffolds are key objectives in this respect. These 3D cell platforms can be used as experimental models, but also as implantable constructs. Recently, several approaches have shown the use of biomaterials for the repair of peripheral nerve injuries, however, it is still quite limited to fabrication of scaffolds for nerve autografts. Bioengineered materials containing acellular nerve matrices and/or synthetic substances, alongside the addition of neurotrophic factors or the seeding of neural cells (e.g. neural stem cells) have shown great promise and could be ideal for repair of nervous system damages.

In the first part of this review article, we have outlined the most well-known techniques that have been developed for scaffold fabrication for tissue engineering applications. The underlying concept during fabrication is the mimicking of the natural neural tissue structure and environment, which not only supports but also directs the development and maturation of tissue. The idea behind this is that if you manage to create the proper structural form, then eventually, the cells and tissue should be able to regain functionality. Some main features that appear in neural tissue are pores, fibers and channels and are all organized in different ways throughout the various layers of the nervous system. As perceived from the literature, there is great effort going into reproducing all these natural structural elements in order to imitate the local microenvironment. The challenge in all these approaches remains the creation of scaffolds with precisely controlled, tunable topography, chemistry and thus surface energy, in order to be able to predict, direct and control the function of neuronal cells [59]. Furthermore, of particular interest in biomimetic scaffold fabrication for NTE, is the ability to spatially pattern growth factors, neurotrophic factors and bioactive proteins, to create concentration gradients of these molecules, so that their spatial distribution and temporal release profile could mimic the complex developmental profiles of native tissues [150, 220, 221].

In both the PNS and CNS, the spatiotemporally dynamic nature of ECM composition and structure helps to regulate many neural processes including cell migration, differentiation and regeneration. Thus, a deeper understanding of the structural anatomy and the micro/nano environment of the nervous system is immensely useful for more informed construct design. The clear relationship between chemical composition and functional architecture still needs to be determined, along with a better understanding of surface topography functions [222]. In spite of what has been achieved so far in the fields of engineering and materials technology, a consistent understanding from a biological point of view is missing. Indeed, the critical question that needs to be answered is “how do neuronal cells sense and interact with micro/nanostructures of well-defined sizes and how do they respond at the cellular and molecular level?”

As an attempt to answer these questions and to unravel the underlying mechanisms of physicochemical cues on adhesion, migration and differentiation, scaffolds with defined chemistry and topographical patterns at the micro- and nanoscale have been used. In this context, the thorough study of the topography-, chemistry- and surface energy-related mechanobiological mechanisms activated in nerves, neuroglia and neural stem cells will contribute to a better understanding of biological processes and trigger the development of advanced biomimetic scaffolds for nerve regeneration.

All in all, the broader insight that will be gained through studying either the functionality of the micro/nano environment of the cells in a spatiotemporal manner on neuronal cell responses or the temporal behavior of topographies on neural stem cells will aid in controlling the outgrowth and patterning of nerve and neuroglial cells, as well as in the development of systems capable of forming functional matured neurons.

## 5. References

[1] WHO, Neurological disorders affect millions globally: WHO report, 2007.
[2] J. FitzGerald, J. Fawcett, Repair in the central nervous system, Journal of Bone and Joint Surgery - Series B, 2007, pp. 1413-1420.
[3] G.D. Mahumane, P. Kumar, L.C. du Toit, Y.E. Choonara, V. Pillay, 3D scaffolds for brain tissue regeneration: architectural challenges, Biomater Sci-Uk, 6 (2018) 2812-2837.
[4] V. Vindigni, R. Cortivo, L. Iacobellis, G. Abatangelo, B. Zavan, Hyaluronan benzyl ester as a scaffold for tissue engineering, International Journal of Molecular Sciences, 2009, pp. 2972-2985.
[5] S.M. Willerth, S.E. Sakiyama-Elbert, Approaches to neural tissue engineering using scaffolds for drug delivery, Advanced Drug Delivery Reviews, 2007, pp. 325-338.
[6] S. Knowlton, S. Anand, T. Shah, S. Tasoglu, Bioprinting for Neural Tissue Engineering, Trends in Neurosciences, Elsevier Ltd, 2018, pp. 31-46.
[7] K.A. van Kampen, R.G. Scheuring, M.L. Terpstra, R. Levato, J. Groll, J. Malda, C. Mota, L. Moroni, Biofabrication: From Additive Manufacturing to Bioprinting, (2019).
[8] J. Groll, T. Boland, T. Blunk, J.A. Burdick, D.W. Cho, P.D. Dalton, B. Derby, G. Forgacs, Q. Li, V.A. Mironov, L. Moroni, M. Nakamura, W.M. Shu, S. Takeuchi, G. Vozzi, T.B.F. Woodfield, T. Xu, J.J. Yoo, J. Malda, Biofabrication: reappraising the definition of an evolving field, Biofabrication, 8 (2016).
[9] H.-Y. Mi, X. Jing, L.-S. Turng, Fabrication of porous synthetic polymer scaffolds for tissue engineering, Journal of Cellular Plastics, 51 (2015) 165-196.
[10] G.R. Guillen, Y. Pan, M. Li, E.M.V. Hoek, Preparation and characterization of membranes formed by nonsolvent induced phase separation: A review, Industrial and Engineering Chemistry Research, 50 (2011) 3798-3817.
[11] T. Lu, Y. Li, T. Chen, Techniques for fabrication and construction of three-dimensional scaffolds for tissue engineering, International Journal of Nanomedicine, 8 (2013) 337.
[12] S. Kyle, A. Aggeli, E. Ingham, M.J. McPherson, Production of self-assembling biomaterials for tissue engineering, Trends in Biotechnology, 2009, pp. 423-433.
[13] C. Tonda-Turo, C. Audisio, S. Gnavi, V. Chiono, P. Gentile, S. Raimondo, S. Geuna, I. Perroteau, G. Ciardelli, Porous Poly(ε-caprolactone) Nerve Guide Filled with Porous Gelatin Matrix for Nerve Tissue Engineering, Advanced Engineering Materials, 13 (2011) B151-B164.
[14] A. Barbetta, A. Gumiero, R. Pecci, R. Bedini, M. Dentini, Gas-in-Liquid Foam Templating as a Method for the Production of Highly Porous Scaffolds, Biomacromolecules, 10 (2009) 3188-3192.
[15] E.R. Aurand, K.J. Lampe, K.B. Bjugstad, Defining and designing polymers and hydrogels for neural tissue engineering, Neuroscience Research, 72 (2012) 199-213.
[16] Y. Xia, G.M. Whitesides, SOFT LITHOGRAPHY, Annual Review of Materials Science, 28 (1998) 153-184.
[17] G.M. Whitesides, E. Ostuni, S. Takayama, X. Jiang, D.E. Ingber, Soft Lithography in Biology and Biochemistry, Annual Review of Biomedical Engineering, 3 (2001) 335-373.
[18] E. Stratakis, A. Ranella, M. Farsari, C. Fotakis, Laser-based micro/nanoengineering for biological applications, Progress in Quantum Electronics, 33 (2009) 127-163.
[19] A.G. Mikos, Y. Bao, L.G. Cima, D.E. Ingber, J.P. Vacanti, R. Langer, Preparation of poly(glycolic acid) bonded fiber structures for cell attachment and transplantation, Journal of Biomedical Materials Research, 27 (1993) 183-189.
[20] J. Wang, J.A. Jansen, F. Yang, Electrospraying: Possibilities and Challenges of Engineering Carriers for Biomedical Applications-A Mini Review, Front Chem, 7 (2019) 258.
[21] M. Nikolaou, T. Krasia-Christoforou, Electrohydrodynamic methods for the development of pulmonary drug delivery systems, European Journal of Pharmaceutical Sciences, 113 (2018) 29-40.
[22] X.-H. Qin, A. Ovsianikov, J. Stampfl, R. Liska, Additive manufacturing of photosensitive hydrogels for tissue engineering applications, BioNanoMaterials, 15 (2014).
[23] A. Mazzoli, Selective laser sintering in biomedical engineering, Med Biol Eng Comput, 51 (2013) 245-256.

[24] V. Gupta, Y. Khan, C.J. Berkland, C.T. Laurencin, M.S. Detamore, Microsphere-Based Scaffolds in Regenerative Engineering, Annual Review of Biomedical Engineering, Vol 19, 19 (2017) 135-161.
[25] I. Zein, D.W. Hutmacher, K.C. Tan, S.H. Teoh, Fused deposition modeling of novel scaffold architectures for tissue engineering applications, Biomaterials, 23 (2002) 1169-1185.
[26] S. Derakhshanfar, R. Mbeleck, K. Xu, X. Zhang, W. Zhong, M. Xing, 3D bioprinting for biomedical devices and tissue engineering: A review of recent trends and advances, Bioact Mater, 3 (2018) 144-156.
[27] I.T. Ozbolat, K.K. Moncal, H. Gudapati, Evaluation of bioprinter technologies, Addit Manuf, 13 (2017) 179-200.
[28] Y. Wan, H. Wu, X. Cao, S. Dalai, Compressive mechanical properties and biodegradability of porous poly(caprolactone)/chitosan scaffolds, Polymer Degradation and Stability, 93 (2008) 1736-1741.
[29] C. Vaquette, C. Frochot, R. Rahouadj, X. Wang, An innovative method to obtain porous PLLA scaffolds with highly spherical and interconnected pores., Journal of biomedical materials research. Part B, Applied biomaterials, 86 (2008) 9-17.
[30] P.X. Ma, J.W. Choi, Biodegradable polymer scaffolds with well-defined interconnected spherical pore network., Tissue engineering, 7 (2001) 23-33.
[31] P. Plikk, S. Målberg, A.-C. Albertsson, Design of resorbable porous tubular copolyester scaffolds for use in nerve regeneration., Biomacromolecules, 10 (2009) 1259-1264.
[32] W.L. Murphy, R.G. Dennis, J.L. Kileny, D.J. Mooney, Salt fusion: an approach to improve pore interconnectivity within tissue engineering scaffolds., Tissue engineering, 8 (2002) 43-52.
[33] J.T. Jung, J.F. Kim, H.H. Wang, E. di Nicolo, E. Drioli, Y.M. Lee, Understanding the non-solvent induced phase separation (NIPS) effect during the fabrication of microporous PVDF membranes via thermally induced phase separation (TIPS), Journal of Membrane Science, 514 (2016) 250-263.
[34] F. Tasselli, Non-solvent Induced Phase Separation Process (NIPS) for Membrane Preparation, (2014) 1-3.
[35] F.C. Pavia, V. La Carrubba, S. Piccarolo, V. Brucato, Polymeric scaffolds prepared via thermally induced phase separation: tuning of structure and morphology., Journal of biomedical materials research. Part A, 86 (2008) 459-466.
[36] G.M. Whitesides, B. Grzybowski, Self-assembly at all scales, Science, 2002, pp. 2418-2421.
[37] S. Takahashi, M. Leiss, M. Moser, T. Ohashi, T. Kitao, D. Heckmann, A. Pfeifer, H. Kessler, J. Takagi, H.P. Erickson, R. Fassler, The RGD motif in fibronectin is essential for development but dispensable for fibril assembly, Journal of Cell Biology, 178 (2007) 167-178.
[38] J.D. Hartgerink, E. Beniash, S.I. Stupp, Self-assembly and mineralization of peptide-amphiphile nanofibers, Science, 294 (2001) 1684-1688.
[39] R.N. Shah, N.A. Shah, M.M.D.R. Lim, C. Hsieh, G. Nuber, S.I. Stupp, Supramolecular design of self-assembling nanofibers for cartilage regeneration, Proceedings of the National Academy of Sciences of the United States of America, 107 (2010) 3293-3298.
[40] E. Kokkoli, A. Mardilovich, A. Wedekind, E.L. Rexeisen, A. Garg, J.A. Craig, Self-assembly and applications of biomimetic and bioactive peptide-amphiphiles, Soft Matter, 2006, pp. 1015-1024.
[41] G. Deidda, S.V.R. Jonnalagadda, J.W. Spies, A. Ranella, E. Mossou, V.T. Forsyth, E.P. Mitchell, M.W. Bowler, P. Tamamis, A. Mitraki, Self-Assembled Amyloid Peptides with Arg-Gly-Asp (RGD) Motifs As Scaffolds for Tissue Engineering, Acs Biomaterials Science & Engineering, 3 (2017) 1404-1416.
[42] L.E. Kokai, Y.-C. Lin, N.M. Oyster, K.G. Marra, Diffusion of soluble factors through degradable polymer nerve guides: Controlling manufacturing parameters, Acta Biomaterialia, 5 (2009) 2540-2550.
[43] S.H. Oh, J.H. Kim, K.S. Song, B.H. Jeon, J.H. Yoon, T.B. Seo, U. Namgung, I.W. Lee, J.H. Lee, Peripheral nerve regeneration within an asymmetrically porous PLGA/Pluronic F127 nerve guide conduit, Biomaterials, 29 (2008) 1601-1609.

[44] C. Sun, X. Jin, J.M. Holzwarth, X. Liu, J. Hu, M.J. Gupte, Y. Zhao, P.X. Ma, Development of Channeled Nanofibrous Scaffolds for Oriented Tissue Engineering, Macromolecular Bioscience, 12 (2012) 761-769.
[45] X. Zhao, S. Zhang, Designer self-assembling peptide materials, Macromol Biosci, 7 (2007) 13-22.
[46] S. Stokols, M.H. Tuszynski, Freeze-dried agarose scaffolds with uniaxial channels stimulate and guide linear axonal growth following spinal cord injury., Biomaterials, 27 (2006) 443-451.
[47] A. Barbetta, G. Rizzitelli, R. Bedini, R. Pecci, M. Dentini, Porous gelatin hydrogels by gas-in-liquid foam templating, Soft Matter, 6 (2010) 1785-1792.
[48] L. Zhang, K.F. Li, W.Q. Xiao, L. Zheng, Y.M. Xiao, H.S. Fan, X.D. Zhang, Preparation of collagen-chondroitin sulfate-hyaluronic acid hybrid hydrogel scaffolds and cell compatibility in vitro, Carbohyd Polym, 84 (2011) 118-125.
[49] T.C. Holmes, S. de Lacalle, X. Su, G.S. Liu, A. Rich, S.G. Zhang, Extensive neurite outgrowth and active synapse formation on self-assembling peptide scaffolds, Proceedings of the National Academy of Sciences of the United States of America, 97 (2000) 6728-6733.
[50] V. Chiono, C. Tonda-Turo, Trends in the design of nerve guidance channels in peripheral nerve tissue engineering,  Progress in Neurobiology, Elsevier Ltd, 2015, pp. 87-104.
[51] N. Zhu, X. Che, Biofabrication of Tissue Scaffolds,  Advances in Biomaterials Science and Biomedical Applications, InTech, 2013.
[52] R.A. Quirk, R.M. France, K.M. Shakesheff, S.M. Howdle, Supercritical fluid technologies and tissue engineering scaffolds, Current Opinion in Solid State and Materials Science, 8 (2004) 313-321.
[53] C.M. Kirschner, K.S. Anseth, Hydrogels in Healthcare: From Static to Dynamic Material Microenvironments., Acta materialia, 61 (2013) 931-944.
[54] N.A. Peppas, J.Z. Hilt, A. Khademhosseini, R. Langer, Hydrogels in biology and medicine: From molecular principles to bionanotechnology, Adv Mater, 18 (2006) 1345-1360.
[55] J.P. Frampton, M.R. Hynd, M.L. Shuler, W. Shain, Fabrication and optimization of alginate hydrogel constructs for use in 3D neural cell culture, Biomed Mater, 6 (2011) 015002.
[56] K.Y. Lee, D.J. Mooney, Hydrogels for Tissue Engineering, Chemical Reviews, 101 (2001) 1869-1880.
[57] O.A. Carballo-Molina, I. Velasco, Hydrogels as scaffolds and delivery systems to enhance axonal regeneration after injuries., Frontiers in cellular neuroscience, 9 (2015) 13.
[58] D. Qin, Y. Xia, G.M. Whitesides, Soft lithography for micro- and nanoscale patterning, Nature Protocols, 5 (2010) 491-502.
[59] A. Ranella, M. Barberoglou, S. Bakogianni, C. Fotakis, E. Stratakis, Tuning cell adhesion by controlling the roughness and wettability of 3D micro/nano silicon structures, Acta Biomater, 6 (2010) 2711-2720.
[60] C. Simitzi, A. Ranella, E. Stratakis, Controlling the morphology and outgrowth of nerve and neuroglial cells: The effect of surface topography, Acta Biomater, 51 (2017) 21-52.
[61] E. Stratakis, A. Ranella, C. Fotakis, Biomimetic micro/nanostructured functional surfaces for microfluidic and tissue engineering applications, Biomicrofluidics, 5 (2011) 013411.
[62] E.L. Papadopoulou, A. Samara, M. Barberoglou, A. Manousaki, S.N. Pagakis, E. Anastasiadou, C. Fotakis, E. Stratakis, Silicon scaffolds promoting three-dimensional neuronal web of cytoplasmic processes, Tissue Eng Part C Methods, 16 (2010) 497-502.
[63] C. Yiannakou, C. Simitzi, A. Manousaki, C. Fotakis, A. Ranella, E. Stratakis, Cell patterning via laser micro/nano structured silicon surfaces, Biofabrication, 9 (2017) 025024.
[64] M. Castillejo, E. Rebollar, M. Oujja, M. Sanz, A. Selimis, M. Sigletou, S. Psycharakis, A. Ranella, C. Fotakis, Fabrication of porous biopolymer substrates for cell growth by UV laser: The role of pulse duration, Appl Surf Sci, 258 (2012) 8919-8927.
[65] E. Stratakis, A. Ranella, C. Fotakis, Laser-Based Biomimetic Tissue Engineering, in: V. Schmidt, M.R. Belegratis (Eds.) Laser Technology in Biomimetics: Basics and Applications, Springer Berlin Heidelberg, Berlin, Heidelberg, 2013, pp. 211-236.

[66] S. Lazare, V. Tokarev, A. Sionkowska, M. Wiśniewski, Surface foaming of collagen, chitosan and other biopolymer films by KrF excimer laser ablation in the photomechanical regime, Applied Physics A, 81 (2005) 465-470.
[67] A. Daskalova, A. Selimis, A. Manousaki, D. Gray, A. Ranella, C. Fotakis, Surface modification of collagen-based biomaterial induced by pulse width variable femtosecond laser pulses, SPIE2013.
[68] A. Daskalova, C.S.R. Nathala, P. Kavatzikidou, A. Ranella, R. Szoszkiewicz, W. Husinsky, C. Fotakis, FS laser processing of bio-polymer thin films for studying cell-to-substrate specific response, Appl Surf Sci, 382 (2016) 178-191.
[69] A. Prigipaki, K. Papanikolopoulou, E. Mossou, E.P. Mitchell, V.T. Forsyth, A. Selimis, A. Ranella, A. Mitraki, Laser processing of protein films as a method for accomplishment of cell patterning at the microscale, Biofabrication, 9 (2017) 045004.
[70] A.M. Martins, Q.P. Pham, P.B. Malafaya, R.A. Sousa, M.E. Gomes, R.M. Raphael, F.K. Kasper, R.L. Reis, A.G. Mikos, The Role of Lipase and α-Amylase in the Degradation of Starch/Poly(ε-Caprolactone) Fiber Meshes and the Osteogenic Differentiation of Cultured Marrow Stromal Cells, Tissue Engineering Part A, 15 (2009) 295-305.
[71] R. Svečko, D. Kusić, T. Kek, A. Sarjaš, A. Hančič, J. Grum, Acoustic emission detection of macro-cracks on engraving tool steel inserts during the injection molding cycle using PZT sensors., Sensors (Basel, Switzerland), 13 (2013) 6365-6379.
[72] T. Freier, R. Montenegro, H. Shan Koh, M.S. Shoichet, Chitin-based tubes for tissue engineering in the nervous system., Biomaterials, 26 (2005) 4624-4632.
[73] E. Babaliari, P. Kavatzikidou, D. Angelaki, L. Chaniotaki, A. Manousaki, A. Siakouli-Galanopoulou, A. Ranella, E. Stratakis, Engineering Cell Adhesion and Orientation via Ultrafast Laser Fabricated Microstructured Substrates, International Journal of Molecular Sciences, 19 (2018).
[74] J. Leijten, L.S. Moreira Teixeira, J. Bolander, W. Ji, B. Vanspauwen, J. Lammertyn, J. Schrooten, F.P. Luyten, Bioinspired seeding of biomaterials using three dimensional microtissues induces chondrogenic stem cell differentiation and cartilage formation under growth factor free conditions, Scientific Reports, 6 (2016).
[75] S. Yang, K.-F. Leong, Z.M.E. Du, C.-K. Chua, The Design of Scaffolds for Use in Tissue Engineering. Part I. Traditional Factors, TISSUE ENGINEERING, Mary Ann Liebert, Inc, 2001.
[76] D.H. Reneker, A.L. Yarin, Electrospinning jets and polymer nanofibers, Polymer, 49 (2008) 2387-2425.
[77] J.M. Deitzel, J. Kleinmeyer, D. Harris, N.C.B. Tan, The effect of processing variables on the morphology of electrospun nanofibers and textiles, Polymer, 42 (2001) 261-272.
[78] F. Yang, R. Murugan, S. Wang, S. Ramakrishna, Electrospinning of nano/micro scale poly(L-lactic acid) aligned fibers and their potential in neural tissue engineering, Biomaterials, 26 (2005) 2603-2610.
[79] P. Eaton, P. Quaresma, C. Soares, C. Neves, M.P. de Almeida, E. Pereira, P. West, A direct comparison of experimental methods to measure dimensions of synthetic nanoparticles, Ultramicroscopy, 182 (2017) 179-190.
[80] A. Taheri, S.M. Jafari, Nanostructures of gums for encapsulation of food ingredients, (2019) 521-578.
[81] S. Park, K. Park, H. Yoon, J. Son, T. Min, G. Kim, Apparatus for preparing electrospun nanofibers: designing an electrospinning process for nanofiber fabrication, Polymer International, 56 (2007) 1361-1366.
[82] J. Xue, J. Xie, W. Liu, Y. Xia, Electrospun Nanofibers: New Concepts, Materials, and Applications, Acc Chem Res, 50 (2017) 1976-1987.
[83] X.M. Shi, W.P. Zhou, D.L. Ma, Q. Ma, D. Bridges, Y. Ma, A.M. Hu, Electrospinning of Nanofibers and Their Applications for Energy Devices, Journal of Nanomaterials, (2015).
[84] J.S. Kim, D.H. Reneker, Polybenzimidazole nanofiber produced by electrospinning, Polymer Engineering and Science, 39 (1999) 849-854.

[85] J. Kameoka, H.G. Craighead, Fabrication of oriented polymeric nanofibers on planar surfaces by electrospinning, Applied Physics Letters, 83 (2003) 371-373.
[86] C.Y. Xu, R. Inai, M. Kotaki, S. Ramakrishna, Aligned biodegradable nanofibrous structure: a potential scaffold for blood vessel engineering, Biomaterials, 25 (2004) 877-886.
[87] D. Yang, B. Lu, Y. Zhao, X. Jiang, Fabrication of Aligned Fibrous Arrays by Magnetic Electrospinning, Adv Mater, 19 (2007) 3702-3706.
[88] D. Li, Y.L. Wang, Y.N. Xia, Electrospinning nanofibers as uniaxially aligned arrays and layer-by-layer stacked films, Adv Mater, 16 (2004) 361-366.
[89] L.D. Sanchez, N. Brack, A. Postma, P.J. Pigram, L. Meagher, Surface modification of electrospun fibres for biomedical applications: A focus on radical polymerization methods, Biomaterials, 106 (2016) 24-45.
[90] H. Ni, W. Kamimura, Y. Uchida, Fabrication of nanofibrous inhaler device, Inhal Toxicol, 19 Suppl 1 (2007) 251-254.
[91] A. Jaworek, A. Krupa, M. Lackowski, A.T. Sobczyk, T. Czech, S. Ramakrishna, S. Sundarrajan, D. Pliszka, Nanocomposite fabric formation by electrospinning and electrospraying technologies, J Electrostat, 67 (2009) 435-438.
[92] S. Miar, A. Shafiee, T. Guda, R. Narayan, Additive Manufacturing for Tissue Engineering, in: A. Ovsianikov, J. Yoo, V. Mironov (Eds.) 3D Printing and Biofabrication, Springer International Publishing, Cham, 2018, pp. 1-52.
[93] F.P.W. Melchels, J. Feijen, D.W. Grijpma, A review on stereolithography and its applications in biomedical engineering, Biomaterials, 31 (2010) 6121-6130.
[94] K.S. Lim, R. Levato, P.F. Costa, M.D. Castilho, C.R. Alcala-Orozco, K.M.A. van Dorenmalen, F.P.W. Melchels, D. Gawlitta, G.J. Hooper, J. Malda, T.B.F. Woodfield, Bio-resin for high resolution lithography-based biofabrication of complex cell-laden constructs, Biofabrication, 10 (2018) 034101.
[95] S.H. Kim, Y.K. Yeon, J.M. Lee, J.R. Chao, Y.J. Lee, Y.B. Seo, M.T. Sultan, O.J. Lee, J.S. Lee, S.I. Yoon, I.S. Hong, G. Khang, S.J. Lee, J.J. Yoo, C.H. Park, Precisely printable and biocompatible silk fibroin bioink for digital light processing 3D printing, Nat Commun, 9 (2018) 1620.
[96] B. Grigoryan, S.J. Paulsen, D.C. Corbett, D.W. Sazer, C.L. Fortin, A.J. Zaita, P.T. Greenfield, N.J. Calafat, J.P. Gounley, A.H. Ta, F. Johansson, A. Randles, J.E. Rosenkrantz, J.D. Louis-Rosenberg, P.A. Galie, K.R. Stevens, J.S. Miller, Multivascular networks and functional intravascular topologies within biocompatible hydrogels, Science, 364 (2019) 458-464.
[97] K. Terzaki, M. Farsari, Polymer Processing Through Multiphoton Absorption, in: J. Van Hoorick, H. Ottevaere, H. Thienpont, P. Dubruel, S. Van Vlierberghe (Eds.) Polymer and Photonic Materials Towards Biomedical Breakthroughs, Springer International Publishing, Cham, 2018, pp. 49-69.
[98] S. Psycharakis, A. Tosca, V. Melissinaki, A. Giakoumaki, A. Ranella, Tailor-made three-dimensional hybrid scaffolds for cell cultures, Biomed Mater, 6 (2011).
[99] K. Parkatzidis, E. Kabouraki, A. Selimis, M. Kaliva, A. Ranella, M. Farsari, M. Vamvakaki, Initiator-Free, Multiphoton Polymerization of Gelatin Methacrylamide, Macromol Mater Eng, 303 (2018).
[100] A. Dobos, J. Van Hoorick, W. Steiger, P. Gruber, M. Markovic, O.G. Andriotis, A. Rohatschek, P. Dubruel, P.J. Thurner, S. Van Vlierberghe, S. Baudis, A. Ovsianikov, Thiol-Gelatin-Norbornene Bioink for Laser-Based High-Definition Bioprinting, Adv Healthc Mater, (2019) e1900752.
[101] J. Torgersen, X.H. Qin, Z.Q. Li, A. Ovsianikov, R. Liska, J. Stampfl, Hydrogels for Two-Photon Polymerization: A Toolbox for Mimicking the Extracellular Matrix, Adv Funct Mater, 23 (2013) 4542-4554.
[102] S.T. You, J.W. Li, W. Zhu, C. Yu, D.Q. Mei, S.C. Chen, Nanoscale 3D printing of hydrogels for cellular tissue engineering, J Mater Chem B, 6 (2018) 2187-2197.
[103] M. Molitch-Hou, Overview of additive manufacturing process, (2018) 1-38.
[104] C. Gayer, J. Ritter, M. Bullemer, S. Grom, L. Jauer, W. Meiners, A. Pfister, F. Reinauer, M. Vucak, K. Wissenbach, H. Fischer, R. Poprawe, J.H. Schleifenbaum, Development of a solvent-free polylactide/calcium carbonate composite for selective laser sintering of bone tissue engineering scaffolds, Mat Sci Eng C-Mater, 101 (2019) 660-673.

[105] S.H.Masood, 10.04 - Advances in Fused Deposition Modeling, Comprehensive Materials Processing, 10 (2014) 69-91.
[106] K.S. Boparai, R. Singh, Advances in Fused Deposition Modeling,  Reference Module in Materials Science and Materials Engineering, Elsevier2017.
[107] N. Ashammakhi, S. Ahadian, C. Xu, H. Montazerian, H. Ko, R. Nasiri, N. Barros, A. Khademhosseini, Bioinks and bioprinting technologies to make heterogeneous and biomimetic tissue constructs, Materials Today Bio, 1 (2019) 100008.
[108] T. Jiang, J.G. Munguia-Lopez, S. Flores-Torres, J. Kort-Mascort, J.M. Kinsella, Extrusion bioprinting of soft materials: An emerging technique for biological model fabrication, Applied Physics Reviews, 6 (2019).
[109] V. Chan, P. Zorlutuna, J.H. Jeong, H. Kong, R. Bashir, Three-dimensional photopatterning of hydrogels using stereolithography for long-term cell encapsulation, Lab Chip, 10 (2010) 2062-2070.
[110] O. Kérourédan, M. Rémy, H. Oliveira, F. Guillemot, R. Devillard, Laser-Assisted Bioprinting of Cells for Tissue Engineering,  Laser Printing of Functional Materials2018, pp. 349-373.
[111] V. Dinca, A. Ranella, M. Farsari, D. Kafetzopoulos, M. Dinescu, A. Popescu, C. Fotakis, Quantification of the activity of biomolecules in microarrays obtained by direct laser transfer, Biomedical Microdevices, 10 (2008) 719-725.
[112] V. Dinca, E. Kasotakis, J. Catherine, A. Mourka, A. Mitraki, A. Popescu, M. Dinescu, M. Farsari, C. Fotakis, Development of peptide-based patterns by laser transfer, Appl Surf Sci, 254 (2007) 1160-1163.
[113] M. Colina, P. Serra, J.M. Fernandez-Pradas, L. Sevilla, J.L. Morenza, DNA deposition through laser induced forward transfer, Biosensors & Bioelectronics, 20 (2005) 1638-1642.
[114] Y.K. Choi, K.W. Kim, Blood-neural barrier: its diversity and coordinated cell-to-cell communication, Bmb Rep, 41 (2008) 345-352.
[115] J.W. McDonald, V. Belegu, D. Becker, Spinal Cord, (2014) 1353-1373.
[116] C.A. Kuliasha, B. Spearman, E.W. Atkinson, A. Furniturewalla, P. Rustogi, S. Mobini, E.B. Nunamaker, A. Brennan, K. Otto, C. Schmidt, J.W. Judy, Robust and Scalable Tissue-Engineerined Electronic Nerve Interfaces (Teeni), (2018) 46-49.
[117] I.V. Yannas, D.P. Orgill, J. Silver, T.V. Norregaard, N.T. Zervas, W.C. Schoene, Regeneration of Sciatic Nerve Across 15mm Gap by Use of a Polymeric Template, (1987) 1-9.
[118] C. Miller, H. Shanks, A. Witt, G. Rutkowski, S. Mallapragada, Oriented Schwann cell growth on micropatterned biodegradable polymer substrates., Biomaterials, 22 (2001) 1263-1269.
[119] V. Chiono, G. Vozzi, F. Vozzi, C. Salvadori, F. Dini, F. Carlucci, M. Arispici, S. Burchielli, F. Di Scipio, S. Geuna, M. Fornaro, P. Tos, S. Nicolino, C. Audisio, I. Perroteau, A. Chiaravalloti, C. Domenici, P. Giusti, G. Ciardelli, Melt-extruded guides for peripheral nerve regeneration. Part I: poly(epsilon-caprolactone). Biomedical microdevices, 11 (2009) 1037-1050.
[120] Y. Katayama, R. Montenegro, T. Freier, R. Midha, J.S. Belkas, M.S. Shoichet, Coil-reinforced hydrogel tubes promote nerve regeneration equivalent to that of nerve autografts., Biomaterials, 27 (2006) 505-518.
[121] J.S. Belkas, C.A. Munro, M.S. Shoichet, R. Midha, Peripheral nerve regeneration through a synthetic hydrogel nerve tube, Restorative Neurology and Neuroscience, 23 (2005) 19-29.
[122] P.D. Dalton, M.S. Shoichet, Creating porous tubes by centrifugal forces for soft tissue application,  Biomaterials, 2001, pp. 2661-2669.
[123] S.-h. Hsu, H.-C. Ni, Fabrication of the Microgrooved/Microporous Polylactide Substrates as Peripheral Nerve Conduits and <i>In Vivo</i> Evaluation, Tissue Engineering Part A, 15 (2009) 1381-1390.
[124] T.B. Bini, S.J. Gao, T.C. Tan, S. Wang, A. Lim, L.B. Hai, S. Ramakrishna, Electrospun poly(L-lactide-co-glycolide) biodegradable polymer nanofibre tubes for peripheral nerve regeneration, Nanotechnology, 15 (2004) 1459-1464.

[125] S. Panseri, C. Cunha, J. Lowery, U. Del Carro, F. Taraballi, S. Amadio, A. Vescovi, F. Gelain, Electrospun micro- and nanofiber tubes for functional nervous regeneration in sciatic nerve transections, Bmc Biotechnology, 8 (2008).
[126] S.Y. Chew, R.F. Mi, A. Hoke, K.W. Leong, Aligned protein-polymer composite fibers enhance nerve regeneration: A potential tissue-engineering platform, Adv Funct Mater, 17 (2007) 1288-1296.
[127] S. Madduri, M. Papaloizos, B. Gander, Trophically and topographically functionalized silk fibroin nerve conduits for guided peripheral nerve regeneration, Biomaterials, 31 (2010) 2323-2334.
[128] U. Assmann, A. Szentivanyi, Y. Stark, T. Scheper, S. Berski, G. Drager, R.H. Schuster, Fiber scaffolds of polysialic acid via electrospinning for peripheral nerve regeneration, Journal of Materials Science-Materials in Medicine, 21 (2010) 2115-2124.
[129] B.S. Jha, R.J. Colello, J.R. Bowman, S.A. Sell, K.D. Lee, J.W. Bigbee, G.L. Bowlin, W.N. Chow, B.E. Mathern, D.G. Simpson, Two pole air gap electrospinning: Fabrication of highly aligned, three-dimensional scaffolds for nerve reconstruction, Acta Biomaterialia, 7 (2011) 203-215.
[130] E.M. Jeffries, Y.D. Wang, Incorporation of parallel electrospun fibers for improved topographical guidance in 3D nerve guides, Biofabrication, 5 (2013).
[131] M.P. Prabhakaran, E. Vatankhah, S. Ramakrishna, Electrospun Aligned PHBV/Collagen Nanofibers as Substrates for Nerve Tissue Engineering, Biotechnology and Bioengineering, 110 (2013) 2775-2784.
[132] R.J. McMurtrey, Patterned and functionalized nanofiber scaffolds in three-dimensional hydrogel constructs enhance neurite outgrowth and directional control, Journal of Neural Engineering, 11 (2014).
[133] J. Hu, D. Kai, H.Y. Ye, L.L. Tian, X. Ding, S. Ramakrishna, X.J. Loh, Electrospinning of poly(glycerol sebacate)-based nanofibers for nerve tissue engineering, Mat Sci Eng C-Mater, 70 (2017) 1089-1094.
[134] W. Jing, Q. Ao, L. Wang, Z.R. Huang, Q. Cai, G.Q. Chen, X.P. Yang, W.H. Zhong, Constructing conductive conduit with conductive fibrous infilling for peripheral nerve regeneration, Chemical Engineering Journal, 345 (2018) 566-577.
[135] H.K. Frost, T. Andersson, S. Johansson, U. Englund-Johansson, P. Ekstrom, L.B. Dahlin, F. Johansson, Electrospun nerve guide conduits have the potential to bridge peripheral nerve injuries in vivo, Scientific Reports, 8 (2018).
[136] J.I. Kim, T.I. Hwang, L.E. Aguilar, C.H. Park, C.S. Kim, A Controlled Design of Aligned and Random Nanofibers for 3D Bi-functionalized Nerve Conduits Fabricated via a Novel Electrospinning Set-up, Scientific Reports, 6 (2016).
[137] K.H. Zhang, C.Y. Wang, C.Y. Fan, X.M. Mo, Aligned SF/P(LLA-CL)-blended nanofibers encapsulating nerve growth factor for peripheral nerve regeneration, Journal of Biomedical Materials Research Part A, 102 (2014) 2680-2691.
[138] T. Wu, D.D. Li, Y.F. Wang, B.B. Sun, D.W. Li, Y. Morsi, H. El-Hamshary, S.S. Al-Deyab, X.M. Mo, Laminin-coated nerve guidance conduits based on poly(L-lactide-co-glycolide) fibers and yarns for promoting Schwann cells' proliferation and migration, J Mater Chem B, 5 (2017) 3186-3194.
[139] Y. Su, X.Q. Li, L.J. Tan, C. Huang, X.M. Mo, Poly(L-lactide-co-epsilon-caprolactone) electrospun nanofibers for encapsulating and sustained releasing proteins, Polymer, 50 (2009) 4212-4219.
[140] B.B. Sun, T. Wu, L.P. He, J.G. Zhang, Y.Y. Yuan, X.J. Huang, H. EI-Hamshary, S.S. Al-Deyab, T. Xu, X.M. Mo, Development of Dual Neurotrophins-Encapsulated Electrospun Nanofibrous Scaffolds for Peripheral Nerve Regeneration, Journal of Biomedical Nanotechnology, 12 (2016) 1987-2000.
[141] M.A. Bhutto, T. Wu, B.B. Sun, H. EI-Hamshary, S.S. Al-Deyab, X.M. Mo, Fabrication and characterization of vitamin B5 loaded poly (L-lactide-co-caprolactone)/silk fiber aligned electrospun nanofibers for schwann cell proliferation, Colloids and Surfaces B-Biointerfaces, 144 (2016) 108-117.
[142] C.M. Valmikinathan, J.J. Tian, J.P. Wang, X.J. Yu, Novel nanofibrous spiral scaffolds for neural tissue engineering, Journal of Neural Engineering, 5 (2008) 422-432.
[143] A. Yamada, F. Niikura, K. Ikuta, A three-dimensional microfabrication system for biodegradable polymers with high resolution and biocompatibility, Journal of Micromechanics and Microengineering, 18 (2008).

[144] T.K. Cui, Y.N. Yan, R.J. Zhang, L. Liu, W. Xu, X.H. Wang, Rapid Prototyping of a Double-Layer Polyurethane-Collagen Conduit for Peripheral Nerve Regeneration, Tissue Engineering Part C-Methods, 15 (2009) 1-9.
[145] D. Radulescu, S. Dhar, C.M. Young, D.W. Taylor, H.J. Trost, D.J. Hayes, G.R. Evans, Tissue engineering scaffolds for nerve regeneration manufactured by ink-jet technology, Materials Science & Engineering C-Biomimetic and Supramolecular Systems, 27 (2007) 534-539.
[146] Y. Hu, Y. Wu, Z.Y. Gou, J. Tao, J.M. Zhang, Q.Q. Liu, T.Y. Kang, S. Jiang, S.Q. Huang, J.K. He, S.C. Chen, Y.A. Du, M.L. Gou, 3D-engineering of Cellularized Conduits for Peripheral Nerve Regeneration, Scientific Reports, 6 (2016).
[147] S. Vijayavenkataraman, S. Thaharah, S. Zhang, W.F. Lu, J.Y.H. Fuh, Electrohydrodynamic jet 3D-printed PCL/PAA conductive scaffolds with tunable biodegradability as nerve guide conduits (NGCs) for peripheral nerve injury repair, Materials & Design, 162 (2019) 171-184.
[148] C.M. Owens, F. Marga, G. Forgacs, C.M. Heesch, Biofabrication and testing of a fully cellular nerve graft, Biofabrication, 5 (2013).
[149] L.Q. Ning, H.Y. Sun, T. Lelong, R. Guilloteau, N. Zhu, D.J. Schreyer, X.B. Chen, 3D bioprinting of scaffolds with living Schwann cells for potential nerve tissue engineering applications, Biofabrication, 10 (2018).
[150] B.N. Johnson, K.Z. Lancaster, G.H. Zhen, J.Y. He, M.K. Gupta, Y.L. Kong, E.A. Engel, K.D. Krick, A. Ju, F.B. Meng, L.W. Enquist, X.F. Jia, M.C. McAlpine, 3D Printed Anatomical Nerve Regeneration Pathways, Adv Funct Mater, 25 (2015) 6205-6217.
[151] M.S. Evangelista, M. Perez, A.A. Salibian, J.M. Hassan, S. Darcy, K.Z. Paydar, R.B. Wicker, K. Arcaute, B.K. Mann, G.R.D. Evans, Single-Lumen and Multi-Lumen Poly(Ethylene Glycol) Nerve Conduits Fabricated by Stereolithography for Peripheral Nerve Regeneration In Vivo, Journal of Reconstructive Microsurgery, 31 (2015) 327-335.
[152] H. Yurie, R. Ikeguchi, T. Aoyama, Y. Kaizawa, J. Tajino, A. Ito, S. Ohta, H. Oda, H. Takeuchi, S. Akieda, M. Tsuji, K. Nakayama, S. Matsuda, The efficacy of a scaffold-free Bio 3D conduit developed from human fibroblasts on peripheral nerve regeneration in a rat sciatic nerve model, Plos One, 12 (2017).
[153] C.J. Pateman, A.J. Harding, A. Glen, C.S. Taylor, C.R. Christmas, P.P. Robinson, S. Rimmer, F.M. Boissonade, F. Claeyssens, J.W. Haycock, Nerve guides manufactured from photocurable polymers to aid peripheral nerve repair, Biomaterials, 49 (2015) 77-89.
[154] Z. Wei, B.T. Harris, L.G. Zhang, Gelatin methacrylamide hydrogel with graphene nanoplatelets for neural cell-laden 3D bioprinting, Conf Proc IEEE Eng Med Biol Soc, 2016 (2016) 4185-4188.
[155] N.I. Moldovan, N. Hibino, K. Nakayama, Principles of the Kenzan Method for Robotic Cell Spheroid-Based Three-Dimensional Bioprinting<sup/>, Tissue Eng Part B Rev, 23 (2017) 237-244.
[156] P. Chrzaszcz, K. Derbisz, K. Suszynski, J. Miodonski, R. Trybulski, J. Lewin-Kowalik, W. Marcol, Application of peripheral nerve conduits in clinical practice: A literature review, Neurologia I Neurochirurgia Polska, 52 (2018) 427-435.
[157] S. Kehoe, X.F. Zhang, D. Boyd, FDA approved guidance conduits and wraps for peripheral nerve injury: A review of materials and efficacy, Injury-International Journal of the Care of the Injured, 43 (2012) 553-572.
[158] E. Cattaneo, C. Zuccato, M. Tartari, Normal huntingtin function: an alternative approach to Huntington's disease, Nature Reviews Neuroscience, 6 (2005) 919-930.
[159] F.L. Maclean, M.K. Horne, R.J. Williams, D.R. Nisbet, Review: Biomaterial systems to resolve brain inflammation after traumatic injury., APL bioengineering, 2 (2018) 021502.
[160] T. Xu, P. Molnar, C. Gregory, M. Das, T. Boland, J.J. Hickman, Electrophysiological characterization of embryonic hippocampal neurons cultured in a 3D collagen hydrogel., Biomaterials, 30 (2009) 4377-4383.
[161] A. Odawara, M. Gotoh, I. Suzuki, A three-dimensional neuronal culture technique that controls the direction of neurite elongation and the position of soma to mimic the layered structure of the brain, RSC Advances, 3 (2013) 23620-23630.

[162] C.E. Semino, J. Kasahara, Y. Hayashi, S. Zhang, Entrapment of migrating hippocampal neural cells in three-dimensional peptide nanofiber scaffold., Tissue engineering, 10 (2004) 643-655.
[163] R.G. Ellis-Behnke, Y.-X. Liang, S.-W. You, D.K.C. Tay, S. Zhang, K.-F. So, G.E. Schneider, Nano neuro knitting: peptide nanofiber scaffold for brain repair and axon regeneration with functional return of vision., Proceedings of the National Academy of Sciences of the United States of America, 103 (2006) 5054-5059.
[164] J.B. Recknor, J.C. Recknor, D.S. Sakaguchi, S.K. Mallapragada, Oriented astroglial cell growth on micropatterned polystyrene substrates, Biomaterials, 25 (2004) 2753-2767.
[165] J.B. Recknor, D.S. Sakaguchi, S.K. Mallapragada, Directed growth and selective differentiation of neural progenitor cells on micropatterned polymer substrates, Biomaterials, 27 (2006) 4098-4108.
[166] Q. Ao, A. Wang, W. Cao, L. Zhang, L. Kong, Q. He, Y. Gong, X. Zhang, Manufacture of multimicrotubule chitosan nerve conduits with novel molds and characterizationin vitro, Journal of Biomedical Materials Research Part A, 77A (2006) 11-18.
[167] S. Möllers, I. Heschel, L.H.H.O. Damink, F. Schügner, R. Deumens, B. Müller, A. Bozkurt, J.G. Nava, J. Noth, G.A. Brook, Cytocompatibility of a novel, longitudinally microstructured collagen scaffold intended for nerve tissue repair., Tissue engineering. Part A, 15 (2009) 461-472.
[168] F. Yang, C.Y. Xu, M. Kotaki, S. Wang, S. Ramakrishna, Characterization of neural stem cells on electrospun poly(L-lactic acid) nanofibrous scaffold, Journal of Biomaterials Science-Polymer Edition, 15 (2004) 1483-1497.
[169] L. Nikkola, J. Seppala, A. Harlin, A. Ndreu, N. Ashammakhi, Electrospun multifunctional diclofenac sodium releasing nanoscaffold, Journal of Nanoscience and Nanotechnology, 6 (2006) 3290-3295.
[170] L. Ghasemi-Mobarakeh, M.P. Prabhakaran, M. Morshed, M.H. Nasr-Esfahani, S. Ramakrishna, Electrospun poly(epsilon-caprolactone)/gelatin nanofibrous scaffolds for nerve tissue engineering, Biomaterials, 29 (2008) 4532-4539.
[171] M.K. Horne, D.R. Nisbet, J.S. Forsythe, C.L. Parish, Three-Dimensional Nanofibrous Scaffolds Incorporating Immobilized BDNF Promote Proliferation and Differentiation of Cortical Neural Stem Cells, Stem Cells and Development, 19 (2010) 843-852.
[172] L.M. He, S.S. Liao, D.P. Quan, K. Ma, C. Chan, S. Ramakrishna, J.A. Lu, Synergistic effects of electrospun PLLA fiber dimension and pattern on neonatal mouse cerebellum C17.2 stem cells, Acta Biomaterialia, 6 (2010) 2960-2969.
[173] D. Fon, K. Zhou, F. Ercole, F. Fehr, S. Marchesan, M.R. Minter, P.J. Crack, D.I. Finkelstein, J.S. Forsythe, Nanofibrous scaffolds releasing a small molecule BDNF-mimetic for the re-direction of endogenous neuroblast migration in the brain, Biomaterials, 35 (2014) 2692-2712.
[174] Z. Alvarez, O. Castano, A.A. Castells, M.A. Mateos-Timoneda, J.A. Planell, E. Engel, S. Alcantara, Neurogenesis and vascularization of the damaged brain using a lactate-releasing biomimetic scaffold, Biomaterials, 35 (2014) 4769-4781.
[175] C. Zanden, N.H. Erkenstam, T. Padel, J. Wittgenstein, J. Liu, H.G. Kuhn, Stem cell responses to plasma surface modified electrospun polyurethane scaffolds, Nanomedicine-Nanotechnology Biology and Medicine, 10 (2014) 949-958.
[176] C.J. Rivet, K. Zhou, R.J. Gilbert, D.I. Finkelstein, J.S. Forsythe, Cell infiltration into a 3D electrospun fiber and hydrogel hybrid scaffold implanted in the brain, Biomatter, 5 (2015) e1005527.
[177] A. Jakobsson, M. Ottosson, M.C. Zalis, D. O'Carroll, U.E. Johansson, F. Johansson, Three-dimensional functional human neuronal networks in uncompressed low-density electrospun fiber scaffolds, Nanomedicine, 13 (2017) 1563-1573.
[178] T. Zhang, Y.N. Yan, X.H. Wang, Z. Xiong, F. Lin, R.D. Wu, R.J. Zhang, Three-dimensional gelatin and gelatin/hyaluronan hydrogel structures for traumatic brain injury, Journal of Bioactive and Compatible Polymers, 22 (2007) 19-29.
[179] F.Y. Hsieh, H.H. Lin, S.H. Hsu, 3D bioprinting of neural stem cell-laden thermoresponsive biodegradable polyurethane hydrogel and potential in central nervous system repair, Biomaterials, 71 (2015) 48-57.

[180] Q. Gu, E. Tomaskovic-Crook, G.G. Wallace, J.M. Crook, 3D Bioprinting Human Induced Pluripotent Stem Cell Constructs for In Situ Cell Proliferation and Successive Multilineage Differentiation, Adv Healthc Mater, 6 (2017).
[181] S.J. Lee, M. Nowicki, B. Harris, L.G. Zhang, Fabrication of a Highly Aligned Neural Scaffold via a Table Top Stereolithography 3D Printing and Electrospinning, Tissue Engineering Part A, 23 (2017) 491-+.
[182] Q. Gu, E. Tomaskovic-Crook, R. Lozano, Y. Chen, R.M. Kapsa, Q. Zhou, G.G. Wallace, J.M. Crook, Functional 3D Neural Mini-Tissues from Printed Gel-Based Bioink and Human Neural Stem Cells, Adv Healthc Mater, 5 (2016) 1429-1438.
[183] V. Melissinaki, A.A. Gill, I. Ortega, M. Vamvakaki, A. Ranella, J.W. Haycock, C. Fotakis, M. Farsari, F. Claeyssens, Direct laser writing of 3D scaffolds for neural tissue engineering applications, Biofabrication, 3 (2011).
[184] H.Y. Yu, Q.M. Zhang, M. Gu, Three-dimensional direct laser writing of biomimetic neuron structures, Optics Express, 26 (2018) 32111-32117.
[185] G.M. Smith, A.E. Falone, E. Frank, Sensory axon regeneration: rebuilding functional connections in the spinal cord, Trends Neurosci, 35 (2012) 156-163.
[186] G.A. Silva, C. Czeisler, K.L. Niece, E. Beniash, D.A. Harrington, J.A. Kessler, S.I. Stupp, Selective Differentiation of Neural Progenitor Cells by High-Epitope Density Nanofibers, Science, 303 (2004) 1352-1355.
[187] V.M. Tysseling-Mattiace, V. Sahni, K.L. Niece, D. Birch, C. Czeisler, M.G. Fehlings, S.I. Stupp, J.A. Kessler, Self-assembling nanofibers inhibit glial scar formation and promote axon elongation after spinal cord injury., The Journal of neuroscience : the official journal of the Society for Neuroscience, 28 (2008) 3814-3823.
[188] Y. Yang, L. De Laporte, C.B. Rives, J.-H. Jang, W.-C. Lin, K.R. Shull, L.D. Shea, Neurotrophin releasing single and multiple lumen nerve conduits, Journal of Controlled Release, 104 (2005) 433-446.
[189] J.B. Scott, M. Afshari, R. Kotek, J.M. Saul, The promotion of axon extension in vitro using polymer-templated fibrin scaffolds., Biomaterials, 32 (2011) 4830-4839.
[190] C. Simitzi, P. Efstathopoulos, A. Kourgiantaki, A. Ranella, I. Charalampopoulos, C. Fotakis, I. Athanassakis, E. Stratakis, A. Gravanis, Laser fabricated discontinuous anisotropic microconical substrates as a new model scaffold to control the directionality of neuronal network outgrowth, Biomaterials, 67 (2015) 115-128.
[191] E.M. Horn, M. Beaumont, X.Z. Shu, A. Harvey, G.D. Prestwich, K.M. Horn, A.R. Gibson, M.C. Preul, A. Panitch, Influence of cross-linked hyaluronic acid hydrogels on neurite outgrowth and recovery from spinal cord injury., Journal of neurosurgery. Spine, 6 (2007) 133-140.
[192] S.V. Kushchayev, M.B. Giers, D. Hom Eng, N.L. Martirosyan, J.M. Eschbacher, M.M. Mortazavi, N. Theodore, A. Panitch, M.C. Preul, Hyaluronic acid scaffold has a neuroprotective effect in hemisection spinal cord injury., Journal of neurosurgery. Spine, 25 (2016) 114-124.
[193] D.Y. Wong, J.-C. Leveque, H. Brumblay, P.H. Krebsbach, S.J. Hollister, F. LaMarca, Macro-Architectures in Spinal Cord Scaffold Implants Influence Regeneration, Journal of Neurotrauma, 25 (2008) 1027-1037.
[194] L. He, Y. Zhang, C. Zeng, M. Ngiam, S. Liao, D. Quan, Y. Zeng, J. Lu, S. Ramakrishna, Manufacture of PLGA Multiple-Channel Conduits with Precise Hierarchical Pore Architectures and <i>In Vitro/Vivo</i> Evaluation for Spinal Cord Injury, Tissue Engineering Part C: Methods, 15 (2009) 243-255.
[195] J.M. Corey, D.Y. Lin, K.B. Mycek, Q. Chen, S. Samuel, E.L. Feldman, D.C. Martin, Aligned electrospun nanofibers specify the direction of dorsal root ganglia neurite growth, J Biomed Mater Res A, 83 (2007) 636-645.
[196] W.N. Chow, D.G. Simpson, J.W. Bigbee, R.J. Colello, Evaluating neuronal and glial growth on electrospun polarized matrices: bridging the gap in percussive spinal cord injuries, Neuron Glia Biol, 3 (2007) 119-126.

[197] E. Schnell, K. Klinkhammer, S. Balzer, G. Brook, D. Klee, P. Dalton, J. Mey, Guidance of glial cell. migration and axonal growth on electrospun nanofibers of poly-epsilon-caprolactone and a collagen/poly-epsilon-caprolactone blend, Biomaterials, 28 (2007) 3012-3025.
[198] J.W. Xie, M.R. MacEwan, S.M. Willerth, X.R. Li, D.W. Moran, S.E. Sakiyama-Elbert, Y.N. Xia, Conductive Core-Sheath Nanofibers and Their Potential Application in Neural Tissue Engineering, Adv Funct Mater, 19 (2009) 2312-2318.
[199] J.W. Xie, M.R. MacEwan, X.R. Li, S.E. Sakiyama-Elbert, Y.N. Xia, Neurite Outgrowth on Nanofiber Scaffolds with Different Orders, Structures, and Surface Properties, Acs Nano, 3 (2009) 1151-1159.
[200] Y.S. Lee, G. Collins, T.L. Arinzeh, Neurite extension of primary neurons on electrospun piezoelectric scaffolds, Acta Biomaterialia, 7 (2011) 3877-3886.
[201] J.L. Bourke, H.A. Coleman, V. Pham, J.S. Forsythe, H.C. Parkington, Neuronal Electrophysiological Function and Control of Neurite Outgrowth on Electrospun Polymer Nanofibers Are Cell Type Dependent, Tissue Engineering Part A, 20 (2014) 1089-1095.
[202] A. Hurtado, J.M. Cregg, H.B. Wang, D.F. Wendell, M. Oudega, R.J. Gilbert, J.W. McDonald, Robust CNS regeneration after complete spinal cord transection using aligned poly-L-lactic acid microfibers, Biomaterials, 32 (2011) 6068-6079.
[203] T. Liu, J.D. Houle, J.Y. Xu, B.P. Chan, S.Y. Chew, Nanofibrous Collagen Nerve Conduits for Spinal Cord Repair, Tissue Engineering Part A, 18 (2012) 1057-1066.
[204] T.L. Downing, A.J. Wang, Z.Q. Yan, Y. Nout, A.L. Lee, M.S. Beattie, J.C. Bresnahan, D.L. Farmer, S. Li, Drug-eluting microfibrous patches for the local delivery of rolipram in spinal cord repair, Journal of Controlled Release, 161 (2012) 910-917.
[205] J. Koffler, W. Zhu, X. Qu, O. Platoshyn, J.N. Dulin, J. Brock, L. Graham, P. Lu, J. Sakamoto, M. Marsala, S.C. Chen, M.H. Tuszynski, Biomimetic 3D-printed scaffolds for spinal cord injury repair, Nature Medicine, 25 (2019) 263-+.
[206] O. Sarig-Nadir, N. Livnat, R. Zajdman, S. Shoham, D. Seliktar, Laser Photoablation of Guidance Microchannels into Hydrogels Directs Cell Growth in Three Dimensions, Biophysical Journal, 96 (2009) 4743-4752.
[207] N. Ashammakhi, H.J. Kim, A. Ehsanipour, R.D. Bierman, O. Kaarela, C. Xue, A. Khademhosseini, S.K. Seidlits, Regenerative Therapies for Spinal Cord Injury, Tissue Eng Part B Rev, 25 (2019) 471-491.
[208] F. Triolo, A.K. Srivastava, Current Approaches to Tissue Engineering of the Nervous System, (2018).
[209] M. Tsintou, K. Dalamagkas, A.M. Seifalian, Advances in regenerative therapies for spinal cord injury: a biomaterials approach, Neural Regen Res, 10 (2015) 726-742.
[210] R. Booth, H. Kim, Characterization of a microfluidic in vitro model of the blood-brain barrier (iBBB) (vol 12, pg 1784, 2012), Lab on a Chip, 12 (2012) 5282-5282.
[211] R. Booth, H. Kim, Permeability Analysis of Neuroactive Drugs Through a Dynamic Microfluidic In Vitro Blood-Brain Barrier Model, Annals of Biomedical Engineering, 42 (2014) 2379-2391.
[212] J. Park, B.K. Lee, G.S. Jeong, J.K. Hyun, C.J. Lee, S.H. Lee, Three-dimensional brain-on-a-chip with an interstitial level of flow and its application as an in vitro model of Alzheimer's disease, Lab on a Chip, 15 (2015) 141-150.
[213] A. Herland, A.D. van der Meer, E.A. FitzGerald, T.E. Park, J.J.F. Sleeboom, D.E. Ingber, Distinct Contributions of Astrocytes and Pericytes to Neuroinflammation Identified in a 3D Human Blood-Brain Barrier on a Chip, Plos One, 11 (2016).
[214] Y.C. Wang, J.Y. Ma, N. Li, L. Wang, L.M. Shen, Y. Sun, Y.J. Wang, J.Y. Zhao, W.J. Wei, Y. Ren, J. Liu, Microfluidic engineering of neural stem cell niches for fate determination, Biomicrofluidics, 11 (2017).
[215] G. Adriani, D.L. Ma, A. Pavesi, R.D. Kamm, E.L.K. Goh, A 3D neurovascular microfluidic model consisting of neurons, astrocytes and cerebral endothelial cells as a blood-brain barrier, Lab on a Chip, 17 (2017) 448-459.

[216] R. van de Wijdeven, O.H. Ramstad, U.S. Bauer, O. Halaas, A. Sandvig, I. Sandvig, Structuring a multi-nodal neural network in vitro within a novel design microfluidic chip, Biomedical Microdevices, 20 (2018).
[217] B.N. Johnson, K.Z. Lancaster, I.B. Hogue, F.B. Meng, Y.L. Kong, L.W. Enquist, M.C. McAlpine, 3D printed nervous system on a chip (vol 16, pg 1393, 2016), Lab on a Chip, 16 (2016) 1946-1946.
[218] H.G. Yi, Y.H. Jeong, Y. Kim, Y.J. Choi, H.E. Moon, S.H. Park, K.S. Kang, M. Bae, J. Jang, H. Youn, S.H. Paek, D.W. Cho, A bioprinted human-glioblastoma-on-a-chip for the identification of patient-specific responses to chemoradiotherapy, Nat Biomed Eng, 3 (2019) 509-519.
[219] I.C. Samper, S.A.N. Gowers, M.L. Rogers, D.R.K. Murray, S.L. Jewell, C. Pahl, A.J. Strong, M.G. Boutelle, 3D printed microfluidic device for online detection of neurochemical changes with high temporal resolution in human brain microdialysate, Lab on a Chip, 19 (2019) 2038-2048.
[220] S.M. Bittner, J.L. Guo, A.G. Mikos, Spatiotemporal Control of Growth Factors in Three-Dimensional Printed Scaffolds, Bioprinting, 12 (2018).
[221] J.F. Mano, I.S. Choi, A. Khademhosseini, Biomimetic Interfaces in Biomedical Devices, Adv Healthc Mater, 6 (2017).
[222] M.H. Kim, M. Park, K. Kang, I.S. Choi, Neurons on nanometric topographies: insights into neuronal behaviors in vitro, Biomater Sci-Uk, 2 (2014) 148-155.